\documentclass[floatfix,a4paper,aps,prd, nofootinbib,twocolumn,preprintnumbers]{revtex4} 

\usepackage{epsfig,ulem}
\usepackage{amsmath}
\usepackage{amssymb}
\usepackage{amsfonts}
\usepackage{axodraw4j}
\usepackage{graphicx}
\usepackage{graphics,subfigure}
\usepackage{pifont}
\usepackage{float}
\usepackage{booktabs}
\usepackage{color}
\usepackage{geometry}
\usepackage{rotating}
\usepackage{psfrag}
\usepackage{bm}
\usepackage{bbm}
\usepackage{multirow}

\newbox\charbox
\newbox\slabox
\def\s#1{{      
 \setbox\charbox=\hbox{$#1$}
 \setbox\slabox=\hbox{$/$}
 \dimen\charbox=\ht\slabox
 \advance\dimen\charbox by -\dp\slabox
 \advance\dimen\charbox by -\ht\charbox
 \advance\dimen\charbox by \dp\charbox
 \divide\dimen\charbox by 2
 \raise-\dimen\charbox\hbox to \wd\charbox{\hss/\hss}
 \llap{$#1$}
}}

\graphicspath{%
{./}%
{figs/}%
}

\newcommand{\newc}{\newcommand}
\newc{\wt}{\widetilde}
\newc{\cL}{{\cal L}}
\newc{\cM}{{\cal M}}
\newc{\ra}{\rightarrow}
\newc{\eps}{\epsilon}
\newc{\bino}{\widetilde{\cal B}}
\newc{\wino}{\widetilde{\cal W}}
\newc{\gluino}{\widetilde{\cal G}}
\newc{\half}{\frac{1}{2}}
\newc{\third}{\frac{1}{3}}
\newc{\fourth}{\frac{1}{4}}
\newc{\eighth}{\frac{1}{8}}
\newc{\gev}{\mbox{~GeV}}
\newc{\lra}{\leftrightarrow}
\newc{\Dslash}{\not\!\! D}
\newc{\sg}{{\cal G}}
\newc{\ovl}{\overline}
\newc{\ok}{$\surd$}
\newc{\etal}{{\it et al.}\ }
\newc{\Hbar}{{\bar H}}
\newc{\hhbar}{{\overline h}}
\newc{\Ubar}{{\bar U}}
\newc{\Dbar}{{\bar D}}
\newc{\Ebar}{{\bar E}}
\newc{\eg}{{\it e.g.}\ }
\newc{\ie}{{\it i.e.}\ }
\newc{\nonum}{\nonumber}
\newc{\kap}{\kappa}
\newc{\Dt}{\frac{d}{dt}}
\newc{\rpv}{{\mbox{${\not\!\!R_p}$}}}
\newc{\bpv}{$\not\!\!B_p$}
\newc{\mpl}{$M_{Pl}$\ }
\newc{\mx}{$M_X$\ }
\newc{\mgut}{M_\mathrm{GUT}}
\newc{\tev}{\mbox{~TeV}}
\newc{\sect}[1]{\ref{sec:#1}}
\newc{\nonr}{\nonum}
\newc{\vev}[1]{\langle{#1}\rangle}
\newc{\eq}[1]{(\ref{eq:#1})}
\newc{\eqs}[2]{(\ref{eq:#1},\ref{eq:#2})}
\newc{\lab}[1]{\label{eq:#1}}
\newc{\Lam}{{\bf \Lambda}}
\newc{\ltau}{\lambda_\tau}
\newc{\lt}{\lambda_t}
\newc{\lb}{\lambda_b}
\newc{\lae}{{\Lam}_E}
\newc{\lad}{{\Lam}_D}
\newc{\lau}{{\Lam}_U}
\newc{\lame}[1]{{\Lam}_{E^{#1}}}
\newc{\lamhe}[1]{{\h}_{E^{#1}}}
\newc{\lamhed}[1]{{\h}_{E^{#1}}^\dagger}
\newc{\lamhd}[1]{{\h}_{D^{#1}}}
\newc{\lamhdd}[1]{{\h}_{D^{#1}}^\dagger}
\newc{\lamhu}[1]{{\h}_{U^{#1}}}
\newc{\lamhud}[1]{{\h}_{U^{#1}}^\dagger}
\newc{\lamd}[1]{{\Lam}_{D^{#1}}}
\newc{\lamu}[1]{{\Lam}_{U^{#1}}}
\newc{\lamet}[1]{{\Lam}_{E^{#1}}^T}
\newc{\lamdt}[1]{{\Lam}_{D^{#1}}^T}
\newc{\lamut}[1]{{\Lam}_{U^{#1}}^T}
\newc{\lames}[1]{{\Lam}_{E^{#1}}^*}
\newc{\lamds}[1]{{\Lam}_{D^{#1}}^*}
\newc{\lamus}[1]{{\Lam}_{U^{#1}}^*}
\newc{\lamed}[1]{{\Lam}_{E^{#1}}^\dagg}
\newc{\lamdd}[1]{{\Lam}_{D^{#1}}^\dagg}
\newc{\lamud}[1]{{\Lam}_{U^{#1}}^\dagg}
\newc{\lam}{{\bf \lambda}}
\newc{\lamp}{{\bf \lambda}^{\prime}}
\newc{\lampp}{{\bf \lambda}^{\prime\prime}}
\newc{\Y}{{\bf Y}}
\newc{\h}{{\bf h}}
\newc{\meee}{{{\rm {\bf  m}}_e}}
\newc{\mdee}{{{\rm {\bf  m}}_d}}
\newc{\myew}{{{\rm {\bf m}}_u}}
\newc{\ye}{{\Y}_E}
\newc{\he}{{\h}_E}
\newc{\hed}{{\h}_E^\dagger}
\newc{\yd}{{\Y}_D}
\newc{\hd}{{\h}_D}
\newc{\hdd}{{\h}_D^\dagger}
\newc{\yu}{{\Y}_U}
\newc{\hu}{{\h}_U}
\newc{\hud}{{\h}_U^\dagger}
\newc{\yes}{{\Y}_E^*}
\newc{\yds}{{\Y}_D^*}
\newc{\yus}{{\Y}_U^*}
\newc{\yet}{{\Y}_E^T}
\newc{\ydt}{{\Y}_D^T}
\newc{\yut}{{\Y}_U^T}
\newc{\yed}{{\Y}_E^\dagg}
\newc{\ydd}{{\Y}_D^\dagg}
\newc{\yud}{{\Y}_U^\dagg}
\newc{\dagg}{\dagger}
\newc{\lp}{\left(}
\newc{\rp}{\right)}
\newc{\inv}{\frac{1}{16\pi^2}}
\newc{\invsq}{\frac{1}{(16\pi^2)^2}}
\newc{\ggam}[2]{\gamma_{#2}^{#1}}
\newc{\yukgam}[2]{\inv \gamma_{#1}^{(1){#2}}+\invsq\gamma_{{#1}}^{(2){#2}}}
\newc{\susyunif}{ohman,nirpaul,marcelacarlos,susyunif}
\newc{\lsim}{\stackrel{<}{\sim}}
\newc{\gsim}{\stackrel{>}{\sim}}
\newc{\Tr}{{~\rm Tr}}
\newc{\me}{{(\bf m_{\tilde{E}}}^2)}
\newc{\mh}[1]{m_{H_{#1}}^2}
\newc{\ml}{{\bf m_{\tilde{L}}}^2}
\newc{\md}{{(\bf m_{\tilde{D}}}^2)}
\newc{\mup}{{(\bf m_{\tilde{U}}}^2)}
\newc{\mq}{{(\bf m_{\tilde{Q}}}^2)}
\newc{\mlh}[1]{{\bf m}_{ \tilde{L}_{#1} H_1}^2}
\newc{\mhl}[1]{{\bf m}_{ H_d \tilde{L}_{#1}}^2}
\newc{\del}{\partial}
\newc{\beq}{\begin{equation}}
\newc{\eeq}{\end{equation}}
\newc{\barr}{\begin{align}}
\newc{\earr}{\end{align}}
\newc{\dspl}{\displaystyle}
\newc{\phmin}{\phantom{-}}
\newc{\stau}{{\tilde\tau}}
\newc{\mnu}{$m_{\nu}$ }
\newc{\AO}{$A_0$ }
\newc{\vd}{$v_d$ }
\newc{\MGUT}{$M_\mathrm{GUT}$}
\newc{\mzero}{M_0}
\newc{\mhalf}{{M_{1/2}}}
\newc{\tanb}{\tan\beta}
\newc{\azero}{A_0}
\newc{\sgnmu}{\textrm{sgn}(\mu)}
\newc{\meff}{M_\mathrm{eff}^\mathrm{vis}}
\def\slashchar#1{\setbox0=\hbox{$#1$}     		
   \dimen0=\wd0                                 	
   \setbox1=\hbox{/} \dimen1=\wd1               	
   \ifdim\dimen0>\dimen1                        	
      \rlap{\hbox to \dimen0{\hfil/\hfil}}      	
      #1                                        	
   \else                                        	
      \rlap{\hbox to \dimen1{\hfil$#1$\hfil}}   	
      /                                         	
   \fi}
\newc{\cf}{\textit{cf.}~}   
\newc{\dzero}{D\O}
\newc{\etmiss}{\slashchar{E}_T}
\newc{\Psix}{{\mathrm{P}_{\!6}}}
\newc{\nPsix}{{\not\!\Psix}}
\newc{\nPsixU}{{\not{\mathrm P}_{\!6}}}
\newc{\Bthree}{{\mathrm{B}_{\,\!3}}}
\newc{\mj}{m_{\tilde{\chi}^0_j}}
\newc{\mk}{m_{\tilde{\chi}^0_k}}
\newc{\slepton}{\tilde \ell}

\makeatletter

\newcommand{\Rmnum}[1]{\expandafter\@slowromancap\romannumeral #1@}
\makeatother

\newc{\squark}{\tilde{q}}
\newc{\ssup}{\tilde{u}}
\newc{\ssdown}{\tilde{d}}
\newc{\ssstrange}{\tilde{s}}
\newc{\sscharm}{\tilde{c}}
\newc{\sstop}{\tilde{t}}
\newc{\ssbottom}{\tilde{b}}
\newc{\sse}{\tilde{e}}
\newc{\ssmu}{\tilde{\mu}}
\newc{\sstau}{\tilde{\tau}}
\newc{\ssnu}{\tilde{\nu}}
\newc{\ssnue}{\tilde{\nu}_{e}}
\newc{\ssnumu}{{\tilde{\nu}_{\mu}}}
\newc{\ssnutau}{{\tilde{\nu}_{\tau}}}
\newc{\ssbnue}{\bar{\tilde{\nu}}_{e}}
\newc{\ssbnumu}{\bar{\tilde{\nu}}_{\mu}}
\newc{\ssbnutau}{\bar{\tilde{\nu}}_{\tau}}
\newc{\neut}{{\tilde{\chi}}^0}
\newc{\charge}{\tilde{\chi}}
\newc{\glu}{\tilde{g}}
\newc{\Higgs}{H^0}

\newc{\nue}{\nu_e}
\newc{\numu}{\nu_{\mu}}
\newc{\nutau}{\nu_{\tau}}
\newc{\bnue}{\bar{\nu}_e}
\newc{\bnumu}{\bar{\nu}_{\mu}}
\newc{\bnutau}{\bar{\nu}_{\tau}}
\newc{\ttext}[1]{{{\color{green}  #1}}  \color{black}}
\newc{\stext}[1]{{{\color{blue}  #1}}  \color{black}}
\newc{\htext}[1]{{{\color{red}  #1}} \color{black}}
\begin{document}

\title{Discovery Potential of Selectron or Smuon as the Lightest Supersymmetric Particle at the LHC}
\author{H.~K.~Dreiner}
\email[]{dreiner@th.physik.uni-bonn.de}
\affiliation{Bethe Center for Theoretical Physics and Physikalisches 
Institut, Universit\"at Bonn, Bonn, Germany}

\author{S.~Grab}
\email[]{sgrab@scipp.ucsc.edu}
\affiliation{SCIPP, University of California Santa Cruz, Santa Cruz, 
CA 95064, USA}

\author{T.~Stefaniak}
\email[]{tim@th.physik.uni-bonn.de}
\affiliation{Bethe Center for Theoretical Physics and Physikalisches 
Institut, Universit\"at Bonn, Bonn, Germany and \Rmnum{2}. Physikalisches Institut, Universit\"at G\"ottingen, G\"ottingen, Germany}

\begin{abstract}
We investigate the LHC discovery potential of $R$-parity violating
supersymmetric models with a right-handed selectron or smuon as the
lightest supersymmetric particle (LSP). These LSPs arise naturally in
$R$-parity violating minimal supergravity models. We classify the
hadron collider signatures and perform for the first time within these
models a detailed signal over background analysis. We develop an
inclusive three-lepton search and give prospects for a discovery at a
center-of-mass energy of $\sqrt{s}=7$ TeV as well as $\sqrt{s}=14$
TeV. There are extensive parameter regions which the LHC can already
test with $\sqrt{s}=7$ TeV and an integrated luminosity of 1
fb$^{-1}$.  We also propose a method for the mass reconstruction of
the supersymmetric particles within our models at $\sqrt{s}=14$~TeV.
\end{abstract}

\preprint{BONN-TH-2011-04, SCIPP 11/01}

\maketitle

\section{Introduction}

Since 2010, the Large Hadron Collider (LHC) is collecting data at a
center of mass energy of $\sqrt{s}=7$~TeV and first searches for
physics beyond the Standard Model (SM) have been published
\cite{Collaboration:2011hh,Collaboration:2010qr,Collaboration:2010eza,
Khachatryan:2010te,atlas:2010bc,Khachatryan:2011ts,Khachatryan:2010fa,
Khachatryan:2010mq,Khachatryan:2010wx,Khachatryan:2010uf,
Khachatryan:2010jd,Collaboration:2011tk}. 
Even with only an integrated luminosity of 35 pb$^{-1}$, the LHC has
already tested supersymmetric models \cite{Haber:1984rc,Martin:1997ns}
beyond the Tevatron searches~\cite{Collaboration:2011tk,Collaboration:2011hh}. 
Furthermore, it is expected that the LHC will collect 1 fb$^{-1}$ of data until the
end of 2011.

One of the most promising LHC signatures for supersymmetry (SUSY) are
multi-lepton final states
\cite{Barbieri:1991vk,Desch:2010gi,Aad:2009wy}. On the one hand,
electrons and muons are easy to identify in the detectors. On the
other hand, the SM background for multi-lepton final states is low. In
this publication, we focus on such signatures.

We consider the supersymmetric extension of the SM with minimal
particle content (SSM) \cite{Haber:1984rc,Martin:1997ns}. Without
further assumptions, the proton usually has a short lifetime in this
model
\cite{Dimopoulos:1981dw,Smirnov:1996bg,Bhattacharyya:1998bx}, in
contradiction with experimental observations \cite{Shiozawa:1998si}.
The proton decays, because renormalizable lepton and baryon number
violating interactions are jointly present. One therefore must
impose an additional discrete symmetry. The most common choice for
this discrete symmetry is $R$-parity, or equivalently at low-energy:
proton-hexality ($\text{P}_6$). Either suppresses all lepton- and
baryon number violating interactions
\cite{Ibanez:1991hv,Ibanez:1991pr,Dreiner:2005rd}. The SSM with 
$R$-parity is usually denoted the minimal supersymmetric SM (MSSM). 

We consider here a different discrete symmetry, baryon-triality
($\text{B}_3$)
\cite{Ibanez:1991hv,Ibanez:1991pr,Dreiner:2005rd,Dreiner:2006xw}, 
which suppresses only the baryon number violating terms, but allows
for lepton number violating interactions.  The $\text{B}_3$ SSM has
the advantage that neutrino masses are generated naturally
\cite{Hall:1983id,Grossman:1998py,Dedes:2006ni,Dreiner:2010ye} without the need to
introduce a new (see-saw) energy scale
\cite{Minkowski:1977sc,Mohapatra:1979ia,seesaw}. 
The lepton number violating interactions can be adjusted, such that
the observed neutrino masses and mixing angles can be explained
\cite{Dreiner:2007uj,Allanach:2007qc}. Note that both P$_6$ and
B$_3$ are discrete gauge anomaly free symmetries
\cite{Ibanez:1991hv,Ibanez:1991pr,Banks:1991xj,Dreiner:2005rd}.

In the $\text{B}_3$ SSM, the lightest supersymmetric particle (LSP)
will decay via the lepton number violating interactions and is thus
not bounded by cosmological observations to be the lightest
neutralino, $\neut_1$ \cite{Ellis:1983ew}. Unlike in the MSSM, the
$\neut_1$ is not a valid dark matter (DM) candidate. However, several
possible DM candidates are easily found in simple extensions of the
$\text{B}_3$ SSM; for example, the axino
\cite{Chun:1999cq,Choi:2001cm,Kim:2001sh,Chun:2006ss}, the gravitino
\cite{Covi:2009pq,Buchmuller:2007ui} or the lightest $U$-parity
particle \cite{Lee:2007qx,Lee:2007fw}.

We consider in this paper the $\text{B}_3$ SSM with a right-handed
scalar electron (selectron, $\sse_R$) or scalar muon (smuon,
$\ssmu_R$) as the LSP. These LSP candidates naturally arise in the
$\text{B}_3$ minimal supergravity (mSUGRA) model
\cite{Allanach:2003eb}, on which we focus in the following. Here, large
lepton number violating interactions at the grand unification (GUT)
scale drive the selectron or smuon mass towards small values at the
electroweak scale via the renormalization group equations (RGEs)
\cite{Dreiner:2008ca}. We describe this effect and the selectron and
smuon LSP parameter space in the next section in more detail. Further
LSP candidates within $\text{B}_3$ mSUGRA (beside the $\neut_1$) are
the lightest stau, $\stau_1$
\cite{Allanach:2003eb,Desch:2010gi,Allanach:2006st}, and the
sneutrino, $\tilde{\nu}_{e,\mu,\tau}$
\cite{Bernhardt:2008jz,Allanach:2003eb}, depending on the dominant 
lepton number violating operator~\cite{Dreiner:2008ca}.

If SUSY exists, the pair production of strongly interacting SUSY
particles (sparticles), like scalar quarks (squarks), is usually the
main source for SUSY events at hadron colliders like the LHC
\cite{Baer:2006rs}.  Furthermore, squarks, $\tilde{q}$, are much
heavier than the $\neut_1$ in most supersymmetric models
\cite{Allanach:2002nj}.  Assuming that we have a right-handed
selectron or smuon, $\slepton_R$, as the LSP, a natural cascade
process at the LHC is
\begin{align}
\tilde{q}\tilde{q} \to qq \neut_1 \neut_1 \to qq \ell \ell \slepton_R 
\slepton_R,
\label{Eqn:intro_cascade}
\end{align}
where the squarks decay into a quark, $q$, and the $\neut_1$. The
$\neut_1$ decays into the $\slepton_R$ LSP and an oppositely
charged lepton, $\ell$, of the same flavor.

The $\slepton_R$ LSP can then decay via the lepton number violating
interactions, for example  
\begin{align}
\slepton_R \to \ell' \nu,
\label{Eqn:intro_LSP_decay}
\end{align}
\textit{i.e.} into another charged lepton $\ell'$ and a neutrino $\nu$.
As we argue in the following, this is the case for large regions of
the $\text{B}_3$ SSM parameter space. We thus obtain from
Eqs.~(\ref{Eqn:intro_cascade}) and (\ref{Eqn:intro_LSP_decay}) an
event with four charged leptons in the final state. Taking into
account that some leptons might not be well identified, we design in
this paper an inclusive three-lepton search for $\slepton_R$--LSP
scenarios. Although we concentrate on the $\text{B}_3$ mSUGRA model,
our results apply also to more general models as long as
Eqs.~(\ref{Eqn:intro_cascade}) and (\ref{Eqn:intro_LSP_decay})
hold. We will show that because of the high lepton multiplicity in
B$_3$ models, the discovery reach at the LHC with $\sqrt{s}=7$ TeV
exceeds searches in the $R$-parity conserving case
\cite{Baer:2010tk}.  We also give prospects for a discovery at
$\sqrt{s}=14$ TeV and propose a method for the reconstruction of
sparticle masses within our model.

The phenomenology of slepton LSPs has mainly been investigated for the
case of a stau LSP. See for example
Refs.~\cite{Allanach:2003eb,Dreiner:2007uj,Dreiner:2008rv,Allanach:2006st,Akeroyd:1997iq,deGouvea:1998yp,Akeroyd:2001pm,Bartl:2003uq,Bernhardt:2008mz,Ghosh:2010tp,Mukhopadhyaya:2010qf,deCampos:2007bn,Hirsch:2002ys,Allanach:2007vi,Desch:2010gi}.
Recently, Ref.~\cite{Desch:2010gi} proposed a tri-lepton search for
stau LSP scenarios, which is similar to our analysis, although the
stau in Ref.~\cite{Desch:2010gi} decays via 4-body decays. LEP II has
searched for slepton LSPs \cite{Heister:2002jc,Abbiendi:2003rn}. No
signals were found and lower mass limits around $90-100$ GeV were
set. Refs.~\cite{Hirsch:2002ys,Bartl:2003uq} investigated the decay
length of slepton LSPs assuming trilinear as well as bilinear
$R$-parity violating interactions. Finally, in
Ref.~\cite{Dreiner:2009fi}, the signature of
Eqs.~(\ref{Eqn:intro_cascade}) and (\ref{Eqn:intro_LSP_decay}) was
pointed out. But in contrast to this work, no signal over background
analysis was performed.

The remainder of this paper is organized as follows. In
Sec.~\ref{Sect:Slep_LSP} we review the $\text{B}_3$ mSUGRA model and
show how a $\slepton_R$ LSP can arise. We present the $\Bthree$ mSUGRA
parameter regions with a $\slepton_R$ LSP and propose a set of
benchmark points for LHC searches. We then classify in
Sec.~\ref{Sect:signatures} the $\slepton_R$ LSP signatures at
hadron colliders as a function of the dominant $R$-parity violating
interaction.  Based on this, we develop in Sec.~\ref{Sect:discovery} a
set of cuts for an inclusive three-lepton search at the LHC and give
prospects for a discovery at $\sqrt{s}=7$ TeV as well as at
$\sqrt{s}=14$ TeV. In Sec.~\ref{Sect:mass_reco} we propose a method
for the reconstruction of the supersymmetric particle masses. We conclude
in Sec.~\ref{Sect:summary}.

Appendix~\ref{App:benchmarks} reviews the mass spectrum and branching
ratios of our benchmark models and Appendix~\ref{App:14TeV} shows the
cutflow for our $\sqrt{s}=14$ TeV analysis.  We give in
Appendix~\ref{App:endpoints} the relevant equations for the kinematic
endpoints for the mass reconstruction of Sec.~\ref{Sect:mass_reco} and
calculate in Appendix~\ref{App:sleptondecays} some missing 3-body
decays of sleptons.

\section{The Selectron and Smuon as the LSP in R-parity Violating mSUGRA}
\label{Sect:Slep_LSP}

\subsection{The $\Bthree$ mSUGRA Model}
\label{Sect:B3model}

In the $\Bthree$ mSUGRA model the boundary conditions 
at the GUT scale ($M_\mathrm{GUT}$) are described by the six parameters
\cite{Allanach:2003eb,Allanach:2006st}
\begin{align}
\mzero,\,\mhalf,\,\azero,\,\tanb,\,\sgnmu,\,\mathbf{\Lambda}. \label{Eqn:B3mSUGRA}
\end{align}
Here, $\mzero$, $\mhalf$ and $\azero$ are the universal scalar mass,
the universal gaugino mass and the universal trilinear scalar
coupling, respectively. $\tanb$ denotes the ratio of the two Higgs
vacuum expectation values (vevs), and $\sgnmu$ fixes the sign of the
bilinear Higgs mixing parameter $\mu$. Its magnitude is derived from
radiative electroweak symmetry breaking \cite{Ibanez:1982fr}.
$\mathbf{\Lambda}$ is described below.

In $\Bthree$ mSUGRA, the superpotential is extended by the
lepton number violating (LNV) terms \cite{Dreiner:1997uz},
\begin{align}
W_\mathrm{LNV} = \frac{1}{2} \lam_{ijk} L_i L_j \bar E_k + 
\lamp_{ijk} L_i Q_j \bar D_k + \kappa_i L_i H_2, 
\label{Eqn:WB3}
\end{align}
which are absent in the MSSM. Here, $L_i$ and $Q_i$ denote the lepton
and quark $SU(2)$ doublet superfields, respectively. $H_2$ is the
Higgs $SU(2)$ doublet superfield which couples to up-type quarks, and
$\bar E_i$ and $\bar D_i$ denote the lepton and down-type quark $SU(2)$ singlet
superfields, respectively. $i,j,k \in \{1,2,3\}$ are generation
indices.  $\lam_{ijk}$ is anti-symmetric in the first two indices ($i
\leftrightarrow j$) and thus denotes nine, $\lamp_{ijk}$ twenty-seven
dimensionless couplings. The bilinear lepton number violating
couplings $\kappa_i$ are three dimensionful parameters, which vanish
in $\Bthree$ mSUGRA at $M_\mathrm{GUT}$ due to a redefinition of the
lepton and Higgs superfields \cite{Allanach:2003eb}. However, they are
generated at lower scales via RGE running with interesting
phenomenological consequences for neutrino masses
\cite{Allanach:2007qc,Dreiner:2010ye}. 

In the $\Bthree$ mSUGRA model, we assume that exactly one of the
thirty-six dimensionless couplings in Eq.~(\ref{Eqn:WB3}) is non-zero
and positive at the GUT scale\footnote{On the one hand, bounds
on products of two different couplings are in general much stronger
than on single couplings \cite{Barbier:2004ez}. On the other hand, one
observes also a large hierarchy between the Yukawa couplings within
the SM.}. The parameter $\mathbf{\Lambda}$ in Eq.~(\ref{Eqn:B3mSUGRA})
refers to this choice, \ie
\begin{align}
\mathbf{\Lambda} \in \{ \lam_{ijk}, \, \lamp_{ijk}\}, \qquad i,j,k = 1,2,3.
\end{align}
Given one coupling at the GUT scale, other couplings that violate
only the same lepton number are generated at the weak scale,
$M_Z$, by the RGEs
\cite{Allanach:2003eb,Dreiner:2008rv,Allanach:1999mh,deCarlos:1996du}.

\subsection{The Selectron and Smuon LSP}
\subsubsection{Renormalization Group Evolution of the $\slepton_R$ Mass}
\label{Sect:RGE}
In order to understand the dependence of the right-handed slepton
\footnote{We consider only the first two generations of sleptons, 
\textit{i.e.} $\slepton_R \in \{\sse_R,\ssmu_R \}$, because 
a stau LSP can also be obtained without (large) $R$-parity violating interactions
\cite{Allanach:2003eb,Allanach:2006st}.},
$\slepton_R$, mass at $M_Z$ on the boundary
conditions at $\mgut$, we have to take a closer look at the relevant
RGEs, which receive additional contributions from the LNV terms in
Eq.~\eqref{Eqn:WB3}. The dominant one-loop contributions to the
running mass of the right-handed slepton of generation $k=1,2$ are
\cite{Allanach:2003eb}
\begin{align}
&16 \pi^2 \frac{d(M_{\slepton^k_R}^2)}{dt}  = - \frac{24}{5}g_1^2|M_1|^2 + 
\frac{6}{5}g_1^2 \mathcal{S} + 2(\mathbf{h_{E^k}})_{ij}^2 \nonumber \\
&\;\;\;\;\;\;+ 4 \lambda_{ijk}^2 \left[ (\mathbf{m_{\tilde L}}^2)_{ii} + 
(\mathbf{m_{\tilde L}}^2)_{jj} + (\mathbf{m_{\tilde E}}^2)_{kk} \right] 
\label{RGE_dmdt}
\end{align}
with
\begin{align}
(\mathbf{h_{E^k}})_{ij} \equiv \lambda_{ijk} \times \azero \qquad \mbox{at}  \quad \mgut,
\label{RGE_hE}
\end{align}
and
\begin{align}
\mathcal{S} =& \Tr[\mathbf{m_{\tilde Q}}^2 - \mathbf{m_{\tilde L}}^2 - 
2 \mathbf{m_{\tilde U}}^2  + \mathbf{m_{\tilde D}}^2 + \mathbf{m_{\tilde E}}^2] \nonumber \\
&+ m_{H_2}^2 - m_{H_1}^2.
\label{RGE_S}
\end{align}
Here, $g_1$ ($M_1$) is the $U$(1) gauge coupling (gaugino mass) and $t
= \ln Q$ with $Q$ the renormalization scale. $(\mathbf{h_{E^k}})_{ij}$
is the trilinear scalar soft breaking coupling corresponding to
$\lambda_{ijk}$. The bold-faced soft mass parameters in
Eq.~\eqref{RGE_dmdt} and Eq.~\eqref{RGE_S} are $3\times3$ matrices in
flavor space: $\mathbf{m_{\tilde Q}}$ and $\mathbf{m_{\tilde L}}$ for
the left-handed doublet squarks and sleptons, $\mathbf{m_{\tilde U}}$,
$\mathbf{m_{\tilde D}}$ and $\mathbf{m_{\tilde E}} $ for the singlet
up-squarks, down-squarks and sleptons, respectively. $m_{H_1}$ and
$m_{H_2}$ are the scalar Higgs softbreaking masses.

The first two terms on the right-hand side in Eq.~\eqref{RGE_dmdt} are
proportional to the gauge coupling squared, $g_1^2$, and also present
in $R$-parity conserving models. The sum of these two terms is {\it
negative} at any scale and thus leads to an {\it increase} of
$M_{\slepton^k_R}$ when running from $\mgut$ down to
$M_Z$. Here, the main contribution comes from the term proportional to
the gaugino mass squared, $M_1^2$, because $\mathcal{S}$ is identical
to zero at $\mgut$ for universal scalar masses. Moreover, the
coefficient of the $M_1^2$ term is larger than that of the
$\mathcal{S}$ term.

The remaining contributions are proportional to $\lambda_{ijk}^2$ and
$(\mathbf{h_{E^k}})^2_{ij}$; the latter implies also a proportionality
to $\lambda_{ijk}^2$ at $\mgut$, \cf Eq.~\eqref{RGE_hE}. These terms
are {\it positive} and will therefore {\it reduce} $M_{\slepton^k_R}$,
when going from $\mgut$ down to $M_Z$. They are new to the
$\mathrm{B}_3$ mSUGRA model compared to $R$-parity conserving
mSUGRA. We can see from Eq.~\eqref{RGE_dmdt}, that if the LNV coupling
is roughly of the order of the gauge coupling $g_1$, \ie $\lam_{ijk}
\gtrsim\mathcal{O}(10^{-2})$, these terms contribute substantially. 
Then, the $\slepton_R$ can be lighter than the lightest neutralino,
$\neut_1$, and lightest stau, $\sstau_1$, at $M_Z$, leading to a
$\slepton_R$ LSP \cite{Dreiner:2008ca}.

\begin{table}[t!]
\centering
\setlength{\tabcolsep}{0.0pc}
\begin{tabular*}{0.48\textwidth}{@{\extracolsep{\fill}}lcc}
\hline\hline
$L_iL_j\bar E_k$ & LSP candidate & $2\sigma$ bound \\
\hline
$\lam_{121}$, $\lam_{131}$		& $\sse_R$	&	$0.020\times (M_{\sse_R}/100\gev)$	\\
 $\lam_{231}$& $\sse_R$	&	$0.033\times (M_{\sse_R}/100\gev)$	\\
$\lam_{132}$			& $\ssmu_R$	& 	
$0.020\times (M_{\ssmu_R}/100\gev)$	\\ 
\hline\hline
\end{tabular*}
\caption{List of $L_iL_j\bar E_k$ couplings (first column) needed to 
generate a $\sse_R$- or $\ssmu_R$-LSP (second column). The third
column gives the most recent experimental bounds [$95\%$ 
confidence level (C.L.)], taken from Ref.~\cite{Kao:2009fg}. The
bounds apply at $M_{\rm GUT}$.  The bounds on $\lam_{212}$ and
$\lam_{232}$ from the generation of too large neutrino masses are in
general too strong to allow for a $\ssmu_R$-LSP
\cite{Allanach:2003eb}, although exceptions might exist
\cite{Dreiner:2010ye}.}
\label{Tab:lamcouplings}
\end{table}

The respective $L_iL_j\bar E_k$ couplings $\mathbf{\Lambda}$, which
can lead to a $\sse_R$ or $\ssmu_R$ LSP, are given in
Table~\ref{Tab:lamcouplings} with their most recent experimental
$2\sigma$ upper bounds at $M_{\rm GUT}$ \cite{Kao:2009fg}. Because of its RGE
running, $\mathbf{\Lambda}$ at $M_Z$ is roughly 1.5 times larger than
at $\mgut$~\cite{deCarlos:1996du,Allanach:1999ic}.

\begin{figure}[t]
\centering
	\includegraphics[scale=1.0]{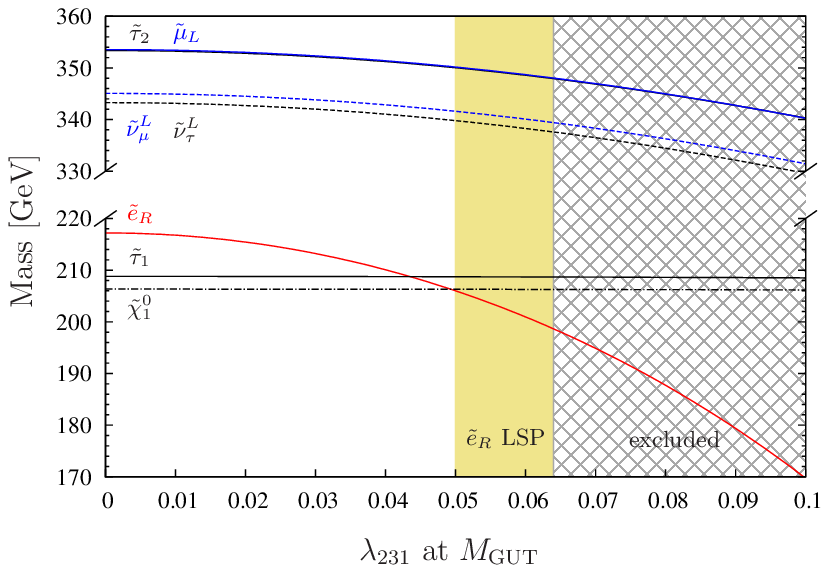}
\caption{Masses of the $\sse_R$, $\neut_1$, $\sstau_1$, $\sstau_2$, 
$\ssnutau$, $\ssmu_L$ and $\ssnumu$ at $M_Z$ as a function of
$\lambda_{231}$ at $\mgut$. The other mSUGRA parameters are $\mzero =
150\gev$, $\mhalf = 500\gev$, $\azero=-1000\gev$, $\tanb = 5$ and $\mu
> 0$. The yellow region corresponds to the experimentally allowed
$\sse_R$ LSP region. The gray patterned region is excluded by the
upper bound on $\lam_{231}$, \cf Table~\ref{Tab:lamcouplings}.}
\label{Fig:massoverlambda}
\end{figure}

As an example, in Fig.~\ref{Fig:massoverlambda}, we demonstrate
the impact of a non-vanishing coupling $\lambda_{231}$ at $M_{\rm GUT}$
on the running of the $\sse_R$ mass. Note that we can obtain a $\sse_R$
LSP ($\ssmu_R$ LSP) with a non-zero coupling $\lam_{121}$ or $\lam_
{131}$ ($\lam_{132}$) at $\mgut$ in a completely analogous way. We
employ {\tt SOFTSUSY3.0.13} \cite{Allanach:2001kg,Allanach:2009bv} for
the evolution of the RGEs. We have chosen a fairly large 
absolute value of $\azero = -1000\gev$ (see the discussion in
Sec.~\ref{Sect:A0dependence}). The other mSUGRA parameters are $\mzero
= 150\gev$, $\mhalf = 500\gev$, $\tanb = 5$ and $\mu > 0$. In the
corresponding $R$-parity conserving case ($\lam_{231}|_{\rm 
GUT} = 0$), the $\neut_1$ is the LSP and the $\sstau_1$ is the next-to
LSP (NLSP).

The $\sse_R$ mass decreases for increasing $\lambda_{231}$, as
described by Eq.~\eqref{RGE_dmdt}. Furthermore, the masses of the
(mainly) left-handed second and third generation sleptons, $\ssmu_L$,
$\sstau_2$, and sneutrinos, $\ssnumu$, $\ssnutau$, decrease\footnote{However,
these (negative) $R$-parity violating contributions are always smaller
than those to the right-handed slepton mass \cite{Allanach:2003eb}.
Thus, the left-handed sleptons and sneutrinos cannot become the LSP
within $\Bthree$ mSUGRA with $\lam_{ijk}|_{\rm GUT} \not = 0$
\cite{Dreiner:2008ca}.}, since
these fields couple directly via $\lambda_{231}$. In contrast, the mass of the $\neut_1$ is not
changed, since it does not couple to the $\lam_{231}$ operator at one
loop level. Also the impact on the mass of the $\sstau_1$, which is
mostly right-handed, is small. We therefore obtain in
Fig.~\ref{Fig:massoverlambda} at $\lambda_{231}|_\mathrm{GUT} \gtrsim
0.05$ a right-handed selectron as the LSP.

Because of the experimental upper bound on $\lam_{231}$ (see Table 
\ref{Tab:lamcouplings}) the gray pattered region in 
Fig.~\ref{Fig:massoverlambda} with $\lam_{231}|_\mathrm{GUT} > 0.064$
is excluded at $95\%$~$\mathrm{C.L.}$.  Note, that the valid parameter
region with a $\sse_R$ LSP becomes larger once we consider scenarios
with heavier sparticles. Moreover, once we go beyond the mSUGRA model
and consider non-universal masses, a $\sse_R$ LSP can also be obtained
with much smaller LNV violating couplings. The collider study that we
present in this publication also applies to these more general
$\slepton_R$ LSP models, provided that we still have a non-vanishing
and dominant $L_i L_j \bar E_k$ operator.

In the following, we investigate which other conditions at $M_{\rm GUT}$ are vital
to obtain a $\slepton_R$ LSP within $\Bthree$~mSUGRA. Especially the dependence on
the trilinear scalar coupling strength $A_0$ plays a crucial role.

\subsubsection{$\azero$ Dependence}
\label{Sect:A0dependence}

According to Eq.~\eqref{RGE_dmdt} and Eq.~\eqref{RGE_hE}, $\azero$
enters the running of $M_{\slepton^k_R}$ via the LNV soft-breaking
trilinear scalar coupling $(\mathbf{h_{E^k}})_{ij}$. As $t=\ln Q$ is
decreased, the $(\mathbf{h_{E^k}})_{ij}$-term gives a negative
contribution to $M_{\slepton^k_R}$. Its full contribution is
proportional to the integral of $(\mathbf{h_{E^k}})_{ij}^2$ over $t$,
from $t_\mathrm{min} = \ln(M_Z)$ to $t_\mathrm{max} = \ln(\mgut)$.
\begin{figure*}[t!]
\centering
\subfigure[\,Running of $(\mathbf{h_{E^k}})_{ij}$.\label{Fig:h}]{
\begin{minipage}{7.9cm}
\centering
	\includegraphics[scale=1.0]{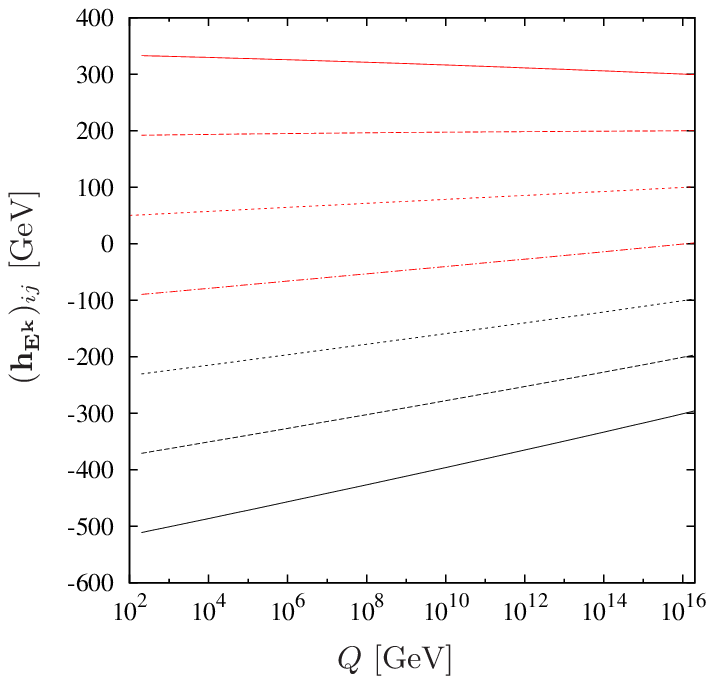}
	\vspace{0.3cm}		
\end{minipage}
}
\subfigure[\,Running of $(\mathbf{h_{E^k}})_{ij}^2$.\label{Fig:hsquared}]{
\begin{minipage}{7.9cm}
\centering
	\includegraphics[scale=1.0]{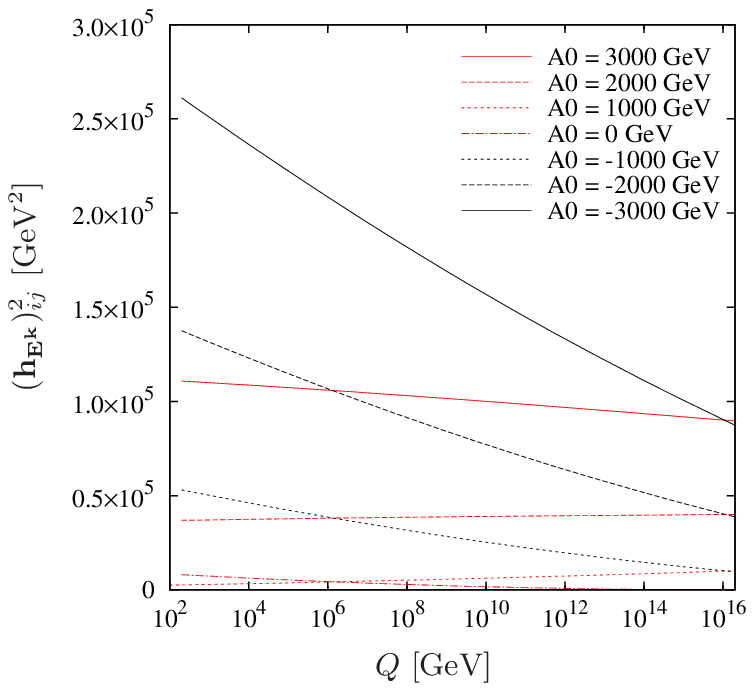}
	\vspace{0.3cm}		
\end{minipage}
}
\caption{Running of $(\mathbf{h_{E^k}})_{ij}$ (left) and $(\mathbf{h_
{E^k}})_{ij}^2$ (right) from $\mgut$ to $M_Z$ for different values of
$\azero$ given in Fig.~\ref{Fig:hsquared}. At $\mgut$, we choose
$\mhalf = 1000\gev$ and $\lambda_{ijk} = 0.1$.}
\label{RGE:hErunning}
\end{figure*}
We now show that a negative $\azero$ with a large magnitude enhances
the (negative) $\lambda_{ijk}$ contribution to the $M_{\slepton^k_R}$
mass. This discussion is similar to the case of a sneutrino LSP
\cite{Bernhardt:2008jz}.
 
In Fig.~\ref{RGE:hErunning} we show the running of the trilinear
coupling $(\mathbf{h_{E^k}})_{ij}$ [Fig.~\ref{Fig:h}] and the
resulting running for $(\mathbf{h_{E^k}})_{ij}^2$
[Fig.~\ref{Fig:hsquared}]. We assume a non-vanishing coupling
$\lambda_{ijk}|_\mathrm{GUT} = 0.1$ and a universal gaugino mass
$\mhalf = 1000\gev$. Different lines correspond to different values of
$\azero$, as indicated in Fig.~\ref{Fig:hsquared}.

The dominant contributions to the RGE of $(\mathbf{h_{E^k}})_{ij}$ are
given by \cite{Allanach:2003eb}
\begin{align}
16 \pi^2 \frac{d(\mathbf{h_{E^k}})_{ij}}{dt} =& - (\mathbf{h_{E^k}})_{ij} \left[ \frac{9}{5} g_1^2 + 3 g_2^2 \right] \nonumber\\
&+ \lambda_{ijk} \left[ \frac{18}{5} g_1^2 M_1 + 6 g_2^2 M_2 \right].
\label{Eqn:RGE_dhdt}
\end{align}
$M_1$ and $M_2$ are the $U(1)$ and $SU(2)$ gaugino masses,
respectively. The running in Eq.~\eqref{Eqn:RGE_dhdt} is governed by
two terms with opposite sign; one proportional to $(\mathbf{h_{E^k}})
_{ij}$ and one proportional to $ \lambda_{ijk}$. In contrast to the 
sneutrino LSP case (\cf Ref.~\cite{Bernhardt:2008jz}) the running is
independent of the strong coupling $g_3$ and the gluino mass $M_3$.

According to Eq.~\eqref{RGE_hE}, the sign of the term proportional to
$(\mathbf{h_{E^k}})_{ij}$ in Eq.~\eqref{Eqn:RGE_dhdt} depends on the
sign of $\azero$. At $\mgut$, this term is positive (negative) for
negative (positive) $\azero$. Hence, for positive $\azero$, the term
proportional to $(\mathbf{h_{E^k}})_{ij}$ increases
$(\mathbf{h_{E^k}})_{ij}$ when we run from $\mgut$ to $M_Z$. Note,
that the gauge couplings $g_1$ and $g_2$ decrease from $\mgut$ to
$M_Z$.

Assuming $ \lambda_{ijk}$ to be positive, the second term is always
positive and thus decreases $(\mathbf{h_{E^k}})_{ij}$ when running
from $\mgut$ to $M_Z$. The $ \lambda_{ijk}$ coupling increases by
roughly a factor of $1.5$ when we run from $\mgut$ to $M_Z$. However,
at the same time, the gaugino masses $M_1$ and $M_2$ as well as the
gauge couplings $g_1$ and $g_2$ decrease. Therefore, this term gets
relatively less important towards lower scales.

Now, we can understand the running of $(\mathbf{h_{E^k}})_{ij}$ in
Fig.~\ref{Fig:h}. Given a positive $\azero$ (red lines), both terms in
Eq.~\eqref{Eqn:RGE_dhdt} have opposite signs and thus partly
compensate each other, resulting only in a small change of
$(\mathbf{h_{E^k}})_{ij}$ during the running. Moreover, due to the
running of the gauge couplings and gaugino masses both terms in
Eq.~\eqref{Eqn:RGE_dhdt} decrease when we run from $\mgut$ to
$M_Z$. In contrast, if we start with a negative $\azero$ (black
lines), both terms give negative contributions to the running of
$(\mathbf{h_{E^k}})_{ij}$. Still, the magnitude of the $\lambda_{ijk}$
term in Eq.~\eqref{Eqn:RGE_dhdt} decreases. However, the contribution
from the term proportional to $(\mathbf{h_{E^k}})_{ij}$ does not
necessarily decrease when running from $\mgut$ to $M_Z$. Thus, for
negative $\azero$, $(\mathbf{h_{E^k}})_{ij}$ decreases with a large
slope.

Recall from Eq.~\eqref{RGE_dmdt}, that $M_{\slepton^k_R}^2$ is reduced proportional to the integral 
of $(\mathbf{h_{E^k}})_{ij}^2$ over $t$. Thus, according to Fig.~\ref{Fig:hsquared}, a negative value 
of $\azero$ leads to a smaller $M_{\slepton^k_R}$ compared to a positive $\azero$ with the same magnitude.

\subsubsection{Selectron and Smuon LSP Parameter Space}
\label{Sect:parameterspace}

In this section, we present two dimensional $\Bthree$ mSUGRA parameter regions which exhibit a 
$\slepton_R$ LSP. As we have seen in Sec.~\ref{Sect:RGE}, the running of the $\slepton_R$ mass 
is analogous for the first and second generation.
Therefore, we only study here the case of a $\sse_R$ LSP with a 
non-vanishing coupling $\lam_{231}$ at $M_{\rm GUT}$. We can obtain 
the $\ssmu_R$ LSP region by replacing coupling $\lam_{231}$ with $\lam_{132}$.

\begin{figure*}[t]
\centering
\subfigure[\,$\azero$--$\mhalf$ plane with $\Bthree$ mSUGRA parameter $\mzero= 90\gev$, $\tanb=4$, 
$\sgnmu = +$ and $\lam_{231}|_\mathrm{GUT} = 0.045$.\label{Fig:m12A0}
]{
	\begin{minipage}{7.8cm}
	   	\includegraphics[scale=1.0]{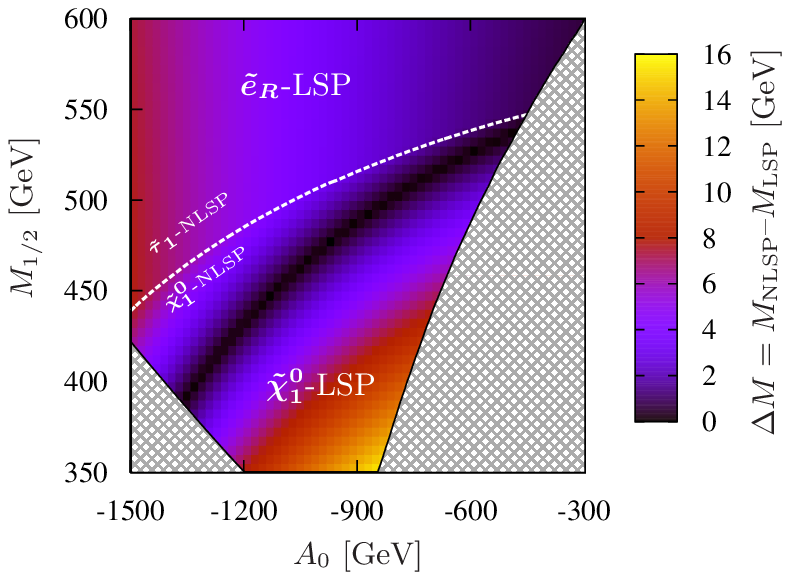}
	\vspace{0.0cm}
	\end{minipage}
}
\subfigure[\,$\mzero$--$\tanb$ plane with $\Bthree$ mSUGRA parameter $\mhalf= 450\gev$, 
$\azero=-1250\gev$, $\sgnmu = +$ and $\lam_{231}|_\mathrm{GUT} = 0.045$.\label{Fig:m0tanb}]{
	\begin{minipage}{7.8cm}
	   	\includegraphics[scale=1.0]{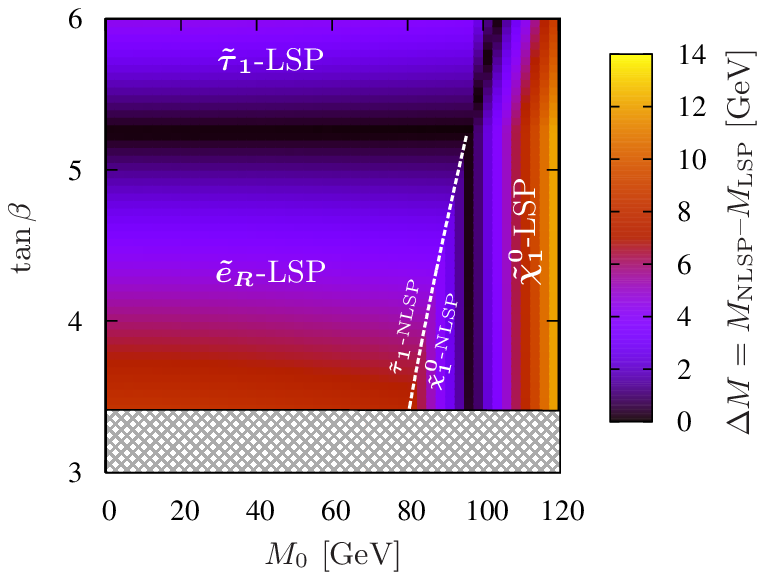}
	\vspace{0.33cm}	
	\end{minipage}
}
\caption{Mass difference, $\Delta M$, between the NLSP and LSP. The LSP candidates are explicitly mentioned. 
The patterned regions correspond to models excluded by the LEP Higgs
bound. The white dotted line separates $\sse_R$-LSP scenarios with
different mass hierarchies: $M_{\sse_R} < M_{\stau_1} < M_{\neut_1}$
(left-hand side) and $M_{\sse_R} < M_{\neut_1} < M_{\stau_1}$
(right-hand side).}
\label{Fig:LSPregions}
\end{figure*}
	
We give in Fig.~\ref{Fig:LSPregions} the $\sse_R$ LSP region in the
$\azero-\mhalf$ plane [Fig.~\ref{Fig:m12A0}] and $\mzero-\tanb$ plane
[Fig.~\ref{Fig:m0tanb}] for a coupling $\lam_{231}|_{\rm GUT} =
0.045$. We show the mass difference, $\Delta M$, between the NLSP and
LSP. For the shown region a lower bound of $135\gev$ on the selectron
mass is employed to fulfill the bound on $\lam_{231}$; \cf
Table~\ref{Tab:lamcouplings}. The pattered regions are excluded by the
LEP bound on the light Higgs mass
\cite{Barate:2003sz,Schael:2006cr}. However, we have reduced this
bound by 3 $\gev$ to account for numerical uncertainties of {\tt
SOFTSUSY}
\cite{Allanach:2003jw,Degrassi:2002fi,Allanach:2004rh} which is 
used to calculate the SUSY and Higgs mass spectrum. 

The entire displayed region fulfills the $2\sigma$ constraints on
the branching ratio of the decay $b\to s\gamma$
\cite{TheHeavyFlavorAveragingGroup:2010qj},
\begin{align}
3.03 \times 10^{-4} < \mathcal{B}(b\to s \gamma) < 4.07 \times 10^{-4},
\end{align}
and the upper limit on the flavor changing neutral current 
(FCNC) decay $B_s^0 \to \mu^+ \mu^-$ \cite{Morello:2009wp}, 
\textit{i.e.}
\begin{align}
\mathcal{B}(B_s^0 \to \mu^+\mu^-) < 3.6 \times 10^{-8},
\end{align}
at $90\%$~$\mathrm{C.L.}$. 

However, the parameter points in Fig.~\ref{Fig:LSPregions} cannot
explain the discrepancy between experiment (using pion spectral
functions from $e^+ e^-$ data) and the SM prediction of the anomalous
magnetic moment of the muon, $a_\mu$; see Ref.~\cite{MALAESCU:2010ne}
and references therein. There exists a $\sse_R$ LSP region consistent
with the measured value of $a_\mu$ at $2\sigma$. But this region is
already excluded by Tevatron tri--lepton SUSY searches
\cite{Dreiner:2010tevatron}. We note however, that the SM prediction
is consistent with the experimental observations at the $2\sigma$
level, if one uses spectral functions from $\tau$ data
\cite{MALAESCU:2010ne}.  We have employed micrOMEGAs2.2
\cite{Belanger:2008sj} to calculate the SUSY contribution to $a_\mu$,
$\mathcal{B}(b\to s\gamma)$ and $\mathcal{B}(B_s^0\to \mu^+\mu^-)$.

We observe in Fig.~\ref{Fig:LSPregions} that the $\sse_R$ LSP lives in
an extended region of the $\Bthree$ mSUGRA parameter space. Competing
LSP candidates are the lightest stau, $\sstau_1$, and the lightest
neutralino, $\neut_1$.

In the $\azero$--$\mhalf$ plane, Fig.~\ref{Fig:m12A0}, we find a
$\sse_R$ LSP for larger values of $\mhalf$, because $\mhalf$ increases
the mass of the (bino-like) $\neut_1$ faster than the mass of the
right-handed sleptons \cite{Drees:1995hj,Ibanez:1984vq}. We can also
see that a $\sse_R$ LSP is favored by a negative $\azero$ with a large
magnitude as discussed in Sec.~\ref{Sect:A0dependence}. In this region
of parameter space the mass difference between the $\sse_R$ LSP and
the $\stau_1$ NLSP increases with increasing $|\azero|$. In principle,
there can also be a $\sse_R$ LSP for a large positive $\azero$, \cf
Fig.~\ref{Fig:hsquared}.  However this configuration is disfavored due
to a too small light Higgs mass \cite{Drees:1995hj}. Note, that a
negative $\azero$ with a large magnitude naturally leads to a light top
squark, $\sstop_1$, since the top Yukawa coupling enters the RGE
running of the $\sstop_1$ mass in a similar way as the $\lam_{ijk}$
Yukawa coupling does for the $\slepton_R$ mass
\cite{Drees:1995hj,Ibanez:1984vq}.  This behavior plays an important
role for the mass reconstruction of the $\sstop_1$, \cf
Sec.~\ref{Sect:mass_reco}.

In the $\mzero$--$\tanb$ plane, Fig. \ref{Fig:m0tanb}, we find a $\sse_R$ LSP for $\tanb\lesssim 5$ 
and $\mzero \lesssim 100\gev$. The mass of the $\stau_1$ decreases with increasing $\tanb$ while the mass 
of the $\sse_R$ is unaffected by $\tanb$. Increasing $\tanb$ increases 
the tau Yukawa coupling and thus its (negative) contribution to the stau mass from RGE running 
\cite{Drees:1995hj,Ibanez:1984vq}. Furthermore, a larger value of $\tanb$ usually 
leads to a larger mixing between the left- and right-handed stau. Thus, $\tanb$ is a 
handle for the mass difference of the $\stau_1$ and $\sse_R$. In contrast, $\mzero$ 
increases the masses of all the scalar particles like the $\stau_1$ and $\sse_R$, while the mass 
of the $\neut_1$ is nearly unaffected by both $\tanb$ and $\mzero$. Therefore, at larger values of 
$\mzero$ we obtain a $\neut_1$ LSP.

We find basically two possible mass hierarchies for the $\sse_R$ LSP parameter space, 
indicated by the white dotted line in Fig.~\ref{Fig:m12A0} and Fig.~\ref{Fig:m0tanb}. 
Close to the $\neut_1$ LSP region, we observe a $\neut_1$ NLSP and a $\sstau_1$ next-to-NLSP (NNLSP), 
\ie
\begin{align}
M_{\sse_R} <  M_{\neut_1} < M_{\stau_1} . \label{Eqn:masshierarchy1}
\end{align}
However, for most of the parameter space, we have
\begin{align}
M_{\sse_R} < M_{\stau_1} <  M_{\neut_1}, \label{Eqn:masshierarchy2}
\end{align}
\ie the $\stau_1$ is the NLSP and the $\neut_1$ is the NNLSP.  
For some regions with a large mass difference between the $\neut_1$ and the 
$\sse_R$ LSP, the $\ssmu_R$ can even be the NNLSP, \ie we have
\begin{align}
M_{\sse_R} < M_{\stau_1} < M_{\ssmu_R} < M_{\neut_1} \label{Eqn:masshierarchy3},
\end{align}
where the $\neut_1$ is the next-to NNLSP (NNNLSP). These three mass hierarchies lead 
to a different collider phenomenology and will be our guideline in the selection of benchmark 
scenarios.

\subsection{Benchmark Scenarios}

\begin{table}[t]
\centering
\setlength{\tabcolsep}{0.0pc}
\begin{tabular*}{0.47\textwidth}{@{\extracolsep{\fill}}lccc}
\hline\hline
$\Bthree$ mSUGRA & \multicolumn{3}{c}{benchmark model}\\
parameter				& BE1	&BE2	& BE3 \\
\hline
$\mzero$ [GeV]			& $0$	&	$90$		&	$90$	\\
$\mhalf$ [GeV]				& $475$	&	$460$	&	$450$	\\
$\azero$ [GeV]				& $-1250$ & $-1400$	&	$-1250$\\
$\tanb$					& $5$	& $4$		&	$4$\\
$\sgnmu$					& $+$	& $+$		&	$+$\\
$\lam_{231}|_\mathrm{GUT}$	& $0.045$	&$0.045$&$0.045$\\
\hline
light sparticles			&		&			&	\\
 (mass/GeV)				&		&			&	\\
\hline
LSP 		&$\sse_R$ ($168.7$)	&$\sse_R$ ($182.3$)	&$\sse_R$ ($182.0$)	\\	
NLSP	&$\stau_1$ ($170.0$)	&$\stau_1$ ($189.0$)	&$\neut_1$ ($184.9$)	\\
NNLSP	&$\ssmu_R$ ($183.6$)	&$\neut_1$ ($189.5$)	&$\stau_1$ ($187.2$)	\\
NNNLSP 	&$\neut_1$ ($195.7$)	&$\ssmu_R$ ($199.0$)	&$\ssmu_R$ ($195.9$)	\\
\hline\hline
\end{tabular*}
\caption{$\Bthree$ mSUGRA parameter and the masses of the four lightest
SUSY particles of the $\sse_R$ LSP benchmark points BE1, BE2 and
BE3. The complete mass spectra and the branching ratios are given in
Appendix~\ref{App:benchmarks}.}
\label{Tab:benchmarks}
\end{table}

In order to investigate the LHC phenomenology of a $\sse_R$ LSP
model in more detail, we select for each mass hierarchy,
Eq.~\eqref{Eqn:masshierarchy1}-\eqref{Eqn:masshierarchy3}, one
representative $\sse_R$ LSP benchmark point. The $\Bthree$ mSUGRA
parameters and the masses of the lightest four sparticles of these
benchmark points, denoted BE1, BE2 and BE3, are given in
Table~\ref{Tab:benchmarks}. All benchmark points exhibit a coupling
$\lam_{231}|_{\rm GUT}= 0.045$ (\cf Table~\ref{Tab:lamcouplings}) and
fulfill the experimental constraints of Sec.~\ref{Sect:parameterspace}
and the constraints from Tevatron tri-lepton SUSY searches
\cite{Dreiner:2010tevatron}.  The supersymmetric mass spectra
and branching ratios are given in Appendix \ref{App:benchmarks}.

The benchmark points BE1 and BE2 both feature a $\stau_1$ NLSP. In
BE1, the $\stau_1$ is nearly mass degenerate with the $\sse_R$ and
decays exclusively via $\lam_{231}$ into an electron and a muon
neutrino. In contrast, in BE2 the (mainly right-handed) $\stau_1$ is
7~GeV heavier than the $\sse_R$ LSP and thus mainly decays via
three-body decays into the $\sse_R$ due to larger phase-space.
Similarly, the $\ssmu_R$ NNLSP in BE1 decays via three-body decays
into the $\sse_R$ or the $\stau_1$. The three-body decays of the
heavier supersymmetric sleptons to the $\sse_R$ LSP are new and are
calculated in Appendix~\ref{App:sleptondecays}.

In BE1, there is a fairly large mass difference between the $\sse_R$
LSP and the $\neut_1$ NNNLSP of about $27\gev$.  The mass difference
between the $\sse_R$ and $\neut_1$ is smaller in BE2 (compared to
BE1), \ie about $7\gev$. The NNNLSP is the $\ssmu_R$.  Finally, the
benchmark point BE3 features a $\neut_1$ NLSP that is $3\gev$ heavier
than the $\sse_R$ LSP. The $\stau_1$ is the NNLSP and decays into the
$\neut_1$ and a $\tau$. The $\ssmu_R$ is the NNNLSP.

\section{Selectron and Smuon LSP Signatures at the LHC}
\label{Sect:signatures}

\begin{table}[t]
\centering
\begin{tabular*}{0.47\textwidth}{@{\extracolsep{\fill}}lcc}
\hline\hline
$\mathbf{\Lambda}$ coupling	&	LSP decay	&	LHC signature		\\
\hline
\multirow{3}{*}{$\lam_{121}$}	&	\multirow{3}{*}{\renewcommand{\arraystretch}{0.1}$\sse_R \to \left\{\begin{array}{l} e \,\numu \\ \mu\, \nue\end{array}\right.$}	& \multirow{3}{*}{\renewcommand{\arraystretch}{0.1}$2j +  2e +\etmiss +  \left\{\begin{array}{l} 2 e \\ e\mu \\ 2 \mu \end{array}\right.$} \\
		&		&		\\
		&		&		\\
\hline
\multirow{3}{*}{$\lam_{131}$}	&	\multirow{3}{*}{\renewcommand{\arraystretch}{0.1}$\sse_R \to \left\{ \begin{array}{l} e \,\nutau \\ \tau\, \nue\end{array}\right.$}	& \multirow{3}{*}{\renewcommand{\arraystretch}{0.1}$2j +  2e +\etmiss +  \left\{\begin{array}{l} 2 e \\ e\tau \\ 2 \tau \end{array}\right.$} \\
		&		&		\\
		&		&		\\
\hline
\multirow{3}{*}{$\lam_{231}$}	&	\multirow{3}{*}{\renewcommand{\arraystretch}{0.1}$\sse_R \to \left\{ \begin{array}{l} \mu \,\nutau \\ \tau\, \numu\end{array}\right.$}	& \multirow{3}{*}{\renewcommand{\arraystretch}{0.1}$2j +  2e +\etmiss +  \left\{\begin{array}{l} 2 \mu \\ \mu\tau \\ 2 \tau \end{array}\right.$} \\
		&		&		\\
		&		&		\\
\hline
\multirow{3}{*}{$\lam_{132}$}	&	\multirow{3}{*}{\renewcommand{\arraystretch}{0.1}$\ssmu_R \to  \left\{\begin{array}{l} e \,\nutau \\ \tau\, \nue\end{array}\right.$}	& \multirow{3}{*}{\renewcommand{\arraystretch}{0.1}$2j +  2\mu +\etmiss +  \left\{\begin{array}{l} 2 e \\ e\tau \\ 2 \tau \end{array}\right.$} \\
		&		&		\\
		&		&		\\
\hline\hline
\end{tabular*}
\caption{LHC signatures (right column) for selectron and smuon LSP 
scenarios (second column) assuming one dominant $L_i L_j \bar E_k$
operator $\mathbf\Lam$ (left column) and the SUSY cascade of
Eq.~\eqref{Eqn:SUSYcascade}. }
\label{Tab:signatures}
\end{table}

We now classify the main LHC signatures of selectron and smuon,
$\tilde{\ell}_R$, LSP models, under the simplifying assumption
that each decay chain of heavy SUSY particles ends in the LSP  
and that the LSP decay is dominated by only
one $R$-parity violating operator $\mathbf\Lambda$, \cf
Table~\ref{Tab:lamcouplings}.  If we assume squark pair production as
the main sparticle production process\footnote{For all of our
benchmark points the gluinos are heavier than the squarks and
dominantly decay to a squark and a quark. Thus for gluino pair
production we simply obtain two jets more per event.}, we obtain as
one of the major cascades
\begin{align}
qq/gg \to \squark \squark \to jj \neut_1 \neut_1 \to jj \ell \ell \slepton_R \slepton_R,
\label{Eqn:SUSYcascade}
\end{align}
where $\squark$ is a squark and $j$ denotes a (parton level) jet. The
two leptons $\ell$ are of the same flavor as the LSP.  The
$\slepton_R$ LSP will promptly decay via the $R$-parity violating
$L_iL_j\bar E_k$ operator into a charged lepton and a neutrino. The resulting
collider signatures are classified in Table~\ref{Tab:signatures}
according to the possible $\slepton_R$ LSP decays.

Assuming the SUSY cascade in Eq.~(\ref{Eqn:SUSYcascade}), the
resulting collider signatures involve two (parton level) jets from
squark decays, two charged leptons from the neutralino decay with the
same flavor as the LSP, as well as additional charged leptons and
missing transverse energy, $\etmiss$, from the LSP decays. Because of
the Majorana nature of the $\neut_1$, every charge combination of the
two $\slepton_R$ LSPs is possible. In what follows, it is important to
note that the transverse momentum, $p_T$, spectrum of the leptons from
the decay $\neut_1 \to \ell \slepton_R$ will depend on the mass
difference between the $\slepton_R$ LSP and the $\neut_1$.  For
smaller mass differences we get on average a smaller lepton
$p_T$.

\begin{figure}
\includegraphics[scale=1.0]{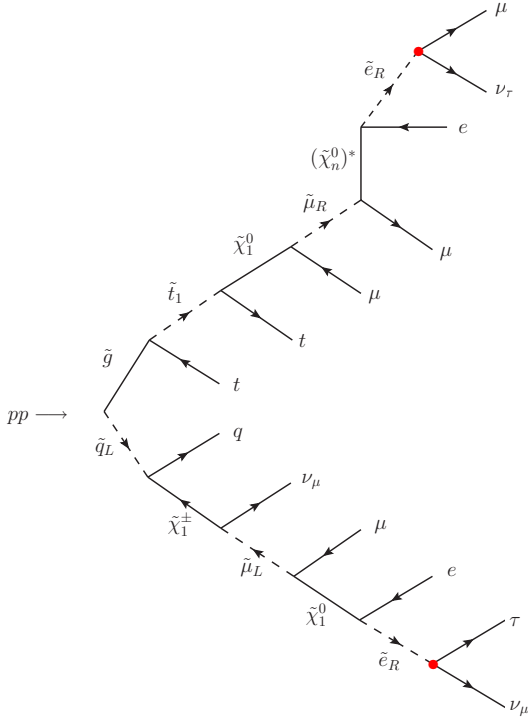}
\caption{Example for squark-gluino production with successive cascade
decay into two $\sse_R$ LSPs.  The $R$-parity violating decays are
marked by red dots. $(\neut_n)^*$ denotes a virtual
neutralino. Note that R-parity violating decays can occur
earlier in the chain. In this case the LSP is not produced. See tables
of the benchmark branching ratios in Appendix \ref{App:benchmarks}.}
\label{Fig:exampleSUSYdecay}
\end{figure}

In general, more complicated SUSY production and decay processes than
Eq.~\eqref{Eqn:SUSYcascade} can occur.
Fig.~\ref{Fig:exampleSUSYdecay} gives an example of (left-handed)
squark-gluino production followed by two lengthy decay chains.
Typically, these processes lead to {\it additional} final state
particles [compared to Eq.~\eqref{Eqn:SUSYcascade} and
Table~\ref{Tab:signatures}], most notably
\begin{itemize}
\item additional jets from the production of gluinos and their 
subsequent decays into squarks and quarks; \cf the upper decay chain
of Fig.~\ref{Fig:exampleSUSYdecay},
\item additional leptons from the decays of heavier neutralinos and 
charginos, which may come from the decay of left-handed squarks, like
in the lower decay chain of Fig.~\ref{Fig:exampleSUSYdecay}, and
\item additional leptons from a $\neut_1$ decay into a non-LSP 
right-handed slepton $\slepton'_R$ (or lightest stau $\stau_1$),
\textit{e.g.} $\neut_1 \to \ell'^- \slepton'^+_R$, followed by the three-body
decay $\slepton'^+_R \to \ell'^+ \ell^\pm \slepton_R^\mp$ via a
virtual neutralino, $(\neut_n)^*$; see the upper decay chain of
Fig.~\ref{Fig:exampleSUSYdecay} for an example.  Here, $\slepton_R$ is
the LSP.
\end{itemize}
These three-body slepton decays are special to $\sse_R$ and $\ssmu_R$
LSP scenarios. The corresponding decay rates are calculated in
Appendix~\ref{App:sleptondecays} and are taken into account in the
following collider analysis.

The coupling $\mathbf\Lambda$ in $\slepton_R$ LSP scenarios is of
similar size as the gauge couplings and thus enables $R$-parity
violating decays with a significant branching ratio of sparticles
which are not the LSP. Thus, not every SUSY decay chain involves the
LSP. Of particular importance are the 2-body $R$-parity violating
decays of the $\stau_1$ \cite{Desch:2010gi}, especially in the case
when a $\stau_1$ NLSP is nearly mass degenerate with the $\slepton_R$
LSP, like for the benchmark point BE1; \cf
Table~\ref{Tab:BE1}. Furthermore, sneutrinos (left-handed charged
sleptons) may decay into two hard charged leptons (one charged lepton
and a neutrino) if they couple directly to the dominant $R$-parity
violating operator.  This leads to a sharp sneutrino mass peak in the
respective dilepton invariant mass distribution as we will show in
Sec.~\ref{Sect:mass_reco}. From the $R$-parity violating left-handed
slepton decays we expect large amounts of missing energy from the
neutrino.

The lightest top squark, $\sstop_1$, is in most $\Bthree$ mSUGRA
scenarios the lightest squark. Thus, $\sstop_1$ pair production forms
a sizable fraction of all SUSY production processes. The decay of each
$\sstop_1$ yields at least one $b$-quark (either directly from the
decay $\sstop_1 \to \charge_1^+ b$ and/or from the top quark decay
after $\sstop_1 \to \neut_1 t$).  We therefore expect an enhanced
$b$-quark multiplicity for $\sstop_1$ pair production. We will use the
$b$-quark multiplicity in Sec.~\ref{Sect:mass_reco} to discriminate
these events from other SUSY processes.

To conclude this discussion, as one can see from
Table~\ref{Tab:signatures}, we expect multi-lepton final states for
$\sse_R$ and $\ssmu_R$ LSP scenarios at the LHC. One the one hand, we
obtain charged leptons from the $\neut_1$ decay into the $\slepton_R$
LSP. On the other hand, each LSP decay involves a charged lepton.
Furthermore, as explained above, also non LSPs can decay via the
dominant $R$-parity violating operator into leptons. Therefore, a
multi-lepton analysis will be the best search strategy for our
$\slepton_R$ LSP scenarios.

Multi charged lepton final states (especially electrons and muons) are
one of the most promising signatures to be tested with early LHC
data. Electrons and muons can be easily identified and the SM
background for high lepton multiplicities is very low
\cite{Aad:2009wy}.  We therefore investigate in the following the
discovery potential of $\sse_R$ LSP scenarios with an inclusive three
lepton search analysis. We will treat electrons and muons equally and
thus expect similar results for $\ssmu_R$ LSP scenarios.

\section{Discovery Potential at the LHC}
\label{Sect:discovery}

In this section, we study the discovery potential of $\sse_R$ and
$\ssmu_R$ LSP models with an inclusive search analysis for tri-lepton
final states at the LHC. Because of the striking multi-leptonic
signature of these models (see Sec.~\ref{Sect:signatures}), a
discovery might be possible with early LHC data. We therefore study
the prospects at the LHC assuming separately a center-of-mass
system (cms) energy of $7\tev$ {\it and} $14\tev$.

\subsection{Major Backgrounds}
In the following Monte Carlo (MC) study, we consider SM backgrounds
that can produce three or more charged leptons (electrons or muons) in
the final state at the particle level, \ie after (heavy
flavor) hadron and tau lepton decays. For the heavy flavor quarks, we
consider bottom, $b$, and charm, $c$, quarks
\cite{Sullivan:2008ki}. Moreover, we expect the SUSY signal events to
contain additional energy from hard jets arising from decays of the
heavier (colored) sparticles.  We thus consider the following SM
processes as the major backgrounds in our analysis:

\begin{itemize}
\item Top production. We consider top pair production ($t\overline t$),
single-top production associated with a $W$ boson ($Wt$) and top pair
production in association with a gauge boson ($Wt\overline t $,
$Zt\overline t$). Each top quark decays into a $W$ boson and a $b$
quark. Leptons may then originate from the $W$ and/or $b$ decay.
\item $Z + \mbox{jets}$, \ie $Z$ boson production in association with
one or two (parton level) jets.  For the associated jet(s) we consider
only $c$- and $b$-quarks. We force the $Z$ boson to decay
leptonically.
\item $W + \mbox{jets}$, \ie $W$ boson production in association with 
two heavy flavor quarks ($c$ or $b$) at parton level. We demand that 
the $W$ decays into a charged lepton and a neutrino. 
\item Di-boson ($WZ$, $ZZ$) and di-boson $+$ jet ($WWj$, $WZj$, $ZZj$)
production. For the $WZ$ and $ZZ$ background, the gauge bosons are 
forced to decay leptonically. For $WWj$, we consider only the heavy 
flavor quarks $c$ and $b$ for the (parton level) jet, while for $WZj$ 
and $ZZj$ every quark flavor is taken into account.
\end{itemize}
We have also included the processes, where we have a virtual gamma
instead of a $Z$ boson.

\begin{table*}[t]
\centering
\begin{tabular*}{\textwidth}{@{\extracolsep{\fill}}llrrrrl}
\hline\hline
Sample	&	Sub-sample	& \multicolumn{2}{c}{LO cross section [pb]}	&	\multicolumn{2}{c}{Simulated events}	&	Generator		\\
		&				&	$7\tev$ &	$14\tev$	&	$7\tev$ &$14\tev$		&	\\		
\hline
top		&	$t\overline t$			&	$86.7$	&	$460$	 &	$ 200~ 000$		& $5~ 000~ 000$	&	{\tt Herwig}		\\
			& $Wt$				&	$10.2$	&	$60.7$ 	&	$100~ 000$		& $1~ 200~ 000$	&	{\tt MadGraph + Herwig}		\\	
			& $Wt\overline t$		&	$0.14$	& 	$0.52$	&	$10~ 000$			& $10~ 000$	&	{\tt MadGraph + Herwig} \\		
			& $Zt\overline t$		&	$0.066$	& 	$0.43$	&	$10~ 000$			& $10~ 000$	&	{\tt Herwig}		\\
\hline
$Z +\mbox{jets}$	& $Zc\overline c$	&	$49.5$	& $187$&	$100~ 000$	& $2~ 000~ 000$	 &	{\tt Herwig}		\\
				& $Zb\overline b$	&	$44.6$	& $171$&	$ 100~ 000$	& $2~ 000~ 000$&	{\tt Herwig}		\\				
				& $Z(\to \ell^+ \ell^-) +  j\,\,(j=c,b)$		& $59.6$ & $203$&	$180~ 000$		& $3~ 700~ 000$&	{\tt MadGraph + Herwig}		\\
\hline
$W +\mbox{jets}$	& $W (\to \ell\nu) + jj\,\,(j=c,b)$	&$38.2$	& $95.2$&	$135~ 000$		& $1~ 400~ 000$&	{\tt MadGraph + Herwig}		\\
\hline
di-boson			& $WZ \to \mbox{leptons}$	& $0.20$	& $0.40$&	$100~ 000$	& $100~ 000$ &	{\tt MadGraph + Herwig}		\\
				& $ZZ \to \mbox{leptons}$	& $0.03$	& $0.06$&	$22~ 000$		& $75~ 000$ &	{\tt MadGraph + Herwig}		\\
				& $WW+j\,\,(j=c,b)$			& $10.9$	& $64.0$&	$120~ 000$	& $1~ 000~ 000$ &	{\tt MadGraph + Herwig}		\\
				& $WZ+j\,\,(j=\mbox{all flavors})$		& $7.0$	& $25.0$ &	$77~ 000$			& $100~ 000$ &	{\tt MadGraph + Herwig}		\\
				& $ZZ+j\,\,(j=\mbox{all  flavors})$		& $3.2$	& $10.2$&	$16~ 000$		& $280~ 000$&	{\tt MadGraph + Herwig}		\\
\hline\hline
\end{tabular*}
\caption{SM background MC samples (first and second column) used for
our analysis. The third and fourth (fifth and sixth) column shows the
leading-order cross section (number of simulated events) for $pp$
collisions at a cms energy of $\sqrt{s}=7\tev$ and $\sqrt{s}=14\tev$,
respectively. For the event simulation we employ the MC generator
listed in the last column. We also have included the processes, where
we have a virtual gamma instead of a $Z$ boson.}
\label{Tab:backgrounds}
\end{table*}

For the backgrounds with heavy flavor quarks, we demand (at parton 
level) a minimal transverse momentum for the $c$ or $b$ quarks of $p_T
\ge 10\gev$ corresponding to our object selection cut for the leptons,
\cf Sec.~\ref{Sect:object_selection}. Table \ref{Tab:backgrounds} 
gives an overview of the background samples used in our analysis. In
principle, QCD production of four heavy flavor quarks, like $b\bar b b
\bar b$ production, can also produce three lepton events. However,
these backgrounds are negligible compared to the other backgrounds in
Table \ref{Tab:backgrounds}, because the probability of obtaining
three {\it isolated} leptons from heavy flavor decay is too low
\cite{Sullivan:2010jk}.

\subsection{Monte Carlo Simulation and Object Selection}
\label{Sect:object_selection}

The $t\overline t$, $Zt\overline t$, $Zc\overline c$ and $Zb\overline
b$ backgrounds are simulated with {\tt Herwig6.510}
\cite{Corcella:2000bw,Corcella:2002jc,Moretti:2002eu}.  For the other
SM processes we employ {\tt MadGraph4.4.30} \cite{Alwall:2007st} for
the generation of the hard process which is then fed into {\tt
Herwig}. The employed MC generators are listed in
Table~\ref{Tab:backgrounds}. We also give the leading-order (LO) cross
section and the number of simulated events for each background sample
for both cms energies. The cross sections are taken from {\tt Herwig}
(for the $t\overline t$, $Zt\overline t$, $Zc\overline c$ and
$Zb\overline b$ backgrounds) or {\tt MadGraph} (else). We only
consider the leading-order cross sections for the signal and
background samples. We note however that the next-to-leading-order
(NLO) corrections can be large, see \textit{e.g.}
Refs.~\cite{Campbell:2006wx,Binoth:2010ra,Bonciani:1998vc,Beenakker:1996ch,Beenakker:1997ut},
and should be included in a more dedicated analysis.  Furthermore, our
simulation does not account for detector effects, \ie we neglect
backgrounds with leptons faked by jets or photons. However, we expect
these backgrounds to be small, because the fake rate for electrons and
muons is quite low \cite{Aad:2009wy}.

The SUSY mass spectra were calculated with {\tt SOFTSUSY3.0.13}
\cite{Allanach:2001kg,Allanach:2009bv}.  The {\tt SOFTSUSY} output was
then fed into {\tt ISAWIG1.200} and {\tt ISAJET7.64}
\cite{Paige:2003mg} in order to calculate the decay widths of the SUSY
particles. We added the missing three-body slepton decays $\slepton'_R
\to\ell' \ell  \slepton_R$ and $\sstau_1 \to \tau  \ell \slepton_R$ to
the {\tt ISAJET} code; see Appendix~\ref{App:sleptondecays} for the
calculation and a discussion of these new slepton decays.  The signal
processes, \ie pair production of {\it all} SUSY particles, were
simulated with {\tt Herwig6.510}.

\begin{table*}[t]
\centering
\renewcommand{\arraystretch}{1.2}\addtolength{\tabcolsep}{0.1cm}
\begin{tabular*}{\textwidth}{@{\extracolsep{\fill}}l|rrr|rrr}
\hline\hline
Production process	& \multicolumn{3}{c}{Cross section [fb] at $\sqrt{s} = 7\tev$} & \multicolumn{3}{c}{Cross section [fb] at $\sqrt{s} = 14\tev$} \\
				&	BE1			&	BE2 			&	BE3		&	BE1			&	BE2 			&	BE3			\\
\hline
$p p \to \mbox{sparton pairs}$	& $86.7$	& \hspace{1cm} $152$	&  $139$	&	$1970$& \hspace{1cm} $2770$& $2760$	\\	
$p p \to \mbox{slepton pairs}$	& $24.0$	& $19.9$	&  $21.1$	&	$96.7$&	$83.9$&	$88.1$\\
\multirow{2}{*}{\renewcommand{\arraystretch}{0.5}$p p \to \begin{array}{l} \mbox{gaugino pairs},\\ \mbox{gaugino+sparton}\end{array}$}	& \multirow{2}{*}{$32.2$}	&	\multirow{2}{*}{$38.6$} &	\multirow{2}{*}{$43.3$}	 &	\multirow{2}{*}{$224$}	 &	\multirow{2}{*}{$259$}	 &	\multirow{2}{*}{$284$}	\\
								&				&				&				&				&				&		\\
\hline
$p p \to \mbox{sparticle pairs}$	& $143$	& $210$		& $203$ &		$2290$	&	$3110$ &	$3130$	\\
\hline\hline
\end{tabular*}
\caption{Total LO cross section (in fb) for the benchmark scenarios 
BE1, BE2 and BE3 for pair production of all SUSY particles (last row)
and three of its subprocesses: sparton (\ie squark and gluino) pair
production (second row), slepton pair production (third row) and
electroweak (EW) gaugino pair or EW gaugino plus sparton production
(fourth row).  We separately assume $pp$ collisions at cms
energies of $\sqrt{s}=7\tev$ and $\sqrt{s}=14\tev$. The cross sections
are calculated with {\tt Herwig}. We have simulated $\approx 15~000$
($\approx 250~000$) SUSY events for the $7\tev$ ($14\tev$) MC signal
sample.}
\label{Tab:signalMC}
\end{table*}

We give in Table~\ref{Tab:signalMC} leading order cross sections for
sparticle pair production at the LHC for the three benchmark scenarios
BE1, BE2 and BE3, \cf Table~\ref{Tab:benchmarks}.  We
separately assume cms energies of $\sqrt{s}=7\tev$ and
$\sqrt{s}=14\tev$. We present the cross sections for the signal (last
row), \ie pair production of all sparticles, and for three of its
subprocesses: The production of sparton pairs (second row), where we
consider squarks and gluinos as spartons, slepton pair production
(third row) and the production of electroweak (EW) gaugino pairs or an
EW gaugino in association with a squark or gluino (fourth row). For
all benchmark points, sparton pair production is the dominant SUSY
production process. Therefore, the majority of the SUSY events will
fulfill our signature expectations including at least two hard jets,
\cf Sec.~\ref{Sect:signatures}. 

For the reconstruction of jets, we employ {\tt FastJet~2.4.1}
\cite{Cacciari:2005hq, Cacciari:web} using the kt-algorithm with cone
radius $\Delta R=0.4$. Here $\Delta R \equiv \sqrt{(\Delta
\phi)^2 + (\Delta \eta)^2}$, where $\eta$ ($\phi$) is the
pseudorapidity (azimuthal angle). We only select jets and leptons (\ie
electrons and muons) if $|\eta|<2.5$ and if their transverse momentum
is larger than $10\gev$. In addition, leptons are rejected, if the
total transverse momentum of all particles within a cone of $\Delta R
< 0.2$ around the lepton three-momentum axis exceeds $1\gev$.

\subsection{Kinematic Distributions}
\label{Sect:kinematic}

In this section we discuss kinematic distributions for the benchmark
points of Table~\ref{Tab:benchmarks} and motivate our cuts of
Sec.~\ref{Sect:cutflow}.  The distributions correspond to our $7\tev$
event sample and are normalized to one.

\begin{figure*}[t]
	\setlength{\unitlength}{1in} \centering
	\subfigure[\,$p_T$ distribution of all
		selected electrons. \label{Fig:lepton_pT_a}]{\includegraphics[scale=0.42]{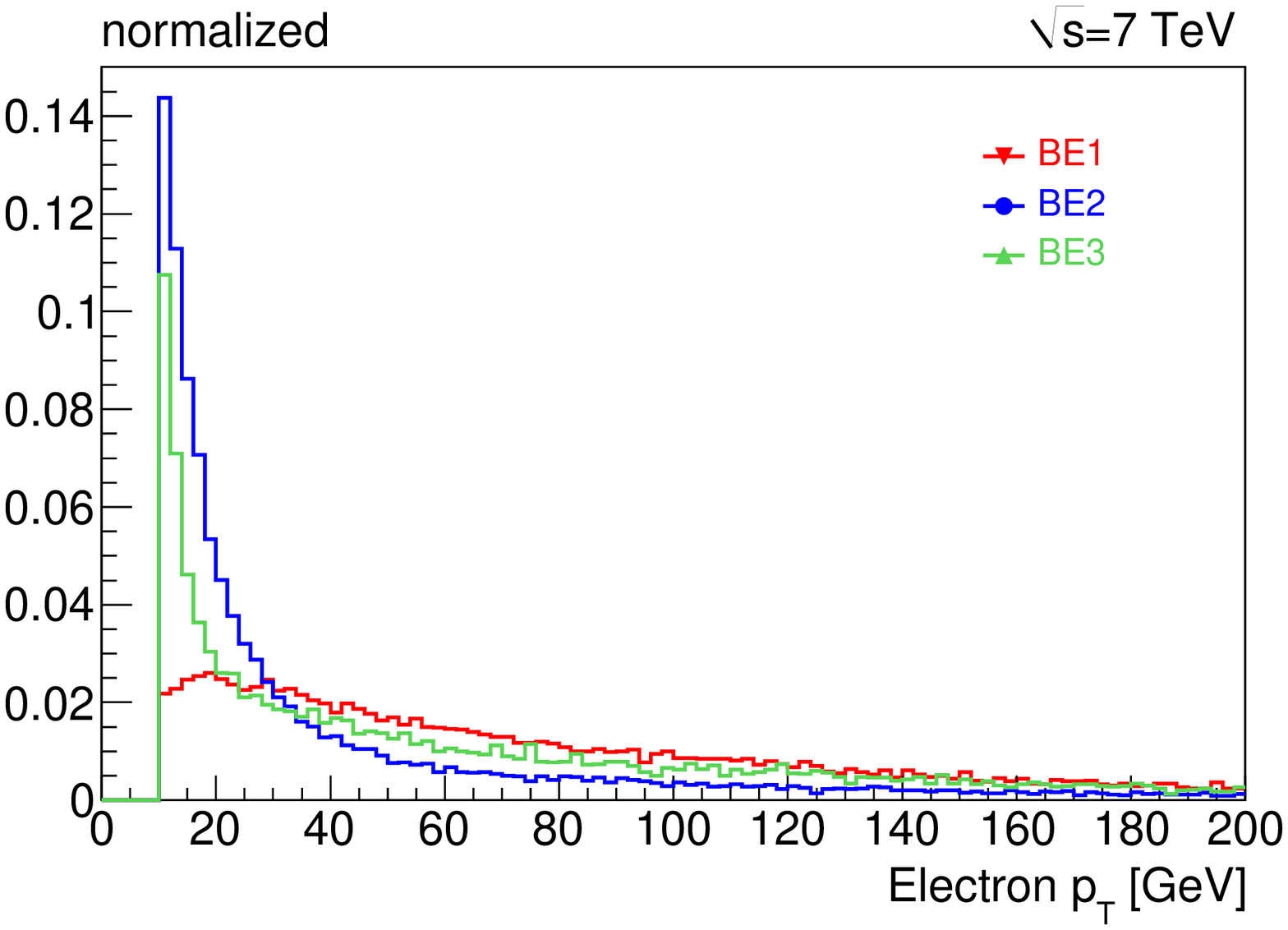}}
	\subfigure[\,$p_T$ distribution of all
		selected muons.\label{Fig:lepton_pT_b}]{\includegraphics[scale=0.42]{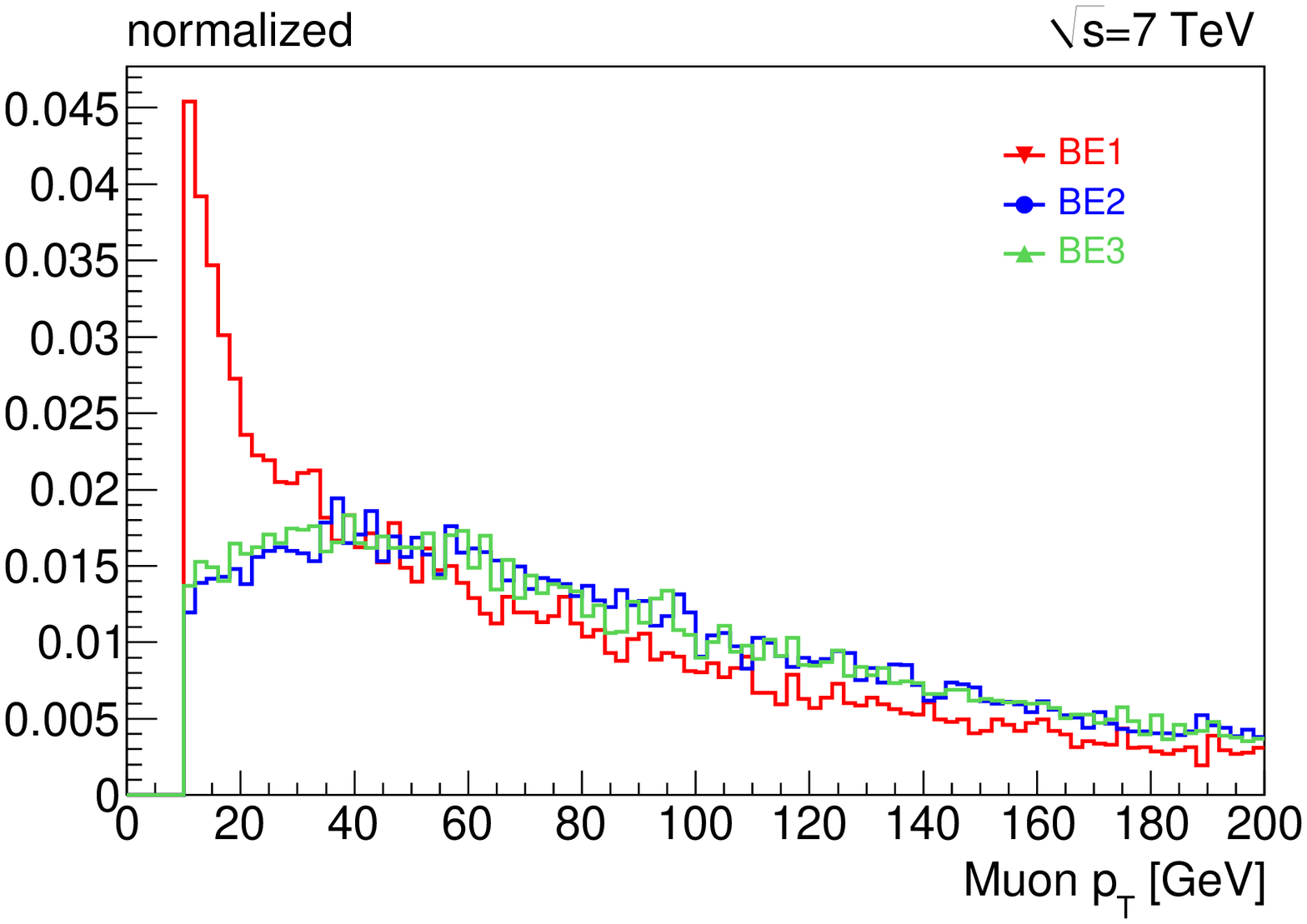}}
	\caption{Electron [Fig.~\ref{Fig:lepton_pT_a}] and
		muon [Fig.~\ref{Fig:lepton_pT_b}] $p_T$ distributions
		at the LHC for the $\Bthree$ mSUGRA benchmark models
		BE1 (red), BE2 (blue) and BE3 (green) for a cms energy
		of $\sqrt{s}=7\tev$ and after object selection
		cuts. The distributions are normalized to one.}
		\label{Fig:lepton_pT}
\end{figure*}	
		
The $p_T$ distribution of all electrons [muons] after object selection
(\cf the last paragraph of Sec.~\ref{Sect:object_selection}) is shown
in Fig.~\ref{Fig:lepton_pT_a} [Fig.~\ref{Fig:lepton_pT_b}] for the
$\Bthree$ mSUGRA benchmark models BE1, BE2 and BE3.  In all scenarios,
the electrons mostly stem from the neutralino decay $\neut_1 \to
\sse_R e$, while many of the muons come from the LSP decay $\sse_R \to
\mu \nu_\tau$, \cf Appendix~\ref{App:benchmarks}.

We observe in Fig.~\ref{Fig:lepton_pT_a} that BE1 leads to the in
average hardest electrons.  In this scenario, the mass difference
between the $\neut_1$ (decaying often via $\neut_1 \to e \sse_R$) and
the $\sse_R$ LSP is about 27 GeV and thus quite large (compared to the
other benchmark points).  Furthermore, the $\stau_1$ NLSP decays
dominantly via the $R$-parity violating decay $\stau_1 \to e
\nu_\mu$. A large fraction of the $\stau_1$ mass is thus transformed
into the 3-momentum of an electron. From both sources, we obtain
electrons with large $p_T$.  For example, $81\%$ of all selected 
electrons have $p_T^\mathrm{el} \gtrsim 25 \gev$ in BE1.

The situation for BE2 and BE3 is different. Because of the smaller
mass difference between the $\neut_1$ and the $\sse_R$ LSP (compared
to BE1), the electrons from $\neut_1$ decay are less energetic. For
instance, the fraction of selected electrons with
$p_T^\mathrm{el}\lesssim 25\gev$ is $55\%$ ($34\%$) for BE2
(BE3). Furthermore, the electron multiplicity is reduced in these
scenarios, because many electrons fail the lower $p_T$ cut
($p_T^\mathrm{el} > 10 \gev$) of the object selection. Due to this,
$30\%$ ($50\%$) of all events do not contain any selected electron in
BE2 (BE3).

In contrast, the situation for the muons, Fig.~\ref{Fig:lepton_pT_b},
is reversed (compared to the electrons).  A large amount of the muons
are soft in BE1, whereas BE2 and BE3 have a harder muon $p_T$
spectrum.  Note that for BE1, a sizable fraction of all muons do
not even fulfill the object selection requirement of $p_T >
10\gev$, so that $34\%$ of all events do not contain any
selected muon. These muons in BE1 stem, for example, from the 3-body
decays of the $\ssmu_R$ into the $\sse_R$ or the $\stau_1$ and are in
general soft due to decreased phase space, \cf
Table~\ref{Tab:BE1}.  In contrast, the muons in BE2 and BE3 are on
average much harder, since the majority of these muons originate from
the $\sse_R$ LSP decay.

We conclude, that the lepton $p_T$ spectrum strongly depends on the
sparticle mass spectrum.  Therefore, we desist from making further
requirements on the lepton $p_T$ since this would imply a strong model
dependence in the event selection. We will only require at least three
charged (and isolated) leptons as one of our cuts in the next section.

\begin{figure*}[t]
		\setlength{\unitlength}{1in} 
		\centering
 		\subfigure[\,$p_T$ distribution of the hardest jet.\label{Fig:jet_pT_a}]{\includegraphics[scale=0.42]{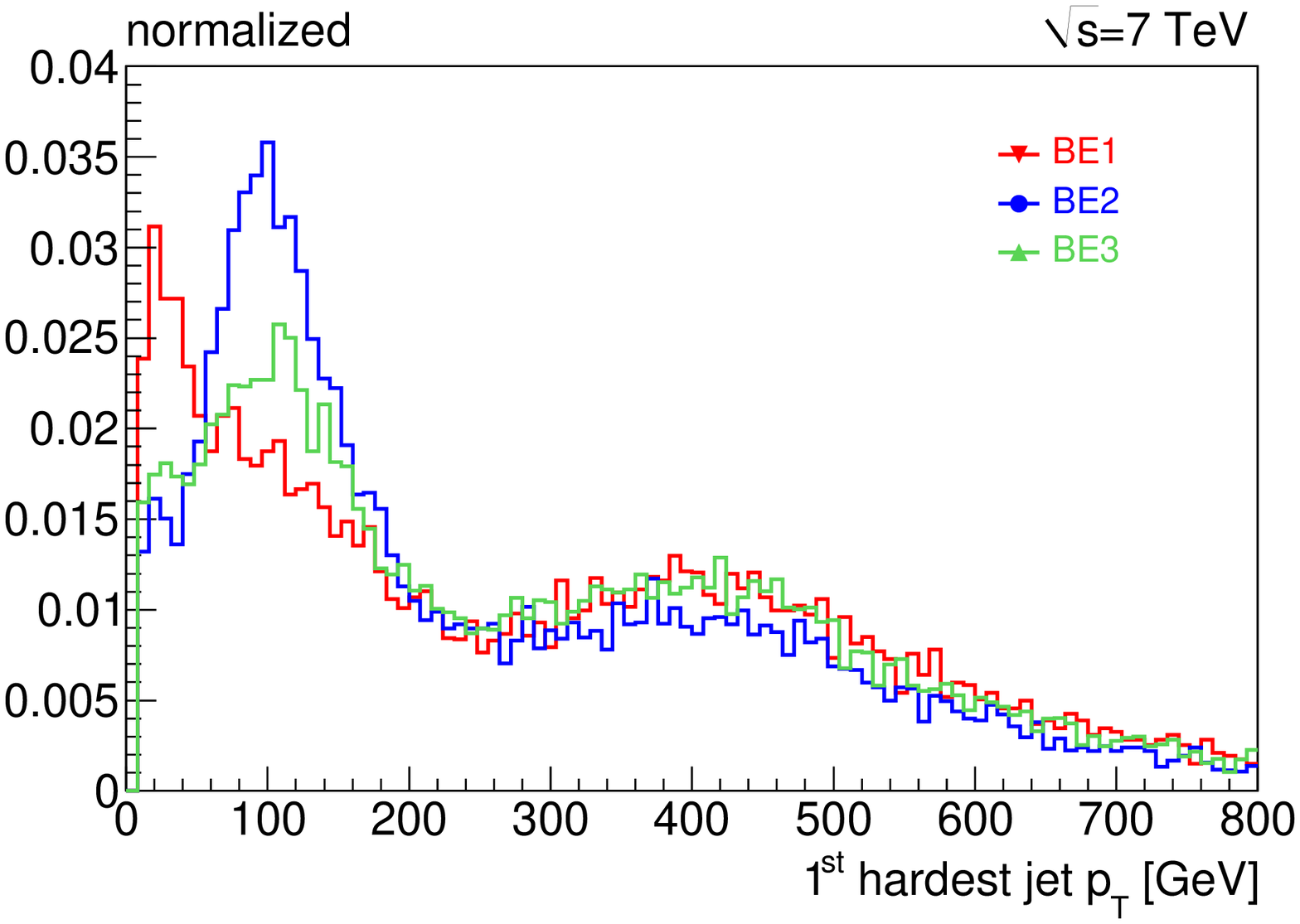}}
		\subfigure[\,$p_T$ distribution of the 2nd hardest jet.\label{Fig:jet_pT_b}]{\includegraphics[scale=0.42]{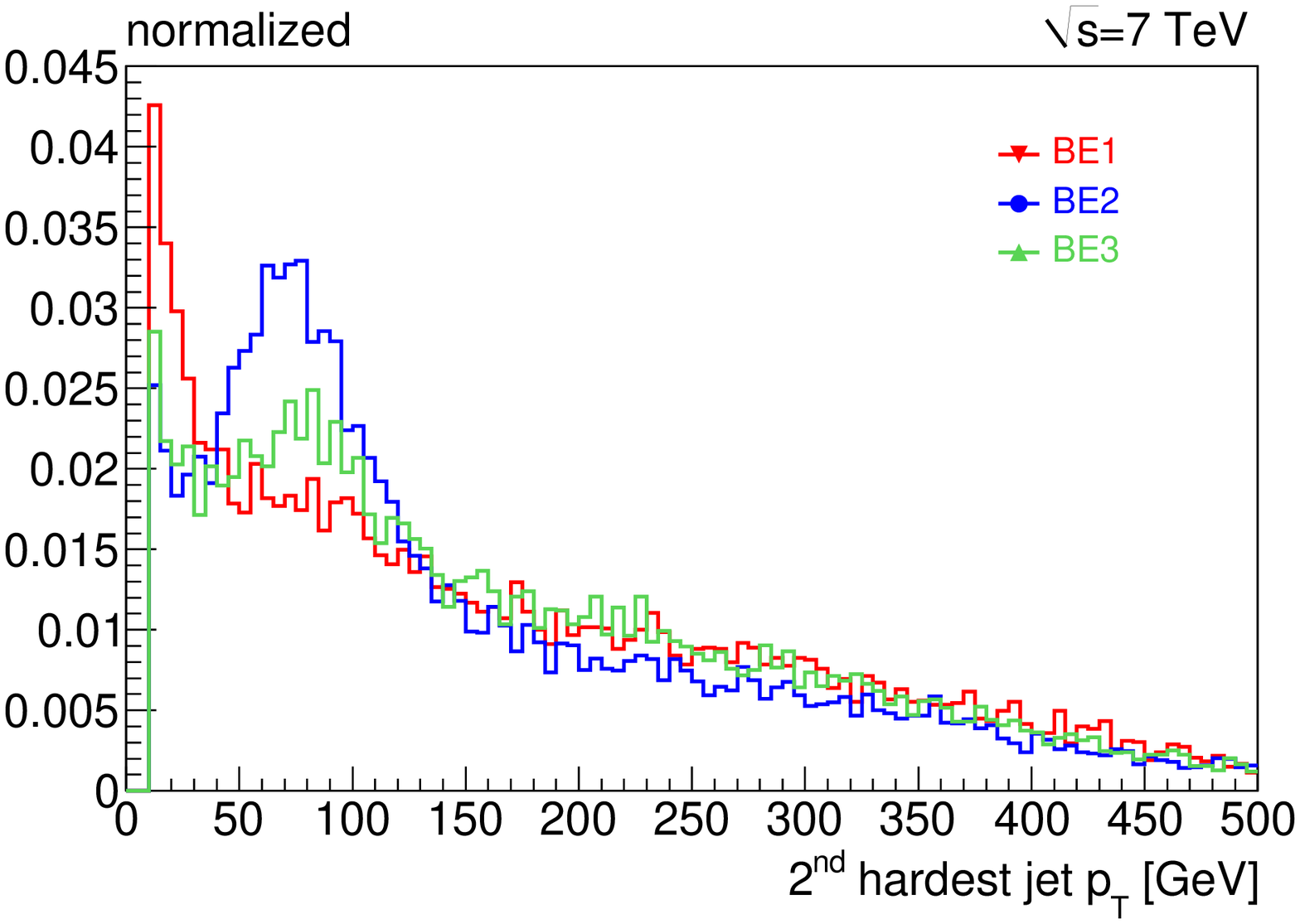}}
	\caption{Jet $p_T$ distributions at the LHC for the $\text{B}_3$ mSUGRA benchmark models BE1 (red), BE2 (blue) 
	and BE3 (green) for a cms energy of $\sqrt{s}=7\tev$ and after object selection cuts. 
	The distributions are normalized to one.}	
	\label{Fig:jet_pT}
\end{figure*}	

We show in Fig.~\ref{Fig:jet_pT_a} [Fig.~\ref{Fig:jet_pT_b}] the $p_T$
distribution of the [second] hardest jet for the benchmark points BE1,
BE2 and BE3. For all scenarios, we observe a broad peak of the hardest
jet $p_T$ at around $400\gev$.  Many of these jets stem from the
decays of first and second generation squarks into the $\neut_1$; \cf
Table~\ref{Tab:BE1}--Tab~\ref{Tab:BE3}.

We find another peak in Fig.~\ref{Fig:jet_pT_a} as well as in
Fig.~\ref{Fig:jet_pT_b} at around $100\gev$. These jets stem mainly
from the $t$ quark decay products from $\sstop_1 \to t \neut_1$
decay. The peak is most pronounced in BE2, since here we have a light
$\sstop_1$ mass, $M_{\sstop_1} = 448\gev$, and thus an enhanced
$\sstop_1$ pair production cross section.  In contrast, the $\sstop_1$
mass is about $80\gev$ heavier in BE1 and therefore, the peak is
hardly visible in Fig.~\ref{Fig:jet_pT}.

For BE1, the $p_T$ distribution of the hardest and second hardest jet
peaks at low values.  These soft jets stem from initial and final
state radiation. They appear as the hardest jets in EW gaugino and
slepton pair production which forms a sizable fraction ($39\%$) of all
SUSY production processes in BE1, \cf Table~\ref{Tab:signalMC}. They
are less important for BE2 and BE3.  However, this picture will change
for a cms energy of $\sqrt{s}=14\tev$, where sparton pair production
is much more dominant in BE1.

Because most events possess at least two jets, we demand in the
following section at least two jets as one of our cuts.  Furthermore,
we take into account that many jets (and some of the leptons) are
hard, \textit{i.e.} we demand the visible effective mass to be larger
than a few 100 GeV; see the next section for details.

\subsection{Event Selection and Cutflow}
\label{Sect:cutflow}

\begin{figure*}[t]
	\setlength{\unitlength}{1in} 
	\centering
	\subfigure[\,Lepton ($e$, $\mu$) multiplicity with no cut applied.\label{Fig:Cut1}]
				{\includegraphics[scale=0.4]{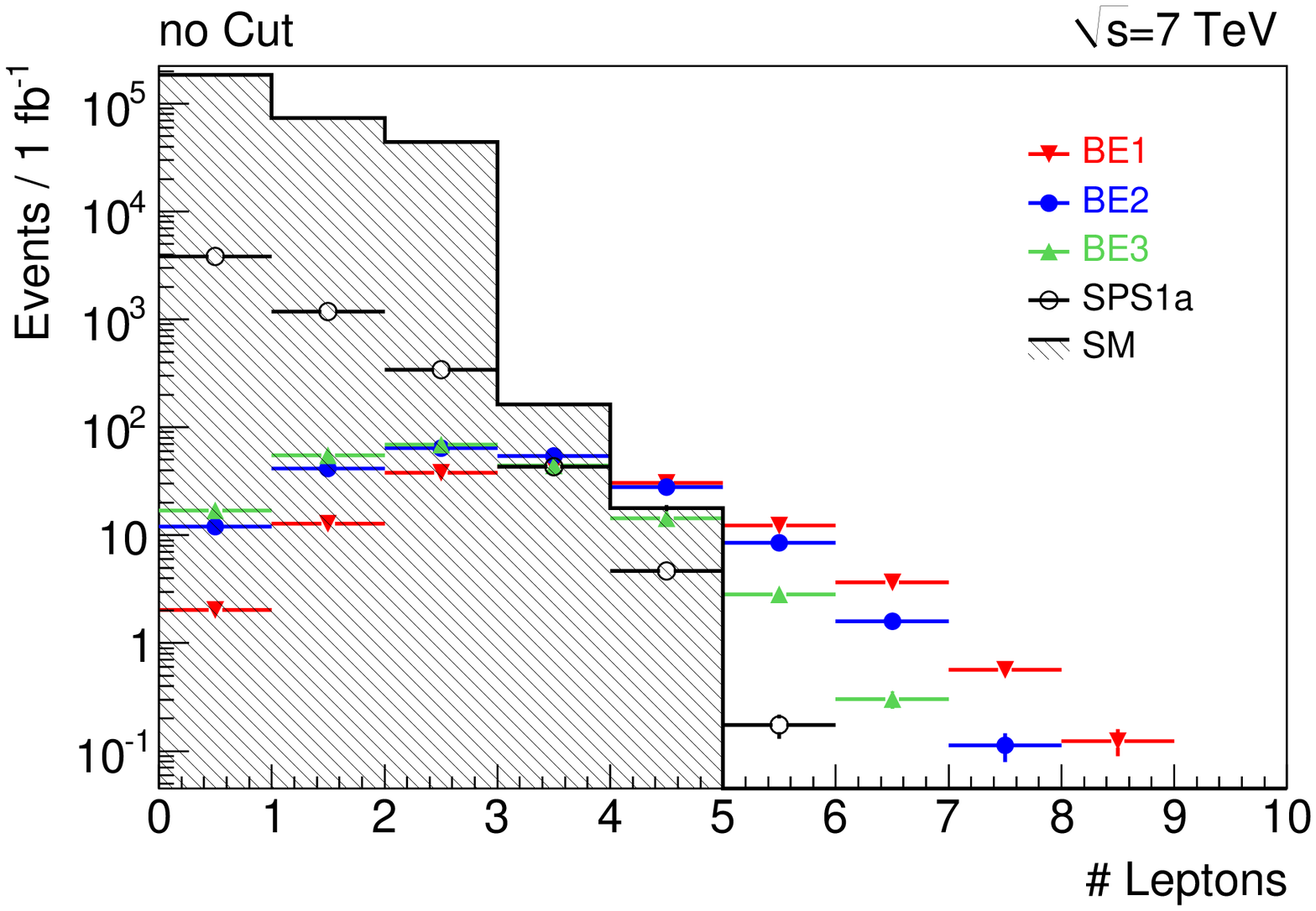}}
	\subfigure[\,Jet multiplicity after the trilepton cut has been applied.\label{Fig:Cut2}]
				{\includegraphics[scale=0.4]{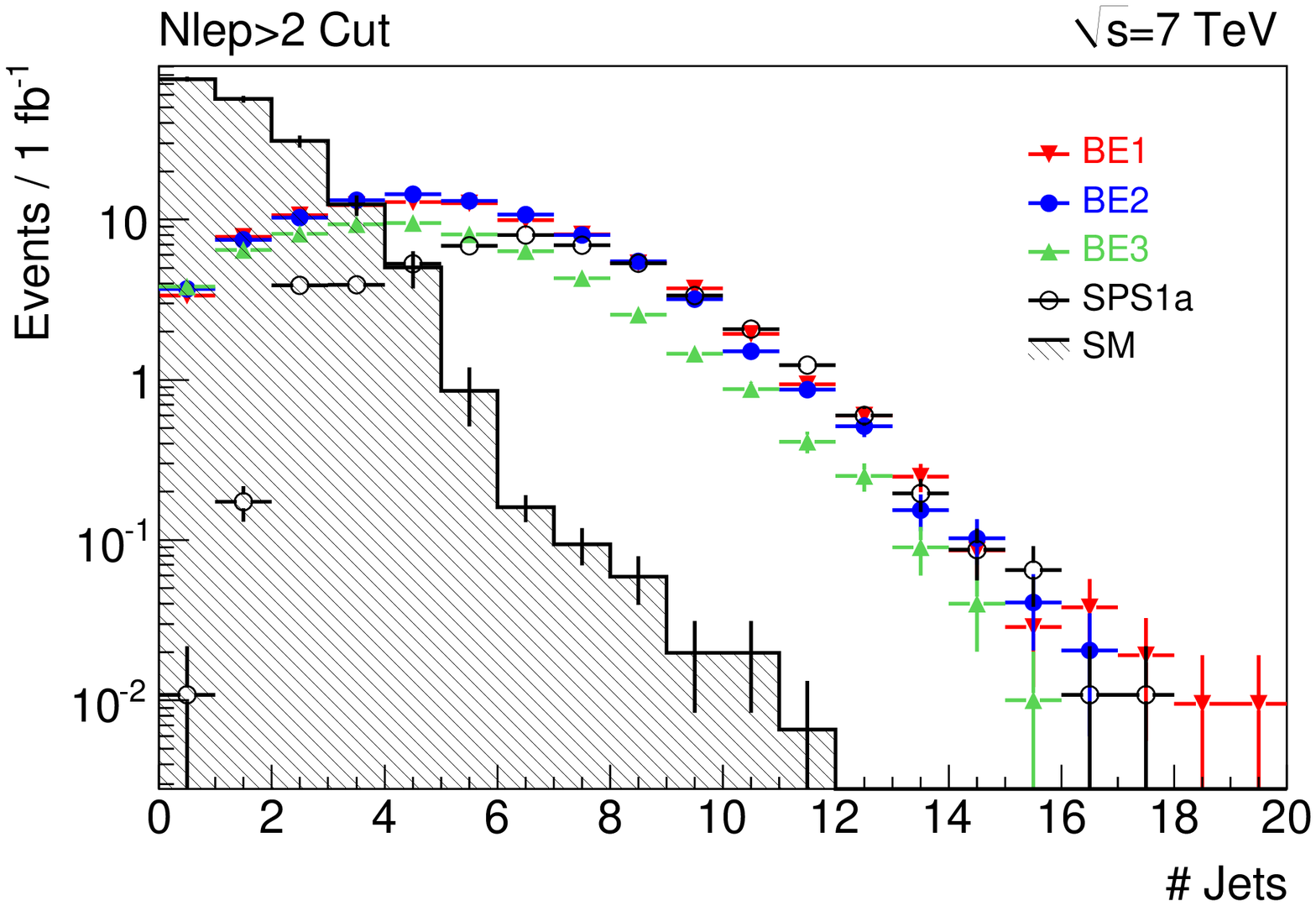}}
	\subfigure[\,Invariant mass distribution of opposite-sign-same-flavor 
                   (OSSF) lepton pairs after the trilepton and jet multiplicity cut have been applied.\label{Fig:Cut3}]
				{\includegraphics[scale=0.4]{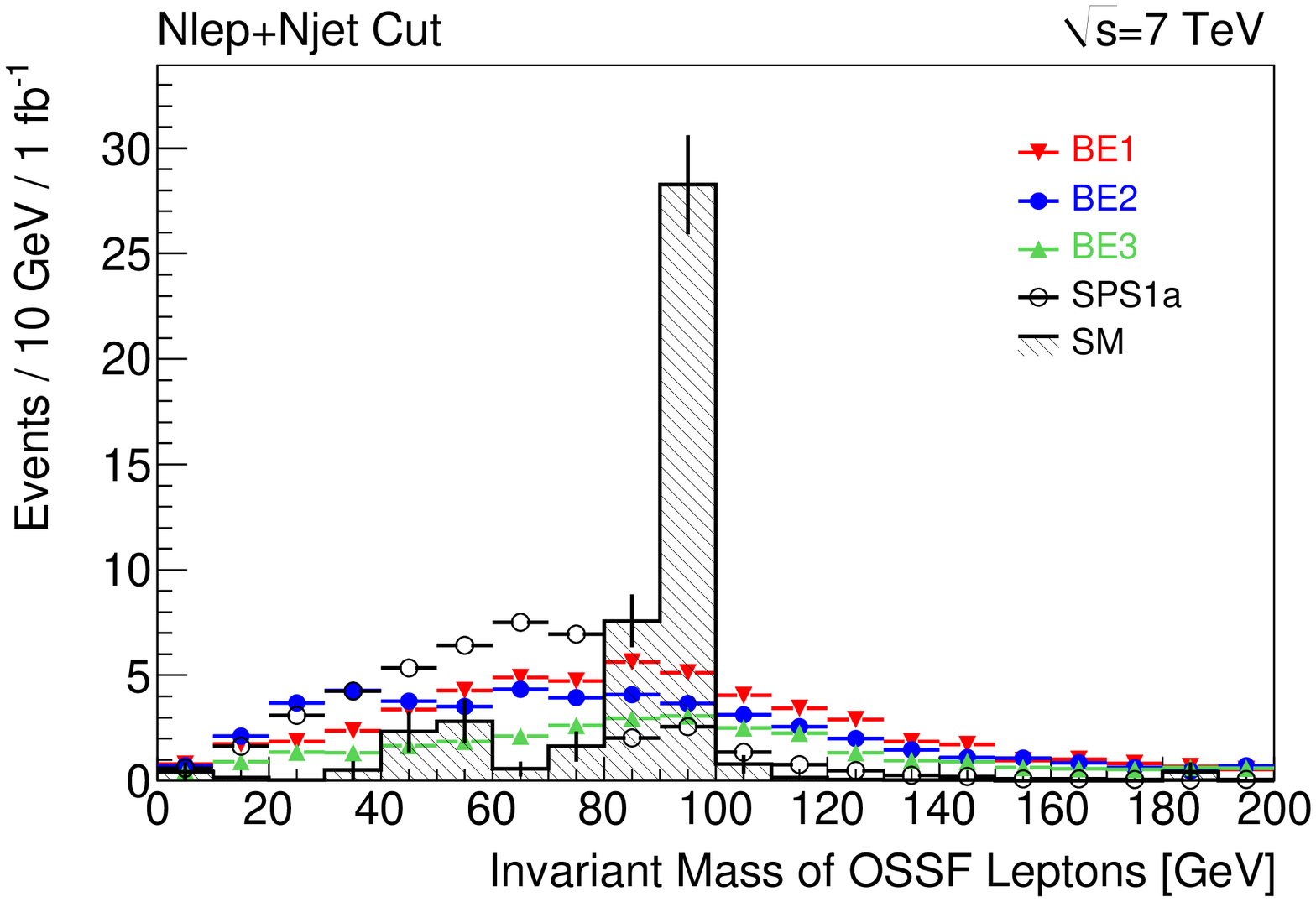}}
	\subfigure[\,Effective visible mass distribution after the trilepton, jet multiplicity and 
				  OSSF mass cut have been applied.\label{Fig:Cut4}]
				{\includegraphics[scale=0.4]{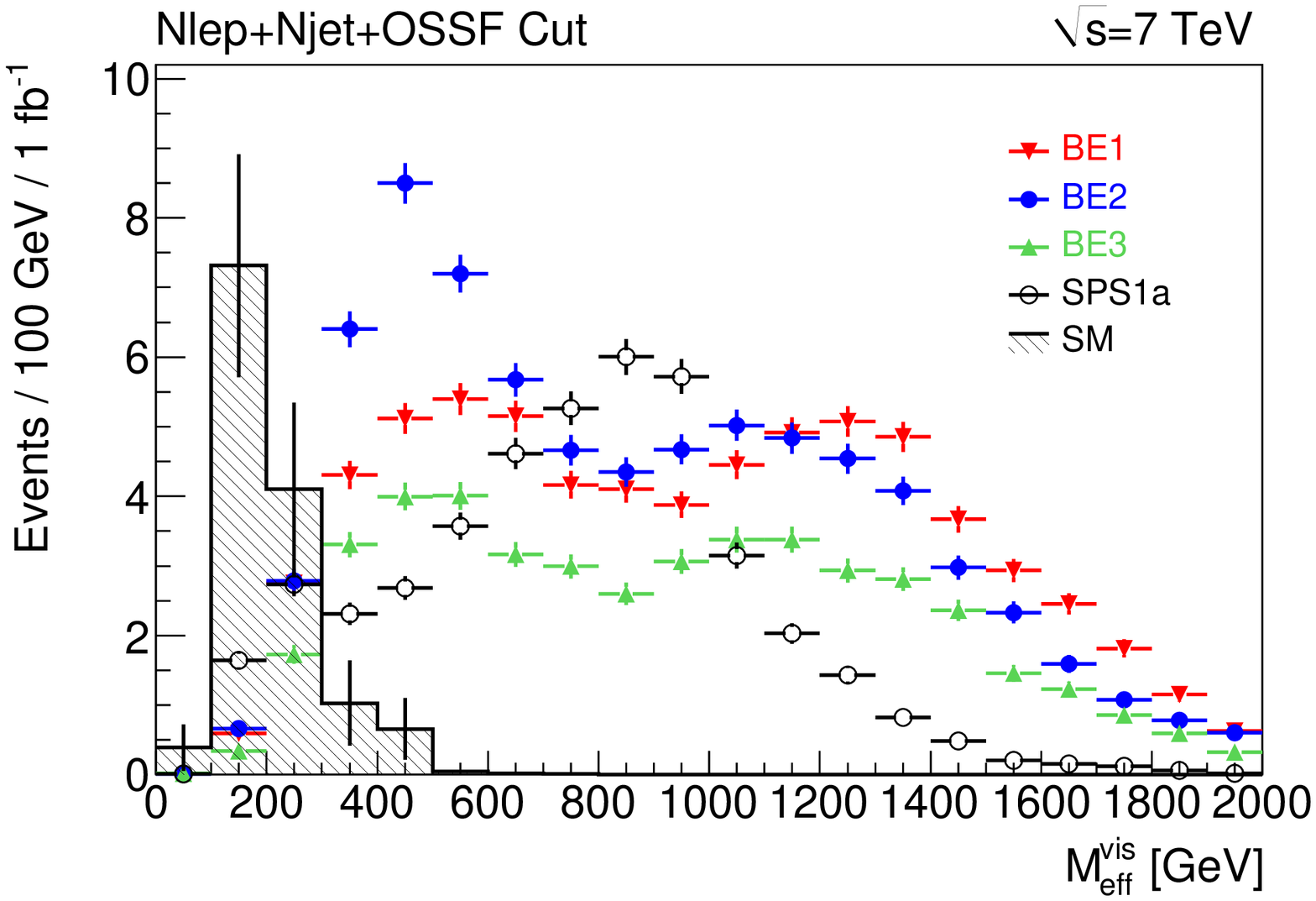}}
	\caption{Event distributions of several cut variables: 
Fig.~\ref{Fig:Cut1} lepton multiplicity, Fig.~\ref{Fig:Cut2} jet
multiplicity, Fig.~\ref{Fig:Cut3} OSSF lepton invariant mass and
Fig.~\ref{Fig:Cut4} visible effective mass for the SM background (gray
patterned) and the SUSY models BE1 (red), BE2 (blue), BE3 (green) and
SPS1a (white). Note that Fig.~\ref{Fig:Cut1} and \ref{Fig:Cut2} are
given on a logarithmic scale. The distributions show the 
number of events before the event selection cut on the 
respective variable (see text) is applied. They are normalized to an 
integrated luminosity of 1 $\mathrm{fb}^{-1}$ at $\sqrt{s}=7\tev$. The
error bars correspond to statistical fluctuations of our MC samples.}
\label{Fig:Cuts}
\end{figure*}

We now develop a set of cuts in order to obtain a
statistically significant signal and a good signal to (SM)
background ratio. To motivate the different selection steps, we show
in Fig.~\ref{Fig:Cuts} the event distributions that correspond to the
different cut variables before the respective cut is applied. We give
distributions for the three $\sse_R$ LSP benchmark models (BE1, BE2,
BE3), for the SM background, and, for comparison, for the $R$-parity
conserving benchmark model SPS1a~\cite{Allanach:2002nj}\footnote{
SPS1a has a mass spectrum similar to BE1, BE2, 
and BE3. The main difference lies in the light part of the spectrum,
where we have in SPS1a a stable and invisible $\neut_1$ LSP. The
$\stau_1$ is the NLSP. Furthermore, the overall mass scale is a bit
lower, e.g. the squark and gluino masses are around 500-600
GeV.}.  The distributions correspond to an integrated luminosity of
1~fb$^{-1}$ at $\sqrt{s}=7\tev$.

\begin{table*}[t]
\centering
\setlength{\tabcolsep}{1pc}
\begin{tabular*}{\textwidth}{@{\extracolsep{\fill}}lrrrrr}
\hline\hline
Sample	&	Before cuts	&	$N_\mathrm{lep}\ge 3$	&	$N_\mathrm{jet} \ge 2$		&	$M_\mathrm{OSSF}$	&	$M_\mathrm{eff}^\mathrm{vis} \ge 300\gev$	\\
\hline
top				&	$97111 \pm 197$	& $14.9 \pm 2.2$	& $13.8 \pm 2.1$	& $12.0 \pm 2.1$	& $2.1\pm 0.8$	\\
$Z +\mbox{jets}$	&	$153591\pm 254$	& $51.9 \pm 4.3$	& $16.6 \pm 2.4$	& $1.0 \pm 0.6$	& $\lesssim 1.0$	\\
$W +\mbox{jets}$	&	$38219 \pm 103$ 	& $\lesssim 1.0$	& $\lesssim 1.0$ 	& $\lesssim 1.0$	& $\lesssim 1.0$	\\
di-boson			&	$21331 \pm 48$	& $179.2 \pm 3.0$	& $53.5 \pm 2.0$	&$2.6 \pm 0.4$		& $0.7 \pm 0.2$ 	\\
\hline
all SM			&	$310252\pm 341$	& $264.0\pm 5.7$	& $83.9 \pm 3.8$	& $15.6\pm 2.2$	& $2.8 \pm 0.8$	\\	
\hline
BE1				& $143.1\pm 1.2$		& $90.5\pm 0.9$	& $79.4 \pm 0.9$	& $68.8\pm 0.8$	& $65.5\pm 0.8$	\\
$S/\sqrt{B}$		& 		-			& $5.6$			& $8.7$			& $17.4$			& $39.1$	\\	
\hline
BE2				& $210.4\pm 1.5$		& $92.6\pm 1.0$	& $81.4\pm 0.9$	& $73.8\pm 0.9$	& $70.4 \pm 0.8$	\\
$S/\sqrt{B}$		&		-			& $5.7$			& $8.9$			& $18.7$			& $42.1$	\\	
\hline
BE3				& $202.7\pm 1.4$		& $61.6\pm 0.8$	& $51.3 \pm 0.7$	& $45.2\pm 0.7$	& $43.2 \pm 0.7$	\\
$S/\sqrt{B}$		&		-			& $3.8$			& $5.6$			& $11.4$			& $25.8$	\\	
\hline\hline
\end{tabular*}
\caption{Number of SM background and signal events after each step in
the event selection corresponding to an integrated luminosity of $1~
\mathrm{fb}^{-1}$ at $\sqrt{s}=7\tev$. For each signal model (BE1, BE2
and BE3, see Table~\ref{Tab:benchmarks}), we show $S/\sqrt{B}$ as 
significance estimator. The uncertainties correspond to statistical
fluctuations.}
\label{Tab:cutflow}
\end{table*}

In Table~\ref{Tab:cutflow}, we give the number of background and
signal events after each cut of the analysis. Furthermore, we provide
for each signal benchmark scenario the significance estimator
$S/\sqrt{B}$, where $S$ ($B$) is the number of signal (SM background)
events. In general, the signal can be defined to be observable
if~\cite{Baer:2010tk}
\begin{align}
S \ge max\left[5\sqrt{B},~5,~0.5 B\right].
\label{Eqn:Sobserved}
\end{align}
The requirement $S\ge 0.5B$ avoids the possibility that a small signal
on top of a large background could otherwise be regarded as
statistically significant, although this would require the background
level to be known to an excellent precision. In the case of a very low
background expectation, $B < 1$, we still require $5$ signal events
for a discovery.

As we have seen in Sec.~\ref{Sect:signatures}, we expect an extensive
number of charged leptons in the final state.  However, the lepton
flavor multiplicity, \textit{i.e.} the multiplicity of electrons and
muons, depends strongly on the LSP flavor as well as on the dominant
$\mathbf{\Lambda}$ coupling, \cf Table~\ref{Tab:signatures}. In
addition, as we have seen in the last section, the $p_T$ spectrum of
the leptons is strongly correlated to the details of the mass
hierarchy.  Therefore, in order to be as model independent as
possible, we simply demand as our first cut three charged leptons
(electrons or muons) in the final state without further requirements
on the $p_T$ (beside the object selection cut of $p_T>10 \gev$).

How useful this cut is, can be seen in Fig.~\ref{Fig:Cut1}, where we
show the lepton multiplicity after object selection cuts. The
distribution for the $\text{B}_3$ benchmark scenarios peaks around 2-3
leptons, whereas most of the SM background events posses less than
three electrons or muons.  In principle, by demanding at least five
charged leptons in the final state, we can already get a (nearly)
background free event sample. However, such a cut would also
significantly reduce the number of signal events and is therefore less
suitable for an analysis of early data.  We also observe in
Fig.~\ref{Fig:Cut1} many more leptons in the $R$-parity
violating scenarios than in SPS1a. This is expected, due to the
additional leptons from the decays of and into the selectron LSP.

As can be seen in the third column of Table~\ref{Tab:cutflow}, after
demanding three leptons, the main SM background comes from di-boson
events. They account for $68 \%$ of the background.  Furthermore, no
$W+\mbox{jets}$ events survive this cut, indicated by ``$\lesssim
1.0$" events in the fourth row of Table~\ref{Tab:cutflow}. At the same
time, the number of signal events is reduced to $63\%$, $44\%$ and
$30\%$ for BE1, BE2 and BE3, respectively. Because of the low mass
difference between the $\neut_1$ and the $\sse_R$ LSP in BE3,
many electrons from $\neut_1$ decay fail the object selection cuts;
\cf the discussion of Fig.~\ref{Fig:lepton_pT_a}.  BE1 and BE2 might
already be observable after the first cut, \textit{i.e.} $S/\sqrt{B}>5$.

Next, we will use the fact that we expect several jets from squark and
gluino decays; see Sec.~\ref{Sect:signatures}. The jet multiplicity
after demanding three leptons is shown in Fig.~\ref{Fig:Cut2}. Because
of the weak object selection criteria for the jets ($p_T>10 \gev$) and
the small radius for the jet reconstruction ($\Delta R = 0.4$), we
observe a high jet multiplicity. As discussed in
Sec.~\ref{Sect:signatures}, we expect at least two jets from squark
and gluino decays. Therefore we demand as our second cut (fourth
column of Table~\ref{Tab:cutflow}) the number of jets to be larger
than two, \textit{i.e.}  $N_\mathrm{jet} \ge 2$. This cut suppresses
roughly two thirds of the di-boson backgrounds $WZ$ and $ZZ$
as well as of the $Z+\mbox{jets}$ background. However, di-boson
production, especially $WZ + j$, still accounts for most of the
background. The number of signal events is only reduced by
12\%-17\%. After this cut, all our benchmark points fulfill the criteria in Eq.~\eqref{Eqn:Sobserved} and are thus observable.

In order to further reduce the SM backgrounds involving $Z$ bosons, we
construct all possible combinations of the invariant mass of
opposite-sign-same-flavor (OSSF) leptons. The distributions (after the
three lepton and $N_\mathrm{jet} \ge 2$ cut) are shown in
Fig.~\ref{Fig:Cut3}.  As expected, the SM background has a large peak
at the $Z$ boson mass $M_Z = 91.2\gev$, while the signal distribution
is mostly flat in that region. Thus as our third cut (fifth column of
Table~\ref{Tab:cutflow}) of our event selection, we reject all events
where the invariant mass of at least one OSSF lepton pair lies
within a $10\gev$ window around the $Z$ boson mass,
\ie we demand
\begin{align}
M_\mathrm{OSSF} \not \in \left[81.2 \gev, 101.2\gev \right].
\end{align}
This cut strongly reduces the $Z+\mbox{jets}$ and di-boson backgrounds,
leaving $t\overline t$ as the dominant SM background. Roughly $90\%$
of the signal events (for all benchmark scenarios) survive this
cut. The statistical significance now lies between 10 and 20 for all
benchmark points.

As we have shown in Sec.~\ref{Sect:kinematic}, our SUSY events contain
a large amount of energy in the form of high-$p_T$ jets and
leptons. Thus, we construct the visible\footnote{We denote this
variable as \textit{visible} effective mass because it does not
include the missing transverse energy as in other definitions of the
effective mass~\cite{Aad:2009wy}.}  effective mass,
\begin{align}
\meff \equiv \sum_{i=1}^{4} p_T^{\mathrm{jet},i} + \sum_{\mathrm{all}} p_T^\mathrm{lep},
\end{align}
\textit{i.e.} the scalar sum of the absolute value of the transverse
momenta of the four hardest jets and all selected leptons in the
event.  The visible effective mass distribution is shown in
Fig.~\ref{Fig:Cut4}. The SM background dominates for $\meff<300\gev$,
while most of the signal events exhibit a visible effective mass above
$300\gev$. This value is slightly higher for the $14 \tev$ dataset.
Therefore, we demand as our last cut of our event selection (last
column of Table~\ref{Tab:cutflow})
\begin{align}
\meff > \left\{ \begin{array}{l} 300 \gev, \quad \mbox{if}\,\sqrt{s} = 7\tev, \\ 400 \gev, \quad \mbox{if}\,\sqrt{s} = 14\tev. \end{array} \right.
\end{align}
After this cut, only $2.8\pm0.8$ SM events remain at $\sqrt{s}
=7\tev$ and an integrated luminosity of 1 fb$^{-1}$. The
background is dominated by $t\overline t$ production. The
signal is nearly unaffected by this cut as can be seen in
Table~\ref{Tab:cutflow}. The statistical significance is now roughly
as large as 25 (40) for the benchmark point(s) BE3 (BE1 and BE2).
Furthermore, the signal to background ratio is now of
$\mathcal{O}(10)$. Therefore, systematic uncertainties of
the SM backgrounds are not problematic. A signal is clearly visible.

We observe in Fig.~\ref{Fig:Cut4} two peaks in the visible effective
mass distributions for our benchmark scenarios. The peak at lower
values of $\meff$ contains mainly events from $\sstop_1 $ pair
production, while events from (right-handed) first and second
generation squark or gluino production build the second peak at higher
$\meff$ values. Because of the large mass difference between the
$\sstop_1$ and the other squarks of about $400\gev - 500\gev$
(depending on the model, see Table~\ref{Tab:BE1}-Table~\ref{Tab:BE3}),
these peaks are clearly separated in the visible effective mass. We
make use of this fact in Sec.~\ref{Sect:mass_reco} when we present a
method to reconstruct the masses of both the $\sstop_1$ and the
right-handed first and second generation squarks.

In order to test the flavor sensitivity of our analysis, we have
applied our cuts to a modified version of the benchmark models
presented in Table~\ref{Tab:benchmarks}. Instead of $\lam_{231}$, we
chose $\lam_{131}$ ($\lam_{132}$) as the dominant $R$-parity coupling
at $M_{\rm GUT}$ to obtain the $\sse_R$ ($\ssmu_R$) as the LSP, while
leaving the other $\Bthree$ mSUGRA parameters unchanged. The results
for the $\ssmu_R$ LSP scenarios are in agreement with the original
benchmark scenarios within statistical fluctuations of the MC samples.

However, for the $\sse_R$ LSP scenarios with a dominant $\lam_{131}$
coupling at $M_{\rm GUT}$, the cut on the invariant mass of OSSF
leptons rejects more signal events than for scenarios with
$\lam_{231}$. For the modified scenario of BE1 (BE2), the number of
signal events passing the $M_\mathrm{OSSF}$ cut is reduced by around
$15\%$ ($3\%$) compared to the original results, \cf
Table~\ref{Tab:cutflow}. This difference is strongest for BE1--like
scenarios, because the endpoint of the di-electron invariant mass
distribution, where one electron comes from the $\neut_1$ decay and
the other from the $\sse_R$ LSP decay, \cf also Eq.~\eqref{Eqn:mll_b},
coincides with the upper value of the $Z$ boson mass window.  However,
this is just a coincidence and a different mass spectrum (compared to
BE1) with a $\sse_R$ LSP and $\lam_{131}$ at $M_{\rm GUT}$ will not
have such a suppression.

We conclude that in most cases, our detailed study of $\sse_R$ LSP
models with a dominant $R$-parity violating coupling $\lam_{231}$ is
representative for all $\Bthree$ mSUGRA models with a $\sse_R$ or
$\ssmu_R$ LSP. 

\begin{figure}[t]
	\setlength{\unitlength}{1in} 
	\centering
	\includegraphics[scale=0.42]{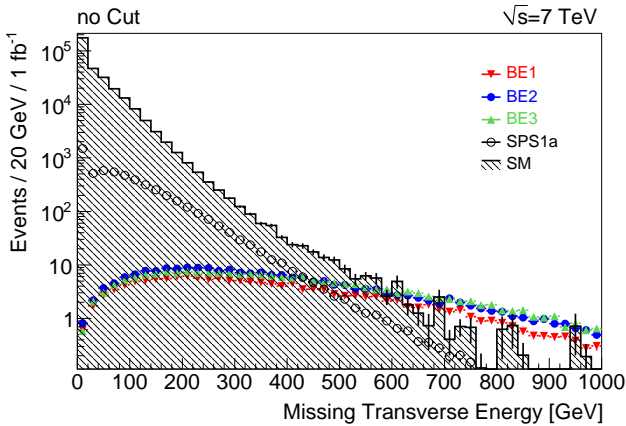}
	\caption{Missing transverse energy, $\etmiss$, distribution 
                 for BE1, BE2, BE3, SPS1a and the SM background. No 
                cuts are applied. The number of events 
                correspond to an integrated luminosity of $1~\mathrm
                {fb}^{-1}$ at $\sqrt{s}=7\tev$.}\label{Fig:etmiss}
\end{figure} 

To end this subsection, we present in
Fig.~\ref{Fig:etmiss} the missing transverse energy, $\etmiss$,
distribution for the benchmark scenarios, for SPS1a and for the
combined SM backgrounds before any cuts are applied. In
$R$-parity conserving scenarios like SPS1a, the $\neut_1$ LSP is
stable and escapes detection leading to large amounts of
$\etmiss$. However, for our benchmark points, even though the $\sse_R$
LSP decays within the detector, we observe a significant amount of
missing energy due to the neutrinos from the LSP decay. Moreover, the
$\etmiss$ distribution for SPS1a falls off more rapidly than in the
$\Bthree$ scenarios. This is because the neutrinos
are quite hard, resulting from a 2--body decay with a large mass
difference. Thus, $\Bthree$ scenarios can lead to even more missing
transverse energy than $R$-parity conserving scenarios. We have not
employed $\etmiss$ in our analysis, because our simple cuts
already sufficiently suppress the SM background. Furthermore, 
it is easier to reconstruct electrons and muons than missing energy, 
especially in the early stages of the experiments.

\subsection{Discovery Potential at the LHC}
\label{Sect:LHC_discovery}

In this subsection, we extend our previous analysis. We perform
a two dimensional parameter scan in the $\mhalf$--$\mzero$ plane
around the benchmark point BE1 (Table~\ref{Tab:benchmarks}).  For each
parameter point, we generate $1000$ signal events, \textit{i.e.} 
the pair production of all SUSY particles. We then apply 
the same cuts developed in the previous section.  We estimate 
the discovery potential of $\Bthree$ mSUGRA models with a $\sse_R$ LSP
for the early LHC run at $\sqrt{s}=7\tev$ and also give prospects for
the design energy of $\sqrt{s}=14\tev$.

\begin{figure*}[t]
\subfigure[\,Signal cross section in pb at the LHC at $\sqrt{s}=7\tev$.
The white dashed contour lines give the (first and second generation) 
$\squark_R$ mass in GeV.\label{Fig:xsec_7TeV}]{
	\begin{minipage}{7.8cm}
	   	\includegraphics[scale=1.0]{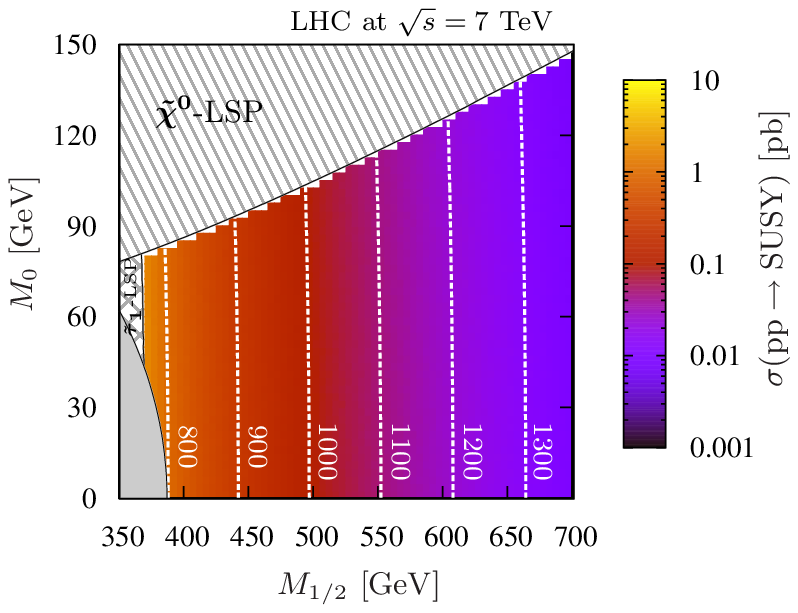}
	\vspace{0.0cm}	
	\end{minipage}
}
\subfigure[\,Selection efficiency for the signal events at the LHC at 
$\sqrt{s}=7\tev$.  The white dashed contour lines give the mass
difference, between the $\neut_1$ and the $\sse_R$ LSP in
GeV.\label{Fig:efficiency_7TeV}]{
\begin{minipage}{7.8cm} 
\includegraphics[scale=1.0]{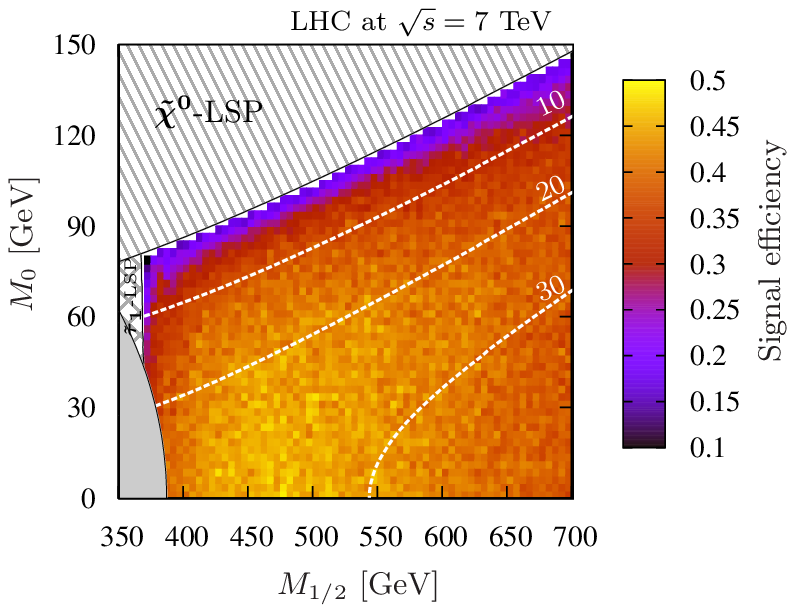} 
\vspace{0.0cm} 
\end{minipage} }
\caption{Signal cross section (in pb) [Fig.~\ref{Fig:xsec_7TeV}] and 
signal efficiency [Fig.~\ref{Fig:efficiency_7TeV}] at the LHC at 
$\sqrt{s}=7\tev$ in the $\mhalf-\mzero$ plane. The other parameters are
those of BE1 ($\azero=-1250\gev$, $\tanb =5$, $\sgnmu=+$,
$\lam_{231}|_\mathrm{GUT}=0.045$). The patterned regions correspond to
scenarios with either a $\stau_1$ or $\neut_1$ LSP. The solid gray
region in the lower left-hand corner is excluded by the bound 
on $\lam_{231}$, \cf Tab~\ref{Tab:lamcouplings}.}
\label{Fig:scan_7TeV}
\end{figure*}
\begin{figure*}[t!]
\subfigure[\,Minimal required integrated luminosity to obtain an 
observable signal, \cf Eq.~\eqref{Eqn:Sobserved}. We show contours 
for $100\,\text{pb}^{-1}$ (blue dashed line), $500\,\text{pb}^{-1}$ 
(green dashed line) and $1\,\text{fb}^{-1}$ (red dashed line), 
respectively. \label{Fig:discover_7TeV}]{
	\begin{minipage}{7.8cm}
	   	\includegraphics[scale=1.0]{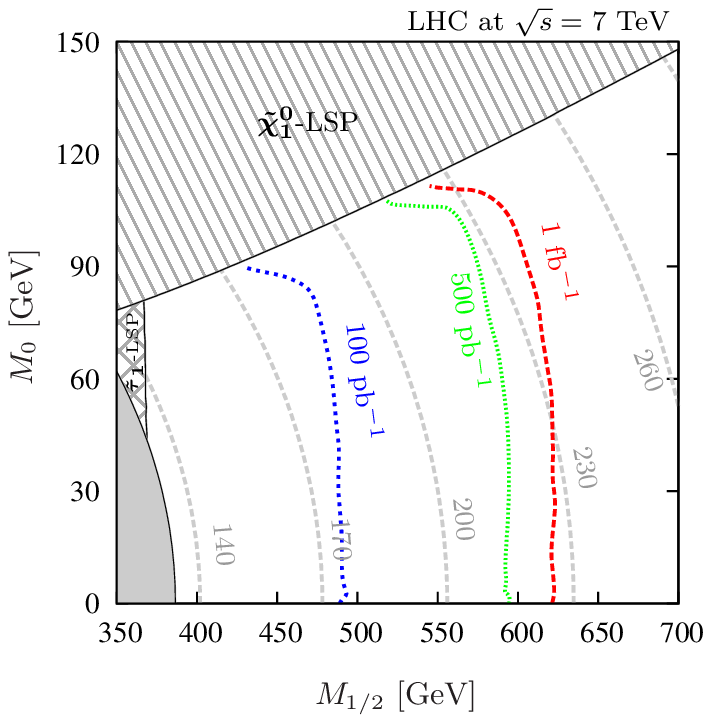}
	\vspace{0.2cm}	
	\end{minipage}
}
\subfigure[\,Contours for the signal over background ratio, $S/B$ (red
solid lines). \label{Fig:SoverB_7TeV}]{
	\begin{minipage}{7.8cm}
	   	\includegraphics[scale=1.0]{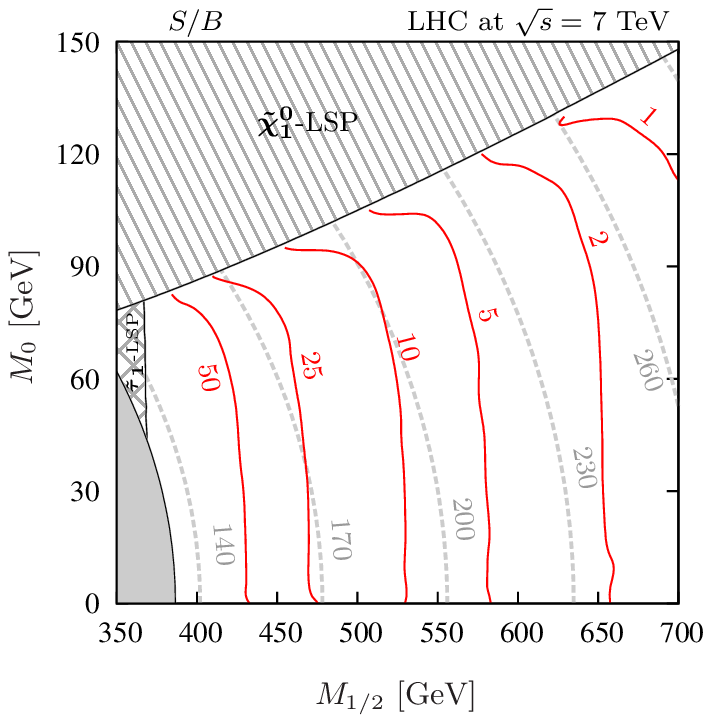}
	\vspace{0.2cm}
	\end{minipage}
}
\caption{Discovery reach at the LHC at $\sqrt{s} = 7\tev$ in the $M_{1/2}$--$M_0$ plane. 
The other $\text{B}_3$ mSUGRA parameters are $\azero = -1250\gev$,
$\tanb = 5$, $\sgnmu = +$ and $\lam_{231}|_\mathrm{GUT} = 0.045$. We
give the minimal required integrated luminosity for a discovery in
Fig.~\ref{Fig:discover_7TeV} and the signal to background ratio,
$S/B$, in Fig.~\ref{Fig:SoverB_7TeV}. The patterned regions correspond
to scenarios with either a $\stau_1$ or $\neut_1$ LSP. The solid gray
region in the lower left-hand corner is excluded by the bound
on $\lam_{231}$, \cf Table~\ref{Tab:lamcouplings}. Gray
dashed contour lines give the $\sse_R$ mass (in GeV) as indicated by
the labels.} \label{Fig:discovery_7TeV}
\end{figure*}

Due to the RGE running, all sparticle masses at the weak scale,
especially those of the strongly interacting sparticles, increase with
increasing $\mhalf$ \cite{Drees:1995hj,Ibanez:1984vq}. Thus, by varying $\mhalf$, we
can investigate the discovery potential as a function of the SUSY mass
scale. Furthermore, as we have seen in the previous two sections, the
discovery potential is quite sensitive to the mass hierarchy of the
lighter sparticles and, in particular, to the mass difference between
the $\neut_1$ and the $\sse_R$ LSP. Increasing $\mzero$ increases the
masses of the scalar particles, while the gaugino masses are nearly
un\-affected. Thus, $\mzero$ provides a handle to control the mass
difference between the $\neut_1$ and the $\sse_R$ (or $\ssmu_R$) LSP.

We show in Fig.~\ref{Fig:xsec_7TeV} the signal cross section (in pb)
for the LHC with $\sqrt{s}=7\tev$ and in Fig.~\ref{Fig:efficiency_7TeV}
the respective signal efficiency, \ie the fraction of signal events
that pass our cuts. The results are given only for models with a
$\sse_R$ LSP, while models with a $\neut_1$ LSP ($\stau_1$ LSP) are
indicated by the striped (checkered) region. The solid gray region
(lower left corner of Fig.~\ref{Fig:scan_7TeV}) is excluded by the
experimental bound on the $\lam_{231}$ coupling,
\cf Tab~\ref{Tab:lamcouplings}.

The signal cross section, Fig.~\ref{Fig:xsec_7TeV}, which is dominated
by the production of colored sparticles, clearly decreases with
increasing $\mhalf$, \ie with an increasing SUSY mass scale. For
instance, increasing $\mhalf$ from $400\gev$ to $500\gev$ reduces the
cross section from $0.6~\mathrm{pb}$ to $0.1~\mathrm{pb}$, while the
right-handed squark (gluino) mass increases from around $820\gev$
($930\gev$) to $1010\gev$ ($1150\gev$). In contrast, the $\mzero$
dependence of the signal cross section is negligible, over the
small range it is varied.

For the benchmark scenario BE1, we find in
Fig.~\ref{Fig:efficiency_7TeV} a signal efficiency of $46\%$. Going
beyond BE1, we observe that the signal efficiency lies between $30\%$
and $50\%$ for most of the $\sse_R$ LSP parameter space. Therefore,
our analysis developed in Sec.~\ref{Sect:cutflow} works also quite
well for a larger set of $\sse_R$ LSP models.

\begin{figure*}[t]
\subfigure[\,Signal cross section in pb at the LHC at $\sqrt{s}=14
\tev$. The white dashed contour lines give the (first and second 
generation) $\squark_R$ mass in GeV.\label{Fig:xsec_14TeV}]{
	\begin{minipage}{7.8cm}
	   	\includegraphics[scale=1.0]{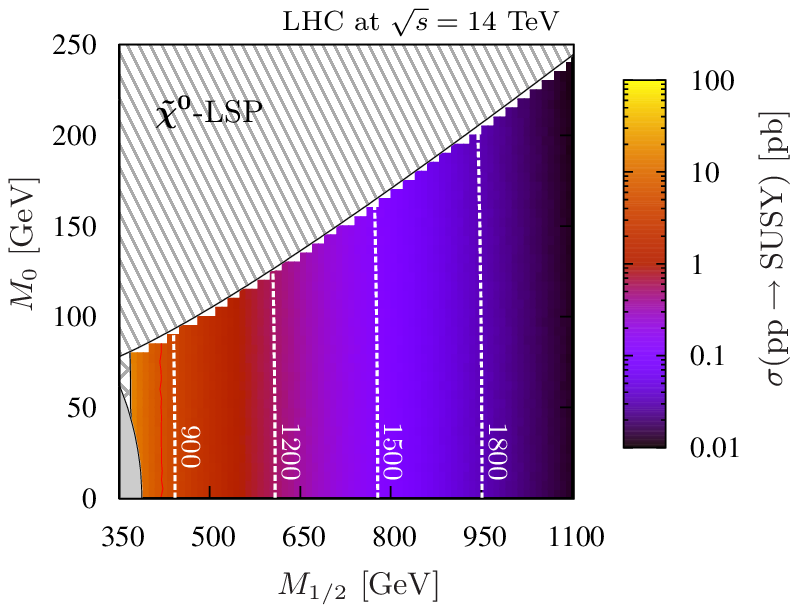}
	\vspace{0.0cm}	
	\end{minipage}
}
\subfigure[\,Selection efficiency for the signal events at the LHC at
$\sqrt{s}=14\tev$. The white dashed contour lines give the mass 
difference between the $\neut_1$ and the $\sse_R$ LSP in GeV.
\label{Fig:efficiency_14TeV}]{
	\begin{minipage}{7.8cm}
	   	\includegraphics[scale=1.0]{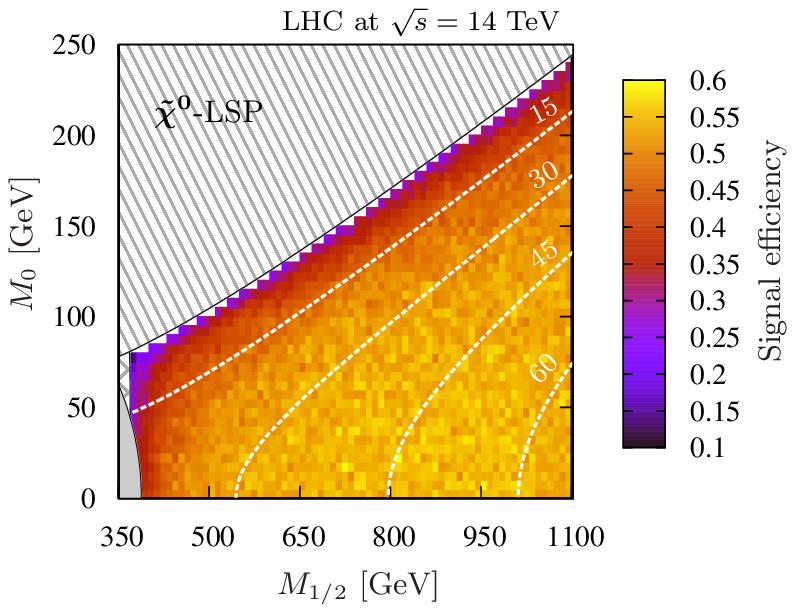}
	\vspace{0.0cm}
	\end{minipage}
}
\caption{Same as Fig.~\ref{Fig:scan_7TeV}, but for a cms energy of $\sqrt{s}=14$ TeV.}
\label{Fig:scan_14TeV}
\end{figure*}

\begin{figure*}[t]
\subfigure[\,Minimal required integrated luminosity to obtain an observable signal, 
\cf Eq.~\eqref{Eqn:Sobserved}. We show contours for $100\,\text{pb}^{-1}$ (blue dashed line), 
$1\,\text{fb}^{-1}$ (green dashed line) and $10\,\text{fb}^{-1}$ (red dashed line), 
respectively.\label{Fig:discover_14TeV}]{
	\begin{minipage}{7.8cm}
	   	\includegraphics[scale=1.0]{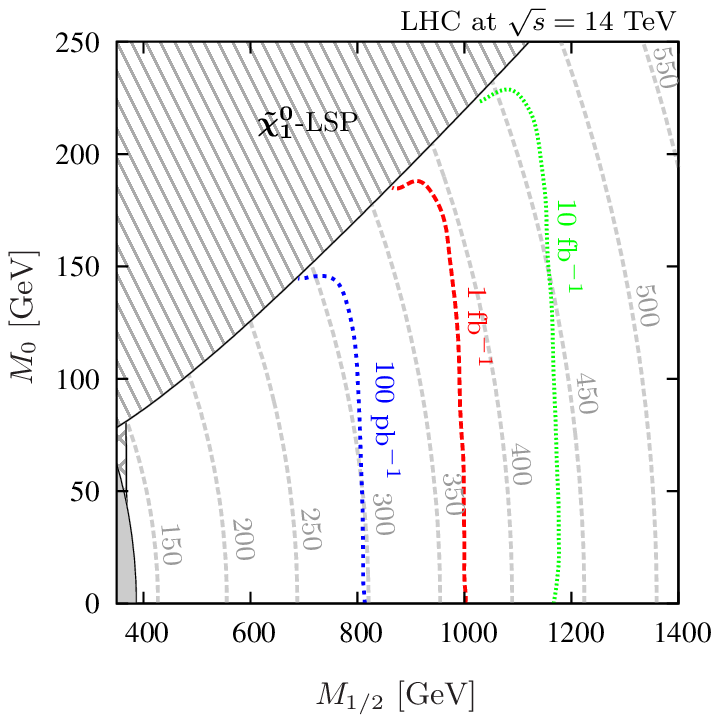}
	\vspace{0.2cm}	
	\end{minipage}
}
\subfigure[\,Contours for the signal over background ratio, $S/B$ (red solid lines).
\label{Fig:SoverB_14TeV}]{
	\begin{minipage}{7.8cm}
	   	\includegraphics[scale=1.0]{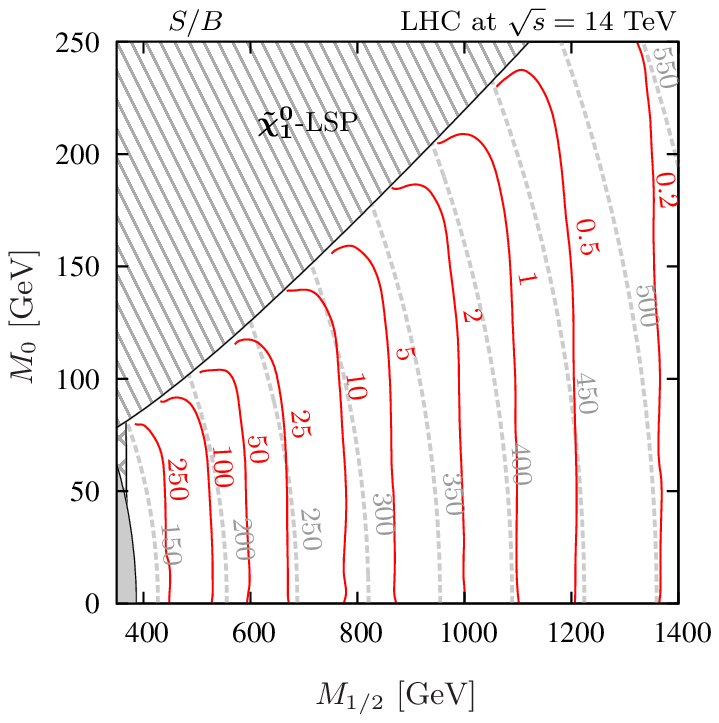}
	\vspace{0.2cm}
	\end{minipage}
}
\caption{Same as Fig.~\ref{Fig:discovery_7TeV}, but for a cms energy of $\sqrt{s}=14$ TeV.}
\label{Fig:discovery_14TeV}
\end{figure*}

However, the signal efficiency decreases dramatically if the mass
difference, $\Delta M$, between the $\neut_1$ and the $\sse_R$ LSP
approaches zero. For models with $\Delta M \lesssim 2.5\gev$, the
signal efficiency lies just around $10\% - 20\%$. As described in
detail in Sec.~\ref{Sect:kinematic}, the electrons in this parameter
region from the decay $\neut_1 \to \sse_R e$ are usually very soft and
thus tend to fail the minimum $p_T$ requirement of the object
selection, \textit{i.e.} $p_T > 10\gev$.  For models with $\Delta M >
10 \gev$, the signal efficiency becomes more or less insensitive
to $\Delta M$.  Note that, if we choose a stronger minimum
lepton $p_T$ requirement in our analysis, the band of low signal
efficiency will become wider.

The signal efficiency depends also slightly on $\mhalf$. At
low values, $\mhalf\lesssim 400\gev$, \ie for models with a light
sparticle mass spectrum, more events are rejected by the cut on the
visible effective mass. Moreover, the SM particles from cascade decays
and LSP decays have in this case on average smaller momenta than in
scenarios with a heavier mass spectrum, and thus may fail to pass the
object selection\footnote{However, due to our rather weak $p_T$
requirements for jets and leptons, this effect does not play a major
role.}.  The signal efficiency is highest for values of $\mhalf$
between $450\gev$ and $550\gev$ and reaches up to $50\%$. However,
when going to very large $\mhalf$, the signal efficiency again
decreases. Here, the production of sparton pairs is suppressed due to
their large masses and the jet multiplicity is reduced. Less events
will then pass the $N_\mathrm{jet} \ge 2$ and $\meff$ cut. For
example, for $M_{1/2}=500\gev$ ($M_{1/2}=700\gev$), sparton pair
production contributes (only) $58\%$ ($24\%$) to the total sparticle
pair production cross section.

We give in Fig.~\ref{Fig:discover_7TeV} the discovery potential of $\sse_R$ LSP scenarios
at the LHC with $\sqrt{s}=7\tev$. The discovery reach for the integrated luminosities 
$100~\mathrm{pb}^{-1}$, $500~\mathrm{pb}^{-1}$ and $1~\mathrm{fb}^{-1}$ is shown. 
We use Eq.~\eqref{Eqn:Sobserved} as criterion for a discovery. Furthermore, we present 
in Fig.~\ref{Fig:SoverB_7TeV} the signal to background ratio, $S/B$, as a measure for
the sensitivity on systematic uncertainties of the SM background. As shown in the 
previous section, the SM background is reduced to $2.8\pm0.8$ events when we employ
the cuts of Table~\ref{Tab:cutflow}. 

Fig.~\ref{Fig:discover_7TeV} suggests that $\sse_R$ LSP scenarios up to $\mhalf \lesssim 620\gev$ 
can be discovered with an integrated luminosity of $1~\mathrm{fb}^{-1}$. This corresponds 
to squark masses of $1.2\tev$ and $\sse_R$ LSP masses of around $230\gev$. For these models, 
we have a signal over background ratio of $S/B \approx 3$ and thus, systematic uncertainties 
of the SM background are not problematic. Furthermore, we see that BE1 ($M_{1/2}=475$~GeV, 
$M_0=0$~GeV) can already be discovered with $\lesssim~100~\mathrm{pb}^{-1}$ of data. We also 
see in Fig.~\ref{Fig:discovery_7TeV} that scenarios with a small mass difference between the 
$\neut_1$ and the $\sse_R$ LSP are more difficult to discover as expected from Fig.~\ref{Fig:efficiency_7TeV}.

We now discuss the prospects of a discovery at the LHC at $\sqrt{s}=14\tev$. In 
Fig.~\ref{Fig:xsec_14TeV}, we give the signal cross section and in 
Fig.~\ref{Fig:efficiency_14TeV} the signal efficiency. We employ the 
cuts developed in Sec.~\ref{Sect:cutflow}. The cutflow at $\sqrt{s}=14\tev$
for the benchmark scenarios can be found in Appendix~\ref{App:14TeV}.

Because of the higher cms energy, the cross section is
$\mathcal{O}(10)$ times larger than for $\sqrt{s}=7\tev$, \cf
Fig.~\ref{Fig:xsec_7TeV}. For instance, at $\mhalf =400\gev$
($500\gev$) the signal cross section at $\sqrt{s}=14\tev$ is now
$7.2~\mathrm{pb}^{-1}$ ($1.7~\mathrm{pb}^{-1}$). Furthermore, the
signal, \textit{i.e.}  sparticle pair production, is now always
dominated by sparton pair production,
\cf also Table~\ref{Tab:signalMC}.

The signal efficiency at $\sqrt{s}=14\tev$ is slightly improved
compared to $\sqrt{s}=7\tev$. Because of the enhanced sparton pair
production cross section, more signal events pass our cut on the jet
multiplicity, $N_\mathrm{jet} \ge 2$, \cf also
Appendix~\ref{App:14TeV}. We now obtain a signal efficiency of about
$51\%$ (compared to $46\%$ at $\sqrt{s}=7$ TeV) for the benchmark
point BE1. Most of the parameter points in
Fig.~\ref{Fig:efficiency_14TeV} exhibit a signal efficiency in the
range of $40\%$ to $60\%$. For the scenarios with low mass difference
between the $\neut_1$ and the $\sse_R$ LSP, $\Delta M\lesssim
2.5\gev$, the signal efficiency is reduced to around $15\% - 25\%$. As
for $\sqrt{s}=7\tev$, the signal efficiency decreases at very large
values of $\mhalf$, because of the increasing sparton mass and the
reduced sparton pair production cross section. Here, this effect
slowly sets in at values $\mhalf \gtrsim 1100\gev$, \ie for scenarios
with squark and gluino masses around $2\tev$. However, even at
$\mhalf=1100\gev$, sparton pair production still forms half of the
total signal cross section.

We show in Fig.~\ref{Fig:discover_14TeV} the discovery potential for
the LHC at $\sqrt{s}=14\tev$. We give the discovery reach for
integrated luminosities of $100~\mathrm{pb}^{-1}$, $1~\mathrm{fb}^{-1}$
and $10~\mathrm{fb}^{-1}$, respectively.  Our cuts of
Sec.~\ref{Sect:cutflow} reduce the SM background to $64.7\pm 7.2$
events for an integrated luminosity of $10~\mathrm{fb}^{-1}$; see
Table~\ref{Tab:cutflow14TeV}.  We observe that scenarios with $\mhalf
\lesssim 1\tev\,(1.15\tev)$ can be discovered with $1~\mathrm{fb}^{-1}$
($10~\mathrm{fb}^{-1}$). This corresponds to squark masses of around
$1.9\tev$ ($2.2\tev$) and LSP masses of roughly $370\gev$ ($450\gev$).
The respective signal over background ratio is $2$ ($0.6$) as can be
seen in Fig.~\ref{Fig:SoverB_14TeV}. Therefore, systematic
uncertainties of the SM background estimate are still not problematic
as long as the SM events can be estimated to a precision of
$\mathcal{O}(10\%)$. This is a reasonable assumption after a few years
of LHC running.

We conclude that due to the striking multi-lepton signature, the
prospects of an early discovery of $\Bthree$ mSUGRA with a
$\slepton_R$ LSP are better than for $R$-parity conserving mSUGRA
models \cite{Baer:2010tk}. Note that the vast reach in $\mhalf$ is
also due to the typically light $\sstop_1$ which has a large
production cross section. For instance, at $\mhalf = 525\gev$, the
$\sstop_1$ mass is around $630\gev$ and thus can still be produced
numerously at the LHC at $\sqrt{s}=7\tev$.

We want to remark that for scenarios with a low mass difference
between the $\neut_1$ and the $\slepton_R$ LSP, $\Delta M \lesssim 2.5
\gev$, the search for like-sign di-lepton final states might be a more
promising approach
\cite{Baer:2010tk,Dreiner:2000vf,Barnett:1993ea,Dreiner:1993ba}. 
However, a detailed analysis of these search channels is beyond the
scope of this paper.

\section{Mass Reconstruction}
\label{Sect:mass_reco}

We have shown in the previous section that large regions of the
$\Bthree$ mSUGRA parameter space with a $\slepton_R$ LSP can already
be tested with early LHC data. If a discovery has been made, the next
step would be to try to determine the sparticle mass spectrum. We
present now a strategy how the sparticle masses can be
reconstructed. We use the benchmark point BE2 as an example.  We
assume an integrated luminosity of 100 fb$^{-1}$ and a cms energy of
$\sqrt{s}=14 \tev$ in order to have enough events for the mass
reconstruction.

The sparticle decay chains cannot be directly reconstructed, because
the $\sse_R$ LSP decays always into an invisible neutrino. Thus, we
focus on the measurement of edges and thresholds of invariant mass
distributions which are a function of the masses of the involved SUSY
particles. Our strategy is analogous to the one, that is widely used to
reconstruct the mass spectrum in $R$-parity conserving SUSY where a
stable $\neut_1$ LSP escapes detection
\cite{Allanach:2000kt,Gjelsten:2004ki,
Gjelsten:2005aw,Barr:2010zj,Bechtle:2009ty,Aad:2009wy}.

\subsection{The Basic Idea}

\begin{figure}
	\begin{center}
	   	\includegraphics[scale=1.0]{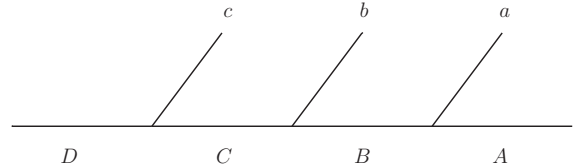}
	\end{center}
\caption{Decay chain assumed for the mass reconstruction.}
\label{Fig:generalcascade}	
\end{figure}

We first discuss the general idea of the method. We assume the decay
chain
\begin{align}
D \to C c \to B b c \to A a b c,
\end{align}
illustrated in Fig.~\ref{Fig:generalcascade}, where the particles $D$,
$C$, $B$, and $A$ are massive\footnote{Particle $A$ does not
necessarily need to be massive. In our case it is a massless
neutrino.} 
and their masses satisfy
\begin{align}
m_D > m_C > m_B > m_A.
\label{Eqn:particlemasses}
\end{align}
The particles $c$, $b$ and $a$ are observable (massless) SM particles. 
Particle A is assumed to be invisible.

From the 4-momenta of the decay products $a$, $b$ and $c$, we can
form the invariant mass combinations $m_{ba}$, $m_{ca}$, $m_{cb}$ and
$m_{cba}$. The maximal (denoted ``max") and minimal (denoted ``min")
endpoints of these distributions,
\begin{align}
m_{ba}^\mathrm{max}, \, m_{ca}^\mathrm{max}, \, m_{cb}^\mathrm{max}, \, 
m_{cba}^\mathrm{max} \,\,\, \text{and} \,\,\, m_{cba}^\mathrm{min},
\label{Eqn:inv_masses}
\end{align}
are functions of the (unknown) particle masses in
Eq.~\eqref{Eqn:particlemasses}\footnote{Another variable which can in
principle be used for our scenarios is the Stransverse mass,
$m_{T2}$~\cite{Lester:1999tx,Barr:2003rg,Barr:2010zj,Barr:2002ex,Burns:2008va,Konar:2009wn}.}.
The respective equations are given in Appendix~\ref{App:endpoints}
\cite{Miller:2005zp}. Note that $m_{ba}^\mathrm{min}$, $m_{ca}^
\mathrm{min}$ and $m_{cb}^\mathrm{min}$ are always equal to zero.

A prominent application of this method is the cascade decay of a left-handed 
squark in $R$-parity conserving SUSY \cite{Aad:2009wy}, 
\begin{align}
\squark_L \to q \neut_2  \to q \ell_n^\pm \slepton^\mp \to q \ell_n^\pm 
\ell_f^\mp \neut_1.
\end{align}
Here, the $\neut_1$ LSP is stable and escapes the detector unseen.
Note that in $R$-parity conserving SUSY, the ``near" lepton, $\ell_n$,
and ``far" lepton, $\ell_f$, are of the same flavor and thus
indistinguishable on an event-by-event basis.  In our scenarios this
is not necessarily the case, as shown below. 

For our $\slepton_R$ LSP scenarios, we investigate the decay chain of
a right-handed squark, \textit{i.e.}
\begin{align}
\squark_R \to q \neut_1 \to q \ell^\pm \slepton_R^\mp \to q \ell^\pm 
\ell'^\mp \nu.
\label{Eqn:massrec_cascade}
\end{align}
The LSP decays into a charged lepton $\ell'$ and a neutrino, where the
flavor depends on the dominant $\mathbf{\Lam}$ coupling, \cf
Table~\ref{Tab:signatures}.  In contrast to the $R$-parity conserving
scenarios, we can actually distinguish the near and far lepton if we
have $\mathbf{\Lambda} \in \{\lam_{231}, \lam_{132}\}$.  The
$\slepton_R$ LSP then decays into a charged lepton of different flavor
from its own. However, we still have to deal with combinatorial
backgrounds, because we might wrongly combine leptons (and jets) from
different cascades within the same event.
\begin{figure}[t]
	\begin{center}
	   	\includegraphics[scale=1.0]{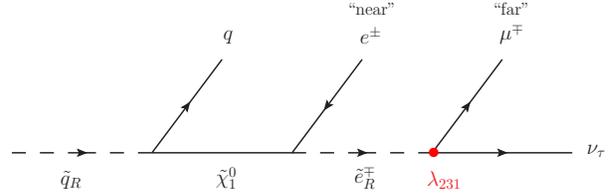}  
	\end{center}
\caption{Relevant decay chain of a right-handed squark, 
Eq.~(\ref{Eqn:massrec_cascade}), for the benchmark scenario BC2. 
The $R$-parity violating decay of the $\sse_R$ LSP via $\lam_{231}$ 
is marked in red.}
\label{Fig:massrec_cascade}	
\end{figure}

In the following, we demonstrate our method for
the $\sse_R$ LSP benchmark model BE2 ($\lam_{231}|_{\rm GUT} \not =
0$), \cf Table~\ref{Tab:benchmarks}.  We focus on the case, where the
$\sse_R$ LSP decays into a muon (instead of a $\tau$) and a
neutrino. On the one hand, muons are much easier to reconstruct than
$\tau$ leptons. On the other hand, muon events have a higher
probability to pass our cuts, \cf Sect.~\ref{Sect:cutflow}.  The
relevant cascade decay, Eq.~\eqref{Eqn:massrec_cascade}, is shown in
Fig.~\ref{Fig:massrec_cascade}.  It yields one jet (at parton level)
and two charged leptons of different flavor and opposite charge. From
these objects, we can form the invariant masses $m_{e\mu}$, $m_{\mu
q}$, $m_{eq}$ and $m_{e\mu q}$.

In the mass determination, one can leave the mass of the neutrino as a
free parameter.  If one measures this parameter consistent with zero,
it would be an important piece of information towards
confirming our model. However, once the $R$-parity violating
decay chain of Fig.~\ref{Fig:massrec_cascade} is experimentally
verified (or assumed), the knowledge of $m_A = 0$,
Eq.~(\ref{Eqn:particlemasses}), simplifies the equations in
Appendix~\ref{App:endpoints} and reduces the number of fit parameters
by one. The endpoints of the invariant mass distributions are then
given by
\begin{subequations}
\begin{align}
(m_{e\mu}^\mathrm{max})^2 = & M_{\neut_1}^2-M_{\sse_R}^2,\label{Eqn:mll_b}\\
(m_{\mu q}^\mathrm{max})^2 = & M_{\squark_R}^2 - M_{\neut_1}^2, 
\label{Eqn:mlqfar_b}\\
(m_{e q}^\mathrm{max})^2 = & (M_{\squark_R}^2 - M_{\neut_1}^2)
(M_{\neut_1}^2-M_{\sse_R}^2)/M_{\neut_1}^2, \label{Eqn:mlqnear_b}\\
(m_{e \mu q}^\mathrm{max})^2 = & M_{\squark_R}^2-M_{\sse_R}^2, 
\label{Eqn:mlmq_max} \\
(m_{e \mu q}^\mathrm{min})^2 = & M_{\squark_R}^2 (M_{\neut_1}^2-
M_{\sse_R}^2)/(2M_{\neut_1}^2). \label{Eqn:mlmq_min}
\end{align}
\label{Eqn:massedges2}
\end{subequations}

In BE2 (and more generally in most $\slepton_R$ LSP models
within $\text{B}_3$ mSUGRA), the (mostly right-handed) $\sstop_1$ is
much lighter than the first and second generation $\squark_R$.
Therefore, we have typically two distinct squark mass scales. This
enables a measurement of the $\sstop_1$ {\it and} (first and second
generation) $\squark_R$ mass simultaneously, if we are able to
separate $\sstop_1$ and $\squark_R$ production from each
other\footnote{From now on, $\squark_R$ stands only for right-handed
squarks of the first and second generation.}. This \textit{is}
possible as we show now.

\subsection{Event Selection}
\label{Sect:mass_eventselection}

For the mass reconstruction, we slightly extend our cuts developed in
Sec.~\ref{Sect:cutflow} for $\sqrt{s}=14\tev$. Each event has to
contain at least one electron and one muon with opposite charge. In
order to enhance the probability of selecting the right muon, \ie the
$\mu$ from the $\sse_R$ LSP decay, we require a minimal transverse
momentum of the muon of $p_T^\mu \ge 25\gev $. We then construct all
possible opposite-sign-different-flavor (OSDF) dilepton invariant
masses, $m_{e\mu}$, of electrons and muons (with $p_T^\mu \ge 25\gev
$). In order to reduce combinatorial backgrounds, we subtract the
dilepton invariant mass distribution of the same-sign-different-flavor
(SSDF) leptons. Note that this also suppresses (R-parity conserving)
SUSY background processes, where the charges of the selected leptons
are uncorrelated, because of an intermediate Majorana particle, \ie a
neutralino. For example, SUSY decay chains involving the cascade
$\ssmu_L^- \to \mu^-\neut_1 \to\mu^- e^\pm\sse_R^\mp$ are thus
suppressed. 

For the invariant mass distributions containing a jet, we design further 
selection cuts to discriminate between $\sstop_1$ and $\squark_R$ 
events. We expect at least two $b$ jets in the $\sstop_1$ events from the 
top quark decays. Thus, we introduce a simple $b$-tagging algorithm in our 
simulation, assuming a $b$-tagging efficiency of $60\%$ \cite{Aad:2009wy}. 
We demand two tagged $b$ jets for the $\sstop_1$ event candidates while 
we require that no $b$ jet must be present for the $\squark_R$
event candidates. Moreover, we use the visible effective mass, $\meff$,
as a handle to discriminate between $\sstop_1$ and $\squark_R$ events,
\ie we impose the cuts
\begin{align}
\begin{array}{l}
400 \gev \le \meff \le 900 \gev \quad \mbox{for $\sstop_1$ events},\\ 
900 \gev \le \meff  \quad \quad \quad \quad \quad \,\quad \mbox{for 
$\squark_R$ events},\end{array}
\label{Eq:mass_meff}
\end{align}
respectively.

For the construction of invariant mass distributions involving quarks,
we consider the hardest and second hardest jet, $j_1$ and $j_2$ in
each event, respectively. Due to the lighter $\sstop_1$ mass, the jets
are expected to be somewhat softer in $\sstop_1$ events than in
$\squark_R$ events.  Therefore, for BE2, we choose the following $p_T$
selection criteria for the jets:
\begin{align}
\begin{array}{l} 
\begin{array}{l} 50 \gev \le p_T^{j_1} \le 250 \gev\\ 
25 \gev \le p_T^{j_2}  \quad \quad \quad \quad \quad \,
\end{array} \Big \}
\quad \mbox{for $\sstop_1$ events}, \\ \\
 \begin{array}{l} 
250 \gev \le p_T^{j_1}  \\ 
100 \gev \le p_T^{j_2}  
\end{array} \Big \} \quad  \mbox{for $\squark_R$ events}. 
\end{array}
\label{Eqn:mass_pT}
\end{align}

The invariant mass distributions $m_{e\mu q}$, $m_{eq}$, and $m_{\mu q}$ are 
now constructed as follows:
\begin{itemize}
\item $m_{e\mu q}$: We take the invariant masses of the opposite sign
electron and muon with $j_1$ and $j_2$. The smaller (larger) value is
taken for the edge (threshold) distribution. Note that we repeat this
procedure for {\it all possible} combinations of electrons and muons.
For the threshold distribution, we demand in addition the dilepton
invariant mass to lie within $m_{e\mu}^\mathrm{max}/\sqrt{2} \le m_{e
\mu} \le m_{e\mu}^\mathrm{max}$, corresponding to the subset of events
in which the angle between the two leptons (in the center of mass
frame of the $\sse_R$ LSP) is greater than $\pi/2$ 
\cite{Allanach:2000kt}.  In the edge distribution, we require $m_{e\mu}
\le m_{e \mu}^\mathrm{max}$ and employ SSDF subtraction to reduce the 
combinatorial background.
\item $m_{e q}$ ($m_{\mu q}$): We construct the invariant mass of all
selected electrons (muons with $p_T^\mu \ge 25\gev $) with $j_1$ and
$j_2$ and take the lower value\footnote{Here, we make use of the fact
that we can distinguish the near and far lepton. However, we have
checked that the model-independent construction of the variables
$m_{\ell q(near/far)}$ as proposed in Ref.~\cite{Allanach:2000kt}
leads to similar results.}. Furthermore, we require $m_{e\mu} \le
m_{e\mu}^\mathrm{max}$.
\end{itemize}
For these constructions, the dilepton invariant mass edge,
$m_{e\mu}^\mathrm{max}$, must have already been fitted. We use the
true value of the dilepton edge, because it can be reconstructed to a
very high precision, \cf Sec.~\ref{Sect:mass_dilepton}.

\subsection{Results}

We now show our results for BE2 for an integrated luminosity of
$100~\mathrm{fb}^{-1}$ at $\sqrt{s}=14\tev$. We assume, that the SM
background can be reduced to a negligible amount (\cf
Appendix~\ref{App:14TeV}) and present only the invariant mass
distributions for the SUSY sample, \textit{i.e.} pair production of
all SUSY particles. We employ the cuts described in the last
section. We give a rough estimate of how accurately the kinematic
endpoints may be determined and investigate whether the result can be
biased due to SUSY background processes or systematical effects of the
event selection. Our discussion should be understood as a
proof-of-principle of the feasibility of the method. It should
be followed by a detailed experimental study including a detector 
simulation.

\subsubsection{Dilepton Invariant Mass}
\label{Sect:mass_dilepton}

\begin{figure*}[t]
	\setlength{\unitlength}{1in} 
	\centering
	\subfigure[\,Dilepton edge, $m_{e\mu}^\mathrm{max}$, 
                   Eq.~(\ref{Eqn:mll_b}). The dashed line gives the 
                   expected value of $51.7\gev$.
		\label{Fig:mll_edge}]{\includegraphics[scale=0.37]{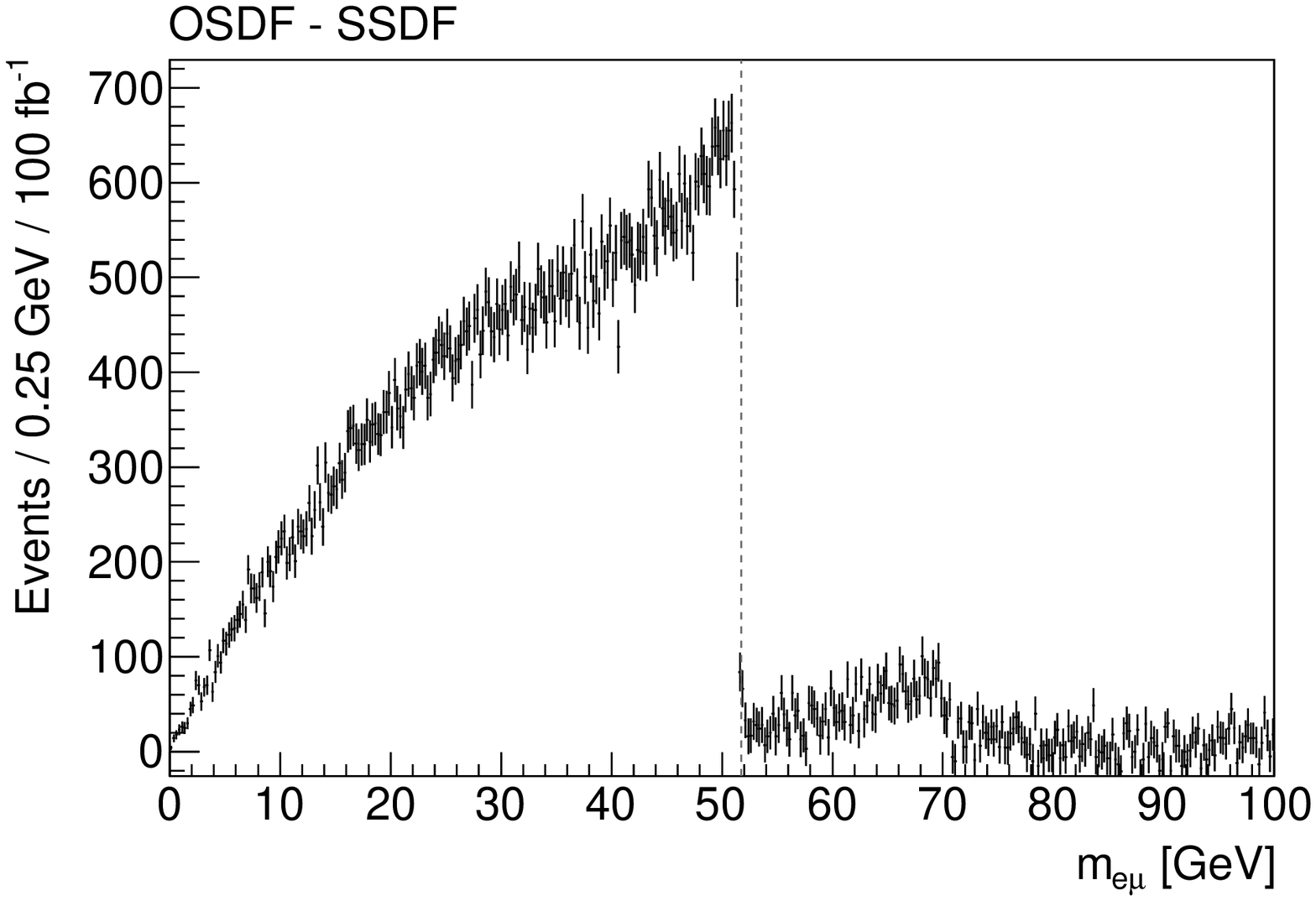}}
		\qquad
	\subfigure[\,Mass peak of the tau sneutrino, $\ssnutau$, due to
                   the $R$-parity violating decay $\ssnutau \to e \mu$.
                   The true mass is $M_\ssnutau = 309.8\gev$, \cf 
                   Table~\ref{Tab:BE2}.
		\label{Fig:mll_ssnu}]{\includegraphics[scale=0.37]{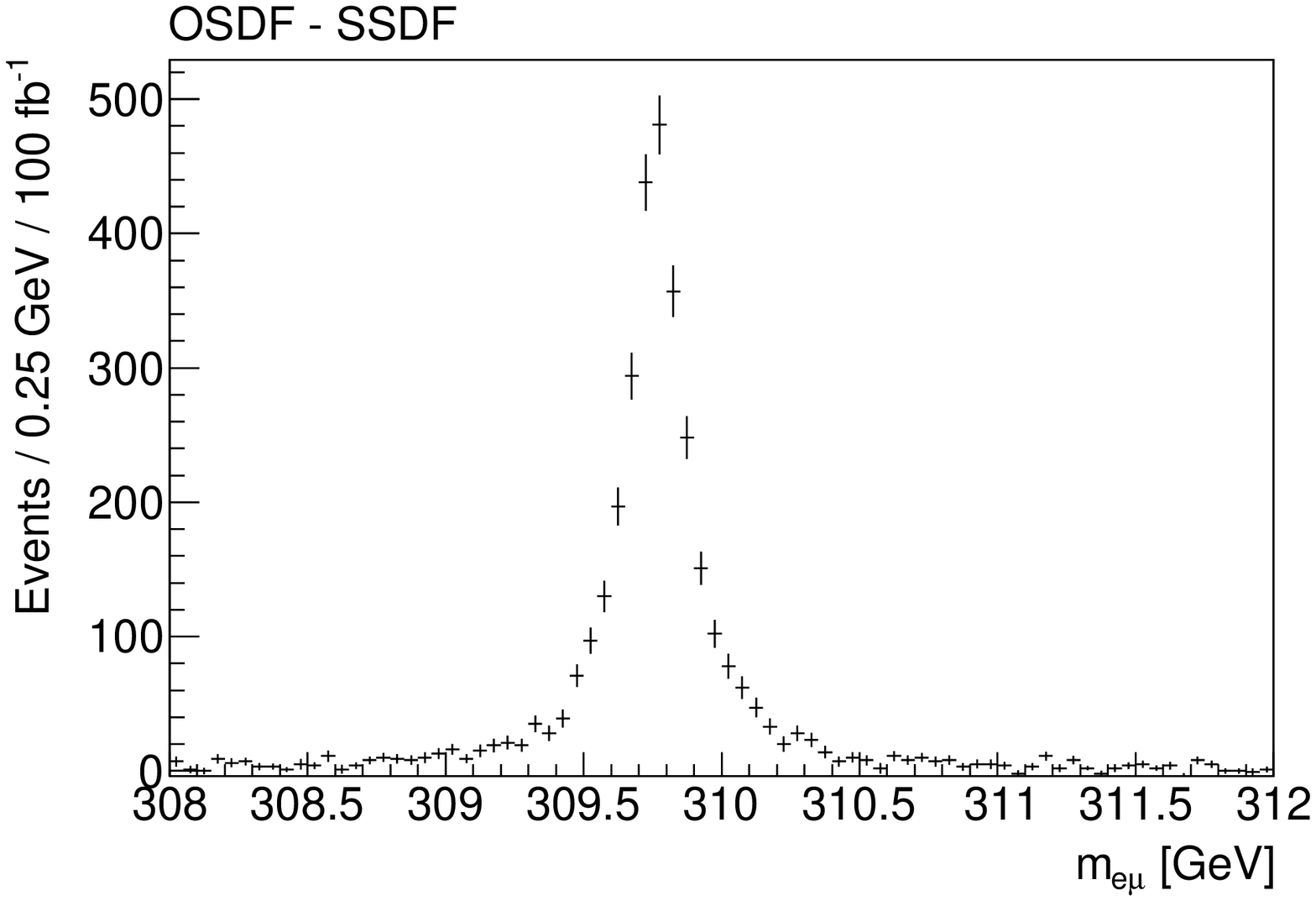}}
	\caption{Dilepton invariant mass distribution, $m_{e\mu}$, for
                 the benchmark point BE2. The distributions are 
                 same-sign-different-flavor (SSDF) subtracted. The error bars 
		correspond to statistical uncertainties at 100 fb$^{-1}$.}
		\label{Fig:mll}
\end{figure*}

We show in Fig.~\ref{Fig:mll} the SSDF subtracted dilepton invariant
mass distribution, $m_{e\mu}$. According to Eq.~(\ref{Eqn:mll_b}), we
expect for the cascade decay in Fig.~\ref{Fig:massrec_cascade} a
dilepton edge at $51.7\gev$ [dashed gray line in
Fig.~\ref{Fig:mll_edge}]. The observed edge quite accurately
matches the expected value and should be observable already
with a few $\mathrm{fb}^{-1}$.

For an invariant mass below the dilepton edge, the distribution shape
slightly deviates from the (expected) triangular shape. This is
because the $\sse_R$ LSP can also decay into a neutrino and a $\tau$
lepton (see Table~\ref{Tab:signatures}), which then decays into a muon
and neutrinos. In this case, the muon only carries a fraction of the
$\tau$ lepton $p_T$ and we obtain an on average
lower $m_{e\mu }$ value compared to the LSP decay $\sse_R\to\mu\nu_
\tau$.

We observe another small edge at about $70\gev$. These events stem
from the decay of a left-handed smuon, \textit{i.e.} $\ssmu_L^\pm \to
\mu^\pm \neut_1\to \mu^\pm e^\mp \sse_R$, \cf Table~\ref{Tab:BE2}. The
true endpoint is $70.7\gev$.

Furthermore, as shown in Fig.~\ref{Fig:mll_ssnu}, we have a sharp peak
at $309.8\gev$ in the di-lepton invariant mass distribution. Here, the
mass of the tau sneutrino, $\ssnutau$, is fully reconstructed. It
decays via the $R$-parity violating decay $\ssnutau \to e^-\mu^+$ with
a branching ratio of 12\%; see Table~\ref{Tab:BE2}.
Analogously, we also expect 
a mass peak in the $e\tau$ invariant mass distribution 
from the respective muon sneutrino decay. However, the observation of
this peak requires the reconstruction of the $\tau$ lepton momentum
which is beyond the scope of this paper.  The sneutrino mass peaks are
expected to be observable with only a few $\mathrm{fb}^{-1}$ of data
and are thus a smoking gun for our scenarios.

\subsubsection{Dilepton plus Jet Invariant Mass}

\begin{figure*}[t]
		\setlength{\unitlength}{1in} \centering
		\subfigure[\,$m_{e\mu q}$ edge-distribution for the
		$\squark_R$ event selection. The dashed line gives the
		expected value of $ 925\gev$, \cf
		Eq.~(\ref{Eqn:mlmq_max}).
		\label{Fig:mllq_edge_squark}]{\includegraphics[scale=0.37]{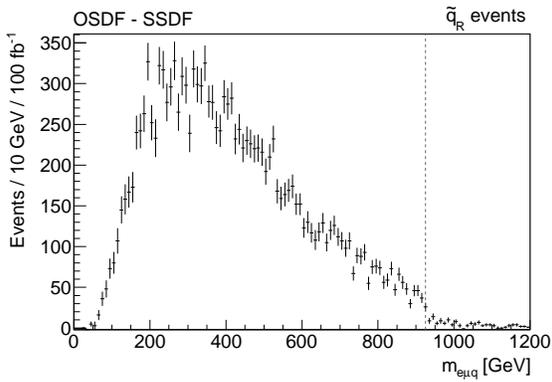}}
		\qquad \subfigure[\,$m_{e\mu q}$ edge-distribution for
		the $\sstop_1$ event selection. The dashed line gives
		the expected value of $410\gev$, \cf
		Eq.~(\ref{Eqn:mlmq_max}) for the $\sstop_1$.
		\label{Fig:mllq_edge_stop}]{\includegraphics[scale=0.37]{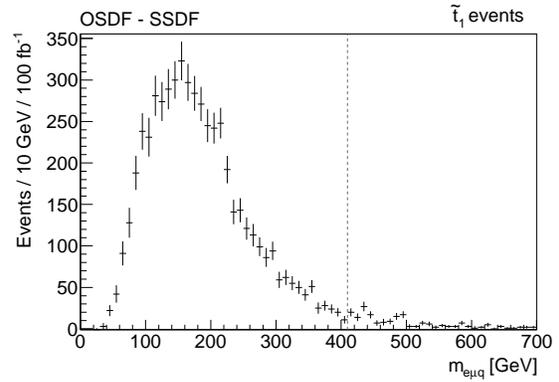}}
	\caption{Dilepton plus jet invariant mass distributions, 
                 $m_{e\mu q}$, for the kinematic edge for the 
                 $\squark_R$ event
		[Fig.~\ref{Fig:mllq_edge_squark}] and $\sstop_1$ event
		[Fig.~\ref{Fig:mllq_edge_stop}] selection. The
		distributions are SSDF subtracted. The errors
		correspond to statistical uncertainties at 100
		fb$^{-1}$.}  \label{Fig:mllq_edge}
\end{figure*}

We show in Fig.~\ref{Fig:mllq_edge} the dilepton plus jet invariant
mass distribution, $m_{e\mu q}$, to obtain the kinematic edge for the
$\squark_R$ event [Fig.~\ref{Fig:mllq_edge_squark}] and $\sstop_1$
event [Fig.~\ref{Fig:mllq_edge_stop}] selection, \cf
Sec.~\ref{Sect:mass_eventselection}.  Recall that we employ different
selection criteria to obtain the edge and the threshold of the
$m_{e\mu q}$ distribution; see the end of
Sec.~\ref{Sect:mass_eventselection} for details.
 
According to Eq.~(\ref{Eqn:mlmq_max}) and Table~\ref{Tab:BE2}, we
expect the edge in Fig.~\ref{Fig:mllq_edge_squark}
[Fig.~\ref{Fig:mllq_edge_stop}] to lie at $925\gev$ [$410\gev$]. For
the $\squark_R$ event selection, this is the case as can be seen\footnote{The endpoint values are usually determined by employing straight line fits, see e.g. Ref.~\cite{Aad:2009wy,Allanach:2000kt,Gjelsten:2004ki,Gjelsten:2005aw}.} by
the dashed gray line in Fig.~\ref{Fig:mllq_edge_squark}.

In contrast, in the $\sstop_1$ event selection the identification of
the endpoint [dashed gray line in Fig.~\ref{Fig:mllq_edge_stop}] is
more difficult. The observable edge is smeared to higher
values. On the one hand, cascade decays of {\it heavier} squarks and
gluinos can leak into the $\sstop_1$ event selection. On the other
hand, the distribution flattens out as it approaches the nominal
endpoint, because the jet (from $t$ decay) carries only a fraction of
the $t$ quark $p_T$. Moreover, the cut imposed on the jet transverse
momentum, $p_T < 250\gev$, Eq.~(\ref{Eqn:mass_pT}), tends to reject
events at high $m_{e\mu q}$ values. Therefore, the endpoint tends to
be smeared. However, the intesection of the x-axis with a 
linear fit on the right flank of Fig.~\ref{Fig:mllq_edge_stop} 
would still provide a quite good estimate of the true edge. Such a 
procedure is also employed for the mass reconstruction of $R$-parity 
conserving models \cite{Allanach:2000kt,Gjelsten:2004ki,
Gjelsten:2005aw,Barr:2010zj,Aad:2009wy}.

\begin{figure*}[t]
		\setlength{\unitlength}{1in} \centering
		\subfigure[\,$m_{e\mu q}$ threshold-distribution for
		the $\squark_R$ event selection. The dashed line gives
		the expected value of $ 181\gev$, \cf
		Eq.~(\ref{Eqn:mlmq_min}).
		\label{Fig:mllq_thres_squark}]{\includegraphics[scale=0.37]{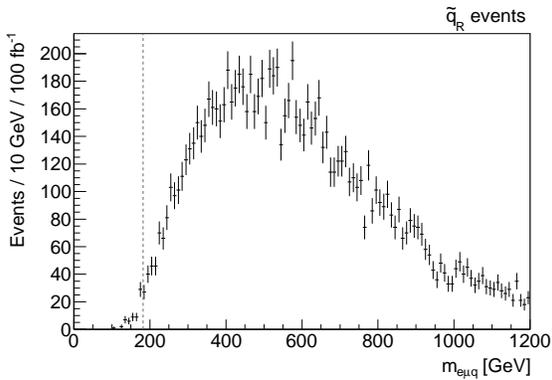}}
		\qquad \subfigure[\,$m_{e\mu q}$
		threshold-distribution for the $\sstop_1$ event
		selection. The dashed line gives the expected value of
		$ 86\gev$, \cf Eq.~(\ref{Eqn:mlmq_min}) for the
		$\sstop_1$.\label{Fig:mllq_thres_stop}]
		{\includegraphics[scale=0.37]{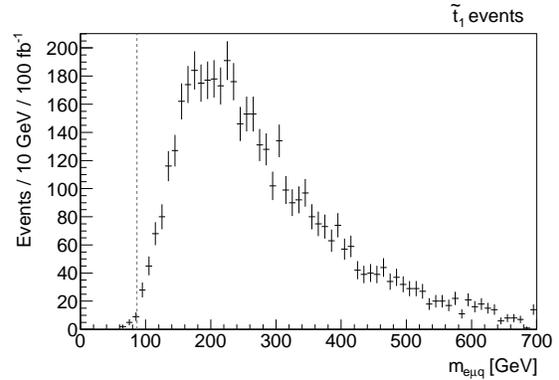}}
		\caption{Dilepton plus jet invariant mass
		distributions, $m_{e\mu q}$, for the kinematic
		threshold for the $\squark_R$ event
		[Fig.~\ref{Fig:mllq_thres_squark}] and $\sstop_1$
		event [Fig.~\ref{Fig:mllq_thres_stop}] selection. The
		errors correspond to statistical uncertainties at 100
		fb$^{-1}$.}  \label{Fig:mllq_thres}
\end{figure*}

In Fig.~\ref{Fig:mllq_thres}, we present the $m_{e\mu q}$
threshold-distribution for the $\squark_R$
[Fig.~\ref{Fig:mllq_thres_squark}] and $\sstop_1$ event
[Fig.~\ref{Fig:mllq_thres_stop}] selection.  In
Fig.~\ref{Fig:mllq_thres_squark}, we observe an edge slightly below
the expected threshold of $181\gev$ (gray dashed line). This shift
towards lower values is mainly due to final state radiation of the
quark from $\tilde{q}_R$ decay \cite{Miller:2005zp}, \textit{i.e.} the
reconstructed jet is less energetic than the original quark. This is
not surprising, because we use a relatively small radius,
$\Delta R=0.4$, for the jet algorithm, \cf
Sec.~\ref{Sect:object_selection}.

In general, the $m_{e\mu q}$ threshold value is set by the lightest 
squark. Therefore, events in Fig.~\ref{Fig:mllq_thres_squark} with 
values far below the endpoint at $181\gev$ usually contain third 
generation squarks. These events can leak into the $\squark_R$ 
event selection when the $b$ quarks are not tagged.

For the $\sstop_1$ event selection, Fig.~\ref{Fig:mllq_thres_stop},
the observed $m_{e\mu q}$ threshold matches quite accurately the 
expected value of $86\gev$ (gray dashed line). We note however,
that detector effects, especially jet miss-measurements, are 
expected to smear the thresholds and edges. But, this lies beyond 
the scope of this paper. 

\subsubsection{Lepton plus Jet Invariant Masses}

We now discuss the invariant mass distributions formed by one charged
lepton and a jet, \textit{i.e.} $m_{eq}$ and $m_{\mu q}$. For these
invariant masses, we generally have larger SUSY backgrounds (compared
to the dilepton and dilepton plus jet invariant mass distributions),
because we cannot employ SSDF subtraction.

\begin{figure*}[t]
	\setlength{\unitlength}{1in} 
	\centering
	\subfigure[\,$m_{eq}$ distribution for the $\squark_R$ event 
              selection. 
		The dashed line gives the expected value of $ 251\gev$, \cf Eq.~(\ref{Eqn:mlqnear_b}).
\label{Fig:meq_squark}]{\includegraphics[scale=0.37]{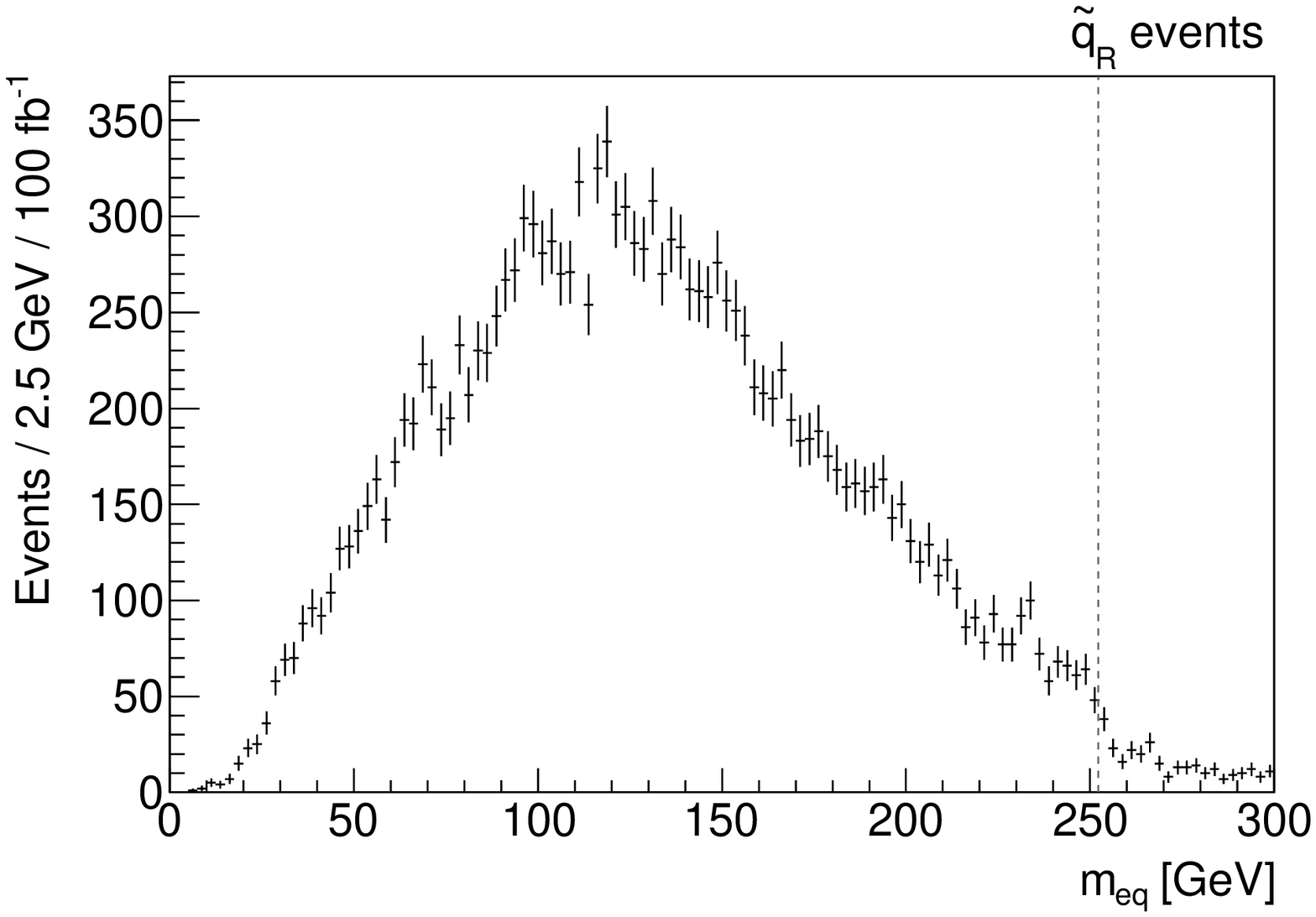}}
		\qquad
	\subfigure[\,$m_{eq}$ distribution for the $\sstop_1$ event selection. 
		The dashed line gives the expected value, $ 111\gev$, \cf Eq.~(\ref{Eqn:mlqnear_b}) for the $\sstop_1$.
\label{Fig:meq_stop}]{\includegraphics[scale=0.37]{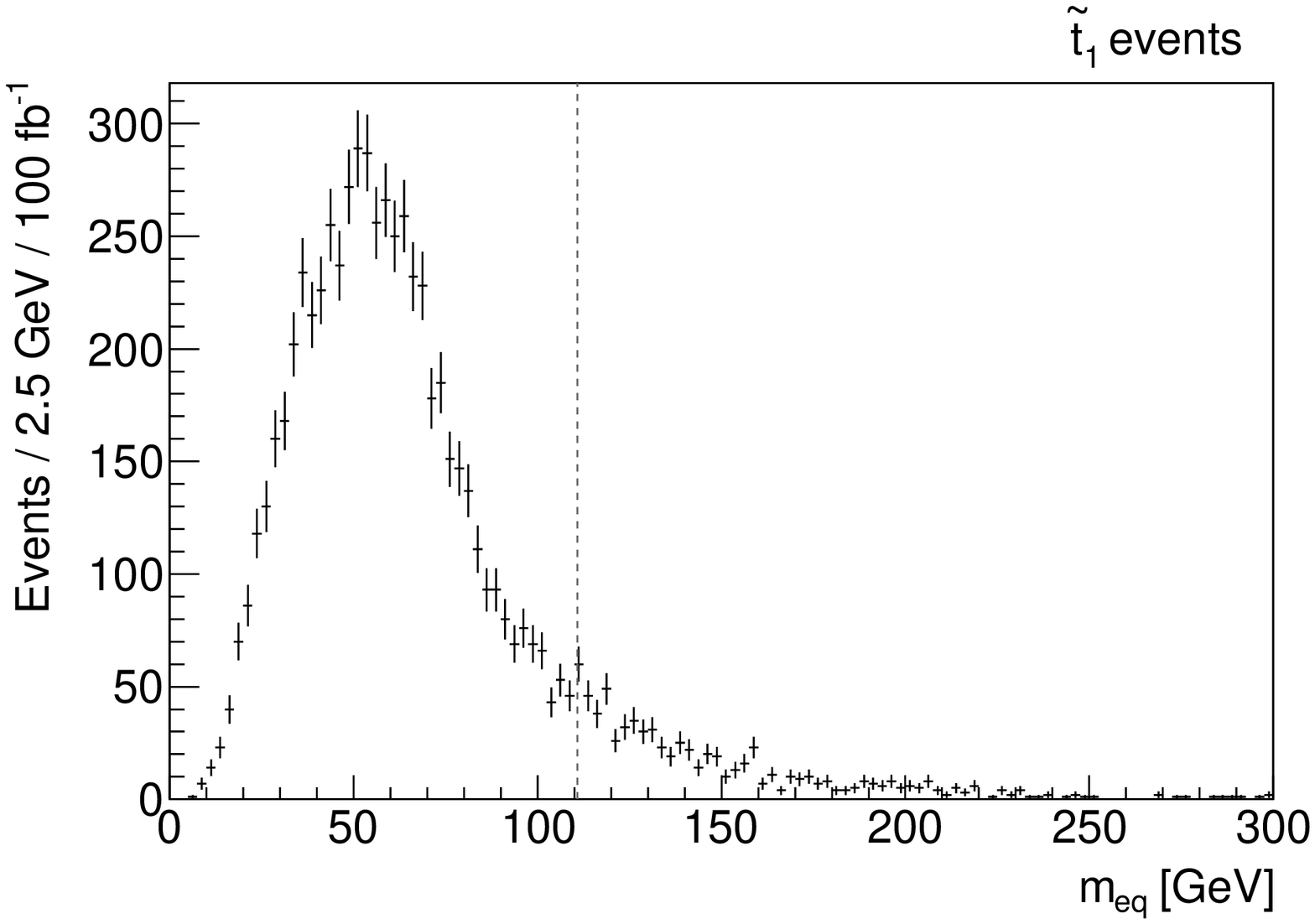}}
	\caption{Electron plus jet invariant mass distribution, $m_{eq}$, for the $\squark_R$ 
		event [Fig.~\ref{Fig:meq_squark}] and $\sstop_1$ event [Fig.~\ref{Fig:meq_stop}] selection.
		The errors correspond to statistical uncertainties at 100 fb$^{-1}$.}
		\label{Fig:meq}
\end{figure*}

The electron-jet invariant mass distributions, $m_{eq}$, are presented
in Fig.~\ref{Fig:meq}. In the $\squark_R$ event selection
[Fig.~\ref{Fig:meq_squark}], we observe an edge near the expected
endpoint of $251\gev$ (gray dashed line). In contrast, in the
$\sstop_1$ event selection [Fig.~\ref{Fig:meq_stop}], the endpoint,
which is expected to lie at $111\gev$, cannot be easily identified.

The jet used for Fig.~\ref{Fig:meq_stop} usually carries only a
fraction of the $t$ quark momentum reducing the invariant mass. In
addition, the $\sstop_1$ cascade decay
\begin{align}
\sstop_1 \overset{28.1\%}{\longrightarrow}  b \charge_1^+ \overset{19.9\%}{\longrightarrow}  
b \mu^+ \ssnumu \overset{14.2\%}{\longrightarrow}  b \mu^+ e^- \tau^+,
\end{align}
possesses an endpoint at $267\gev$ in $m_{eq}$ which produces
events beyond the expected endpoint. As a result, a measurement of the
$111\gev$ endpoint will be difficult. 

\begin{figure*}[t]
		\setlength{\unitlength}{1in} 
		\centering
	\subfigure[$m_{\mu q}$ distribution for the $\squark_R$ event selection. 
		The dashed line gives the expected value of $ 921\gev$, \cf Eq.~(\ref{Eqn:mlqfar_b}).
\label{Fig:mmuq_squark}]{\includegraphics[scale=0.37]{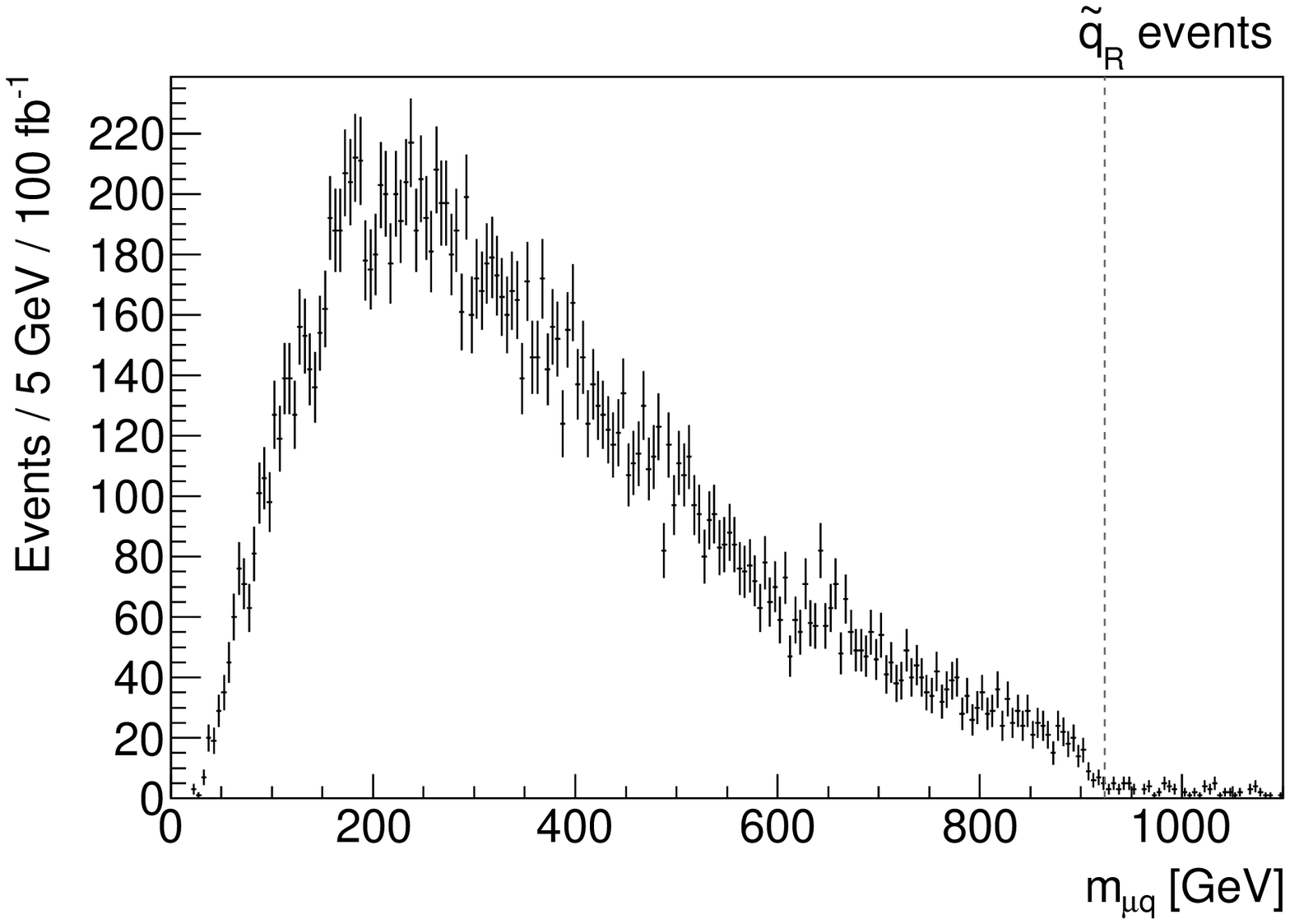}}
		\qquad
	\subfigure[$m_{\mu q}$ distribution for the $\sstop_1$ event 
                   selection. The dashed line gives the expected value
                   of $ 406\gev$, \cf Eq.~(\ref{Eqn:mlqfar_b}) for 
                   the $\sstop_1$.
\label{Fig:mmuq_stop}]{\includegraphics[scale=0.37]{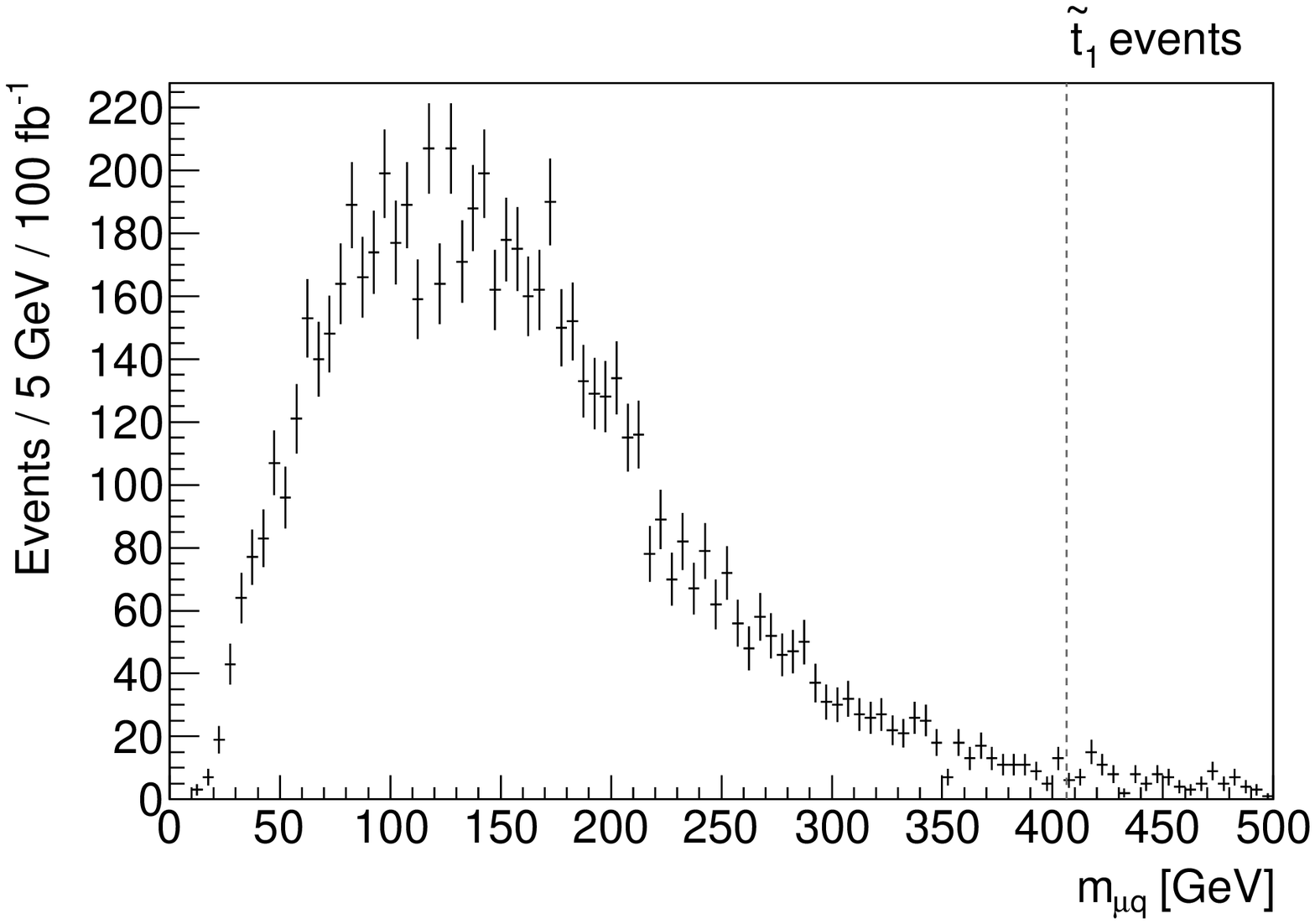}}
	\caption{Muon plus jet invariant mass distribution, $m_{\mu q}
              $, for the $\squark_R$ event [Fig.~\ref{Fig:mmuq_squark}] 
              and $\sstop_1$ event [Fig.~\ref{Fig:mmuq_stop}] selection.
	The errors correspond to statistical uncertainties at 100 fb$^{-1}$.}
		\label{Fig:mmuq}
\end{figure*}

In Fig.~\ref{Fig:mmuq} we show the muon-jet invariant mass
distributions for the $\squark_R$ event [Fig.~\ref{Fig:mmuq_squark}]
and $\sstop_1$ event [Fig.~\ref{Fig:mmuq_stop}] selection. Assuming
the $\squark_R$ cascade decay of Fig.~\ref{Fig:massrec_cascade}, the
$m_{\mu q}$ distribution, Fig.~\ref{Fig:mmuq_squark}, has an expected
endpoint at $921\gev$, Eq.~\eqref{Eqn:mlqfar_b}. We can clearly
observe an endpoint in Fig.~\ref{Fig:mmuq_squark}. However, in general
it might be slightly underestimated, due to final state radiation of
the quark from squark decay.

In the $\sstop_1$ event selection, the endpoint is again more difficult 
to observe, \cf Fig.~\ref{Fig:mmuq_stop}. For $m_{\mu q} \gtrsim 300 \gev$, 
the distribution approaches the endpoint with a very flat slope. Thus, the 
determination of the endpoint requires high statistics. Moreover, 
we have background events beyond the endpoint from heavier squark 
cascade decays or combinations with a jet from a decaying gluino.

We conclude that the standard method that is used to reconstruct
sparticle masses in $R$-parity conserving SUSY works also very well
for our $\slepton_R$ LSP models, where the LSP decays
semi-invisibly. We therefore expect that most of the SUSY masses
in our model can be reconstructed with a similar precision as in $R$-parity conserving models~\cite{Allanach:2000kt,
Gjelsten:2004ki,Gjelsten:2005aw,Barr:2010zj,Bechtle:2009ty,Aad:2009wy},
\ie we expect for our benchmark model a relative error 
of about 10\% or less. We have not calculated the sparticle masses from the
kinematic edges, because for a reliable estimate of the errors, one
has to include detector effects. However, this lies beyond the scope
of this work.

\section{Summary and Conclusion}
\label{Sect:summary}

If $R$-parity is violated, new lepton number violating interactions
can significantly alter the renormalization group running of SUSY
particle masses if the coupling strength is of the order of the gauge
couplings. Within the framework of the $\Bthree$ mSUGRA model, we
showed that a selectron and smuon LSP can arise in large regions of
the SUSY parameter space (\cf Fig.~\ref{Fig:LSPregions}) if a
non-vanishing lepton number violating coupling $\lambda_{ijk}$ with
$k=1,2$ is present at the GUT scale; see Table~\ref{Tab:lamcouplings}
for a list of all allowed couplings.

The selectron or smuon LSP decays mainly into a charged lepton and a
neutrino. Additional charged leptons are usually produced via cascade
decays of heavier sparticles into the LSP. Keeping in mind that
sparticles at the LHC are mostly produced in pairs, we end up with
roughly four charged leptons in each event at
parton-level. Furthermore, two or more jets are expected from decays
of strongly interacting SUSY particles. Table~\ref{Tab:signatures}
gives an overview of the expected LHC signatures.

Based on this, we have developed in Sec.~\ref{Sect:discovery} a
dedicated trilepton search for our SUSY scenarios.  We found that
demanding three charged leptons and two jets in the final state as
well as employing a $Z$-veto and a lower cut on the visible effective mass is
sufficient to obtain a good signal to background ratio.  For example,
for an integrated luminosity of 1 fb$^{-1}$ at $\sqrt{s}=7$ TeV, only
approximately three SM events survive whereas the number of SUSY
events passing our cuts can be of $\mathcal{O}(10-100)$,
\cf Table~\ref{Tab:cutflow}.

We found within the $\Bthree$ mSUGRA model that scenarios with squark
(selectron or smuon LSP) masses up to 1.2 TeV (230 GeV) can be
discovered with an integrated luminosity of 1 fb$^{-1}$ at
$\sqrt{s}=7$ TeV, thus exceeding the discovery reach of $R$-parity
conserving models. Our scenarios are therefore well suited for an
analysis with early LHC data. Going to a cms energy of $\sqrt{s}=14$
TeV and assuming an integrated luminosity of 10 fb$^{-1}$, allows a
discovery of 2.2 TeV (450 GeV) squarks (selectron and smuon LSPs).

After a discovery has been made, a next step would be the
reconstruction of the SUSY mass spectrum. Unfortunately, although the
LSPs decay, a direct mass reconstruction is often not possible (see
Fig.~\ref{Fig:mll_ssnu} for an exception), because (invisible)
neutrinos are always part of the LSP decays. We therefore proposed in
Sec.~\ref{Sect:mass_reco} a method relying on the measurement of
kinematic edges of invariant mass distributions. This method is
analogous to the one usually used for $R$-parity conserving
models, although different SUSY particles are involved in the decay
chain. For example, the neutrino from the LSP decay in our
models plays the role of the lightest neutralino in $R$-parity
conserving models. We also showed that decay chains from heavier
(first and second generation) squarks can be distinguished from those
of the lighter (third generation) top-squarks. Therefore, a
measurement of both squark mass scales is possible.

\begin{acknowledgments}
We thank Ben Allanach, Klaus Desch, Sebastian Fleischmann and Peter Wienemann for helpful discussions. 
S.G. thanks the Alexander von Humboldt Foundation for financial support. 
The work of S.G. was also partly financed by the DOE grant
DE-FG02-04ER41286. The work of
H.K.D. was supported by the BMBF ``Verbundprojekt HEP--Theorie''
under the contract 05H09PDE and the Helmholtz Alliance ``Physics at
the Terascale''.
\end{acknowledgments}

\appendix 

\section{Properties of the Benchmark Models}
\label{App:benchmarks}

{\squeezetable
\begin{table}
\begin{tabular}{| lc | ll | ll |}
\hline
				&	Mass [GeV]			&	Channel				&	BR				&	Channel				 &	BR		\\ 
\hline
$\sse_R^-$		&	$\mathbf{168.7}$		&	$\mu^-\nu_\tau$		&	$\mathbf{50\%}$	&	$\tau^- \nu_\mu$		&	$\mathbf{50\%}$	\\
\hline
$\sstau_1^-$		&	$170.0$				&	$e^- \bnumu$		&	$\mathbf{100\%}$			&		&		\\
\hline
$\ssmu_R^-$		&	$183.6$				&	$\sse_R^+ e^- \mu^-$		&	$34.6\%$		&$\sse_R^- e^+ \mu^-$		&	$28.3\%$		\\ 
				&						&	$\stau_1^+ \tau^- \mu^-$		&	$20.4\%$		& $\stau_1^- \tau^+ \mu^-$	& $16.7\%$		\\
\hline
$\neut_1$			&	$195.7$				&	$\sse_R^- e^+$	&	$23.8\%$			&	$\sse_R^+ e^-$	&	$23.8\%$	\\
				&						&	$\stau_1^- \tau^+$	&	$21.0\%$			&	$\stau_1^+ \tau^-$	&	$21.0\%$	\\
				&						&	$\ssmu_R^- \mu^+$	&	$5.1\%$			&	$\ssmu_R^+ \mu^-$	&	$5.1\%$	\\
\hline
$\ssnutau$		&	$\mathbf{306.5}$		&	$\neut_1 \nu_\tau$		&	$60.2\%$			&	$W^+\sstau_1^-$		&	$28.4\%$		\\
				&						&	$e^- \mu^+$			&	$\mathbf{11.4\%}$		&						&			\\
\hline
$\ssnumu$		&	$\mathbf{309.4}$		&	$\neut_1 \nu_\mu$		&	$84.4\%$			&	$e^- \tau^+$			&	$\mathbf{15.6\%}$	\\
\hline
$\ssnue$			&	$313.5$				&	$\neut_1 \nu_e$		&	$100\%$			&						&			\\
\hline
$\sstau_2^-$		&	$\mathbf{318.4}$		&	$\neut_1 \tau^-$		&	$59.0\%$			&	$\Higgs \sstau_1^-$	&	$16.5\%$	\\
				&						&	$Z^0 \sstau_1^-$		&	$14.1\%$			&	$e^- \bnumu$		&	$\mathbf{10.4\%}$			\\
\hline
$\ssmu_L^-$		&	$\mathbf{318.7}$		&	$\neut_1 \mu^-$		&	$84.1\%$			&	$e^- \bnutau$		&	$\mathbf{15.9\%}$	\\
\hline
$\sse_L^-$		&	$322.8$				&	$\neut_1 e^-$		&	$100\%$			&			&		\\
\hline
$\neut_2$			&	$372.0$				&	$\ssbnutau \nutau$		&	$10.0 \%$			&	$\ssnutau \bnutau$		&	$10.0 \%$		\\
				&						&	$\ssbnumu \numu$		&	$9.2 \%$			&	$\ssnumu \bnumu$		&	$9.2 \%$		\\
				&						&	$\ssbnue\nue$			&	$8.1 \%$			&	$\ssnue \bnue$			&	$8.1 \%$		\\
				&						&	$\ssmu_L^- \mu^+$		&	$7.2\%$			&	$\ssmu_L^+ \mu^-$		&	$7.2\%$		\\
				&						&	$\sstau_2^- \tau^+$		&	$7.1\%$			&	$\sstau_2^+ \tau^-$		&	$7.1\%$		\\
				&						&	$\sse_L^- e^+$			&	$6.2\%$			&	$\sse_L^+ e^-$			&	$6.2\%$		\\		
				&						&	$\sstau_1^- \tau^+$		&	$1.6\%$			&	$\sstau_1^+ \tau^-$		&	$1.6\%$		\\
\hline
$\charge_1^-$		&	$372.0$				&	$\ssbnutau \tau^-$		&	$20.6\%$			&	$\ssbnumu \mu^-$		&	$19.0\%$		\\
				&						&	$\ssbnue e^-$			&	$16.8\%$			&	$\ssmu_L^- \bnumu$	&	$13.9\%$		\\
				&						&	$\sstau_2^- \bnutau$	&	$13.7\%$			&	$\sse_L^- \bnue$		&	$12.0\%$		\\
				&						&  	$\sstau_1^- \bnutau$	&	$3.1\%$			&						&				\\
\hline
$\sstop_1$		&	$531.1$				&	$\neut_1 t$			&	$62.2\%$			&	$\charge_1^+ b$		&	$37.8\%$		\\
\hline
$\ssbottom_1$		&	$847.3$				&	$W^- \sstop_1$		&	$71.5\%$			&	$\charge_1^- t$		&	$17.5\%$		\\
				&						&	$\neut_2 b$			&	$10.4\%$			&						&				\\
\hline
$\neut_3$			&		$898.0$			&	$\sstop_1 \bar{t}$		&	$19.7\%$			&	$\sstop_1^* t$			&	$19.7\%$		\\
				&						&	$\charge_1^- W^+$		&	$18.4\%$			&	$\charge_1^+ W^-$		&	$18.4\%$		\\
				&						&	$\neut_2 Z^0$			&	$16.5\%$			&	$\neut_1 Z^0$			&	$4.8\%$		\\
				&						&	$\neut_2 \Higgs$		&	$1.2\%$			&						&				\\
\hline
$\charge_2^-$		&		$906.0$			&	$\sstop_1^* b$			&	$47.6\%$			&	$\neut_2 W^-$			&	$15.9\%$		\\
				&						&	$\charge_1^- Z^0$		&	$15.4\%$			&	$\charge_1^- \Higgs$	&	$14.6\%$		\\
				&						&	$\neut_1 W^-$			&	$4.2\%$			&						&				\\
\hline
$\neut_4$			&		$906.4$			&	$\sstop_1 \bar{t}$		&	$29.6\%$			&	$\sstop_1^* t$			&	$29.6\%$		\\
				&						&	$\charge_1^- W^+$		&	$12.1\%$			&	$\charge_1^+ W^-$		&	$12.1\%$		\\
				&						&	$\neut_2 \Higgs$		&	$10.3\%$			&	$\neut_1 \Higgs$		&	$2.9\%$		\\
\hline
$\sstop_2$		&	$919.4$				&	$Z^0 \sstop_1$			&	$49.1\%$			&	$\Higgs \sstop_1$		&	$24.6\%$		\\
				&						&	$\charge_1^+ b$		&	$17.3\%$			&	$\neut_2 t$			&	$7.6\%$		\\
				&						&	$\neut_1 t$			&	$1.5\%$			&						&				\\
\hline
$\ssbottom_2$		&	$959.5$				&	$\neut_1 b$			&	$67.0\%$			&	$W^- \sstop_1$		&	$28.9\%$		\\
				&						&	$\charge_1^- t$		&	$2.1\%$			&	$\neut_2 b$			&	$1.2\%$		\\
\hline
$\ssdown_R\,(\ssstrange_R)$	&	$962.3$		&	$\neut_1 d (s) $		&	$100\%$			&						&				\\
\hline
$\ssup_R\,(\sscharm_R)$	&	$965$			&	$\neut_1 u (c) $		&	$100\%$			&						&				\\
\hline
$\ssup_L\,(\sscharm_L)$	&	$1001.8$			&	$\charge_1^+ d (s) $	&	$65.9\%$			&	$\neut_2 u (c)$			&	$32.9\%$		\\
				&						&	$\neut_1 u (c)$			&	$1.2\%$			&						&				\\
\hline
$\ssdown_L\,(\ssstrange_L)$	&	$1004.7$		&	$\charge_1^- u (c)$		&	$65.5\%$			&	$\neut_2 d (s)$			&	$32.8\%$		\\
						&				&	$\neut_1 d (s) $		&	$1.7\%$			&						&				\\
\hline
$\glu$			&		$1093.7$			&	$\sstop_1 \bar{t}$		&	$20.9\%$			&	$\sstop_1^*  t $			&	$20.9\%$		\\
				&						&	$\ssbottom_1 \bar{b}$	&	$8.5\%$			&	$\ssbottom_1^* b $		&	$8.5\%$		\\
				&						&	$\ssbottom_2 \bar{b}$	&	$2.9\%$			&	$\ssbottom_2^* b $		&	$2.9\%$		\\
				&						&	$\ssdown_R \bar{d} (\ssstrange_R \bar{s})$	&	$2.7\%$	&	$\ssdown_R^* d (\ssstrange_R^* s ) $&	$2.7\%$	\\
				&						&	$\ssup_R \bar{u} (\sscharm_R \bar{c})$	&	$2.6\%$	&	$\ssup_R^* u (\sscharm_R^* c ) $&	$2.6\%$	\\
				&						&	$\sstop_2 \bar{t}$		&	$1.6\%$			&	$\sstop_2^* t$			&	$1.6\%$		\\
				&						&	$\ssup_L \bar{u} (\sscharm_L \bar{c})$	&	$1.4\%$	&	$\ssup_L^* u (\sscharm_L^* c ) $&	$1.4\%$	\\
				&						&	$\ssdown_L \bar{d} (\ssstrange_L \bar{s})$	&	$1.3\%$	&	$\ssdown_L^* d (\ssstrange_L^* s ) $&	$1.3\%$	\\
\hline
\end{tabular}
\caption{Branching ratios (BRs) and sparticle masses for the benchmark
scenario BE1; see Table~\ref{Tab:benchmarks}. BRs smaller than $1\%$
are neglected. $R$-parity violating decays and masses which are
reduced by more than 5 GeV (compared to the $R$-parity conserving
case) are shown in bold-face.}
\label{Tab:BE1}
\end{table}
}
{\squeezetable
\begin{table}
\begin{tabular}{| lc | ll | ll |}
\hline
				&	Mass [GeV]			&	Channel				&	BR				&	Channel				 &	BR		\\ 
\hline
$\sse_R^-$		&	$\mathbf{182.3}$		&	$\mu^-\nu_\tau$		&	$\mathbf{50\%}$	&	$\tau^- \nu_\mu$		&	$\mathbf{50\%}$	\\
\hline
$\sstau_1^-$		&	$189.0$				&	$\sse_R^+ e^- \tau^-$	&	$50.2\%$			&	$\sse_R^- e^+ \tau^-$	&	$49.5\%$	\\
\hline
$\neut_1$			&	$189.5$				&	$\sse_R^- e^+$	&	$50\%$			&	$\sse_R^+ e^-$	&	$50\%$	\\
\hline
$\ssmu_R^-$		&	$199.0$				&	$\neut_1 \mu^-$	&	$100\%$		&						&				\\ 
\hline
$\ssnutau$		&	$\mathbf{309.8}$		&	$\neut_1 \nu_\tau$		&	$71.0\%$			&	$W^+\sstau_1^-$		&	$17.0\%$		\\
				&						&	$e^- \mu^+$			&	$\mathbf{12.0\%}$		&						&			\\
\hline
$\ssnumu$		&	$\mathbf{312.0}$		&	$\neut_1 \nu_\mu$		&	$85.8\%$			&	$e^- \tau^+$			&	$\mathbf{14.2\%}$	\\
\hline
$\ssnue$			&	$317.0$				&	$\neut_1 \nu_e$		&	$100\%$			&						&			\\
\hline
$\sstau_2^-$		&	$\mathbf{320.8}$		&	$\neut_1 \tau^-$		&	$69.9\%$			&	$e^- \bnumu$		&	$\mathbf{11.3\%}$	\\
				&						&	$\Higgs \sstau_1^-$		&	$10.2\%$			&	$Z^0 \sstau_1^-$		&	$8.6\%$		\\	
\hline
$\ssmu_L^-$		&	$\mathbf{320.8}$		&	$\neut_1 \mu^-$		&	$85.2\%$			&	$e^- \bnutau$		&	$\mathbf{14.8\%}$	\\
\hline
$\sse_L^-$		&	$325.7$				&	$\neut_1 e^-$		&	$100\%$			&			&		\\
\hline
$\neut_2$			&	$360.1$				&	$\ssbnutau \nutau$		&	$10.5 \%$			&	$\ssnutau \bnutau$		&	$10.5 \%$		\\
				&						&	$\ssbnumu \numu$		&	$9.7 \%$			&	$\ssnumu \bnumu$		&	$9.7 \%$		\\
				&						&	$\ssbnue\nue$			&	$7.9 \%$			&	$\ssnue \bnue$			&	$7.9 \%$		\\
				&						&	$\ssmu_L^- \mu^+$		&	$7.0\%$			&	$\ssmu_L^+ \mu^-$		&	$7.0\%$		\\
				&						&	$\sstau_2^- \tau^+$		&	$6.8\%$			&	$\sstau_2^+ \tau^-$		&	$6.8\%$		\\
				&						&	$\sse_L^- e^+$			&	$5.4\%$			&	$\sse_L^+ e^-$			&	$5.4\%$		\\		
				&						&	$\sstau_1^- \tau^+$		&	$2.0\%$			&	$\sstau_1^+ \tau^-$		&	$2.0\%$		\\
				&						&	$\neut_1 \Higgs$		&	$1.3\%$			&						&				\\
\hline
$\charge_1^-$		&	$360.2$				&	$\ssbnutau \tau^-$		&	$21.7\%$			&	$\ssbnumu \mu^-$		&	$19.9\%$		\\
				&						&	$\ssbnue e^-$			&	$16.3\%$			&	$\ssmu_L^- \bnumu$	&	$13.4\%$		\\
				&						&	$\sstau_2^- \bnutau$	&	$13.2\%$			&	$\sse_L^- \bnue$		&	$10.5\%$		\\
				&						&  	$\sstau_1^- \bnutau$	&	$3.8\%$			&	$\neut_1 W^-$			&	$1.3\%$		\\
\hline
$\sstop_1$		&	$448.3$				&	$\neut_1 t$			&	$71.9\%$			&	$\charge_1^+ b$		&	$28.1\%$		\\
\hline
$\ssbottom_1$		&	$809.1$				&	$W^- \sstop_1$		&	$78.1\%$			&	$\charge_1^- t$		&	$13.3\%$		\\
				&						&	$\neut_2 b$			&	$8.1\%$			&						&				\\
\hline
$\sstop_2$		&	$887.0$				&	$Z^0 \sstop_1$			&	$52.7\%$			&	$\Higgs \sstop_1$		&	$25.9\%$		\\
				&						&	$\charge_1^+ b$		&	$14.1\%$			&	$\neut_2 t$			&	$6.1\%$		\\
				&						&	$\neut_1 t$			&	$1.2\%$			&						&				\\
\hline
$\neut_3$			&		$936.7$			&	$\sstop_1 \bar{t}$		&	$26.0\%$			&	$\sstop_1^* t$			&	$26.0\%$		\\
				&						&	$\charge_1^- W^+$		&	$14.6\%$			&	$\charge_1^+ W^-$		&	$14.6\%$		\\
				&						&	$\neut_2 Z^0$			&	$13.3\%$			&	$\neut_1 Z^0$			&	$3.8\%$		\\
\hline
$\ssbottom_2$		&	$937.7$				&	$\neut_1 b$			&	$67.9\%$			&	$W^- \sstop_1$		&	$26.0\%$		\\
				&						&	$\charge_1^- t$		&	$2.7\%$			&	$\neut_2 b$			&	$1.5\%$		\\
\hline
$\ssdown_R\,(\ssstrange_R)$	&	$939.8$		&	$\neut_1 d (s) $		&	$100\%$			&						&				\\
\hline
$\ssup_R\,(\sscharm_R)$	&	$942.9$			&	$\neut_1 u (c) $		&	$100\%$			&						&				\\
\hline
$\charge_2^-$		&		$944.5$			&	$\sstop_1^* b$			&	$55.6\%$			&	$\neut_2 W^-$			&	$13.5\%$		\\
				&						&	$\charge_1^- Z^0$		&	$13.1\%$			&	$\charge_1^- \Higgs$	&	$12.4\%$		\\
				&						&	$\neut_1 W^-$			&	$3.4\%$			&						&				\\
\hline
$\neut_4$			&		$945.1$			&	$\sstop_1 \bar{t}$		&	$33.3\%$			&	$\sstop_1^* t$			&	$33.3\%$		\\
				&						&	$\charge_1^- W^+$		&	$10.0\%$			&	$\charge_1^+ W^-$		&	$10.0\%$		\\
				&						&	$\neut_2 \Higgs$		&	$8.5\%$			&	$\neut_1 \Higgs$		&	$2.4\%$		\\
\hline
$\ssup_L\,(\sscharm_L)$	&	$977.6$			&	$\charge_1^+ d (s) $	&	$65.9\%$			&	$\neut_2 u (c)$			&	$32.9\%$		\\
					&					&	$\neut_1 u (c)$			&	$1.2\%$			&						&				\\
\hline
$\ssdown_L\,(\ssstrange_L)$	&	$980.4$		&	$\charge_1^- u (c)$		&	$65.6\%$			&	$\neut_2 d (s)$			&	$32.8\%$		\\
						&				&	$\neut_1 d (s) $		&	$1.6\%$			&						&				\\
\hline
$\glu$			&		$1063.1$			&	$\sstop_1 \bar{t}$		&	$23.0\%$			&	$\sstop_1^*  t $			&	$23.0\%$		\\
				&						&	$\ssbottom_1 \bar{b}$	&	$8.7\%$			&	$\ssbottom_1^* b $		&	$8.7\%$		\\
				&						&	$\ssbottom_2 \bar{b}$	&	$2.4\%$			&	$\ssbottom_2^* b $		&	$2.4\%$		\\
				&						&	$\ssdown_R \bar{d} (\ssstrange_R \bar{s})$	&	$2.4\%$	&	$\ssdown_R^* d (\ssstrange_R^* s ) $&	$2.4\%$	\\
				&						&	$\ssup_R \bar{u} (\sscharm_R \bar{c})$	&	$2.3\%$	&	$\ssup_R^* u (\sscharm_R^* c ) $&	$2.3\%$	\\
				&						&	$\sstop_2 \bar{t}$		&	$2.0\%$			&	$\sstop_2^* t$			&	$2.0\%$		\\
				&						&	$\ssup_L \bar{u} (\sscharm_L \bar{c})$	&	$1.2\%$	&	$\ssup_L^* u (\sscharm_L^* c ) $&	$1.2\%$	\\
				&						&	$\ssdown_L \bar{d} (\ssstrange_L \bar{s})$	&	$1.1\%$	&	$\ssdown_L^* d (\ssstrange_L^* s ) $&	$1.1\%$	\\
\hline
\end{tabular}
\caption{Same as Table~\ref{Tab:BE1}, but for the benchmark scenario BE2.}
\label{Tab:BE2}
\end{table}
}
{\squeezetable
\begin{table}
\begin{tabular}{| lc | ll | ll |}
\hline
				&	Mass [GeV]			&	Channel				&	BR				&	Channel				 &	BR		\\ 
\hline
$\sse_R^-$		&	$\mathbf{182.0}$		&	$\mu^-\nu_\tau$		&	$\mathbf{50\%}$	&	$\tau^- \nu_\mu$		&	$\mathbf{50\%}$	\\
\hline
$\neut_1$			&	$184.9$				&	$\sse_R^- e^+$	&	$50\%$			&	$\sse_R^+ e^-$	&	$50\%$	\\
\hline
$\sstau_1^-$		&	$187.2$				&	$\neut_1 \tau^-$		&	$64.5\%$			&	$e^-\bnumu$			&	$\mathbf{35.5\%}$		\\
\hline
$\ssmu_R^-$		&	$195.9$				&	$\neut_1 \mu^-$	&	$100\%$		&						&				\\ 
\hline
$\ssnutau$		&	$\mathbf{304.3}$		&	$\neut_1 \nu_\tau$		&	$73.6\%$			&	$W^+\sstau_1^-$		&	$14.2\%$		\\
				&						&	$e^- \mu^+$			&	$\mathbf{12.2\%}$		&						&			\\
\hline
$\ssnumu$		&	$\mathbf{306.2}$		&	$\neut_1 \nu_\mu$		&	$86.0\%$			&	$e^- \tau^+$			&	$\mathbf{14.0\%}$	\\
\hline
$\ssnue$			&	$310.4$				&	$\neut_1 \nu_e$		&	$100\%$			&						&			\\
\hline
$\ssmu_L^-$		&	$\mathbf{315.2}$		&	$\neut_1 \mu^-$		&	$85.2\%$			&	$e^- \bnutau$		&	$\mathbf{14.8\%}$	\\
\hline
$\sstau_2^-$		&	$\mathbf{315.3}$		&	$\neut_1 \tau^-$		&	$72.5\%$			&	$e^- \bnumu$		&	$\mathbf{11.7\%}$	\\
				&						&	$\Higgs \sstau_1^-$		&	$8.5\%$			&	$Z^0 \sstau_1^-$		&	$7.3\%$		\\	
\hline
$\sse_L^-$		&	$319.3$				&	$\neut_1 e^-$		&	$100\%$			&			&		\\
\hline
$\neut_2$			&	$351.2$				&	$\ssbnutau \nutau$		&	$10.5 \%$			&	$\ssnutau \bnutau$		&	$10.5 \%$		\\
				&						&	$\ssbnumu \numu$		&	$9.7 \%$			&	$\ssnumu \bnumu$		&	$9.7 \%$		\\
				&						&	$\ssbnue\nue$			&	$8.1 \%$			&	$\ssnue \bnue$			&	$8.1 \%$		\\
				&						&	$\ssmu_L^- \mu^+$		&	$6.8\%$			&	$\ssmu_L^+ \mu^-$		&	$6.8\%$		\\
				&						&	$\sstau_2^- \tau^+$		&	$6.6\%$			&	$\sstau_2^+ \tau^-$		&	$6.6\%$		\\
				&						&	$\sse_L^- e^+$			&	$5.4\%$			&	$\sse_L^+ e^-$			&	$5.4\%$		\\		
				&						&	$\sstau_1^- \tau^+$		&	$2.0\%$			&	$\sstau_1^+ \tau^-$		&	$2.0\%$		\\
				&						&	$\neut_1 \Higgs$		&	$1.6\%$			&						&				\\
\hline
$\charge_1^-$		&	$351.2$				&	$\ssbnutau \tau^-$		&	$21.7\%$			&	$\ssbnumu \mu^-$		&	$20.1\%$		\\
				&						&	$\ssbnue e^-$			&	$16.8\%$			&	$\ssmu_L^- \bnumu$	&	$13.0\%$		\\
				&						&	$\sstau_2^- \bnutau$	&	$12.7\%$			&	$\sse_L^- \bnue$		&	$10.4\%$		\\
				&						&  	$\sstau_1^- \bnutau$	&	$3.8\%$			&	$\neut_1 W^-$			&	$1.6\%$		\\
\hline
$\sstop_1$		&	$481.7$				&	$\neut_1 t$			&	$62.1\%$			&	$\charge_1^+ b$		&	$37.9\%$		\\
\hline
$\ssbottom_1$		&	$805.4$				&	$W^- \sstop_1$		&	$73.9\%$			&	$\charge_1^- t$		&	$15.9\%$		\\
				&						&	$\neut_2 b$			&	$9.7\%$			&						&				\\
\hline
$\sstop_2$		&	$881.7$				&	$Z^0 \sstop_1$			&	$51.3\%$			&	$\Higgs \sstop_1$		&	$24.2\%$		\\
				&						&	$\charge_1^+ b$		&	$16.1\%$			&	$\neut_2 t$			&	$7.0\%$		\\
				&						&	$\neut_1 t$			&	$1.4\%$			&						&				\\
\hline
$\neut_3$			&		$884.0$			&	$\sstop_1 \bar{t}$		&	$22.1\%$			&	$\sstop_1^* t$			&	$22.1\%$		\\
				&						&	$\charge_1^- W^+$		&	$17.0\%$			&	$\charge_1^+ W^-$		&	$17.0\%$		\\
				&						&	$\neut_2 Z^0$			&	$15.4\%$			&	$\neut_1 Z^0$			&	$4.5\%$		\\
				&						&	$\neut_2 \Higgs$		&	$1.0\%$			&						&				\\
\hline
$\charge_2^-$		&		$892.4$			&	$\sstop_1^* b$			&	$50.8\%$			&	$\neut_2 W^-$			&	$15.1\%$		\\
				&						&	$\charge_1^- Z^0$		&	$14.6\%$			&	$\charge_1^- \Higgs$	&	$13.7\%$		\\
				&						&	$\neut_1 W^-$			&	$3.8\%$			&						&				\\
\hline
$\neut_4$			&		$893.1$			&	$\sstop_1 \bar{t}$		&	$31.4\%$			&	$\sstop_1^* t$			&	$31.4\%$		\\
				&						&	$\charge_1^- W^+$		&	$11.1\%$			&	$\charge_1^+ W^-$		&	$11.1\%$		\\
				&						&	$\neut_2 \Higgs$		&	$9.4\%$			&	$\neut_1 \Higgs$		&	$2.7\%$		\\
\hline
$\ssbottom_2$		&	$919.3$				&	$\neut_1 b$			&	$70.5\%$			&	$W^- \sstop_1$		&	$26.0\%$		\\
				&						&	$\charge_1^- t$		&	$1.8\%$			&	$\neut_2 b$			&	$1.1\%$		\\
\hline
$\ssdown_R\,(\ssstrange_R)$	&	$921.1$		&	$\neut_1 d (s) $		&	$100\%$			&						&				\\
\hline
$\ssup_R\,(\sscharm_R)$	&	$923.8$			&	$\neut_1 u (c) $		&	$100\%$			&						&				\\
\hline
$\ssup_L\,(\sscharm_L)$	&	$957.9$			&	$\charge_1^+ d (s) $	&	$65.9\%$			&	$\neut_2 u (c)$			&	$33.0\%$		\\
					&					&	$\neut_1 u (c)$			&	$1.1\%$			&						&				\\
\hline
$\ssdown_L\,(\ssstrange_L)$	&	$961.0$		&	$\charge_1^- u (c)$		&	$65.5\%$			&	$\neut_2 d (s)$			&	$32.7\%$		\\
						&				&	$\neut_1 d (s) $		&	$1.7\%$			&						&				\\
\hline
$\glu$			&		$1041.8$			&	$\sstop_1 \bar{t}$		&	$22.6\%$			&	$\sstop_1^*  t $			&	$22.6\%$		\\
				&						&	$\ssbottom_1 \bar{b}$	&	$8.9\%$			&	$\ssbottom_1^* b $		&	$8.9\%$		\\
				&						&	$\ssbottom_2 \bar{b}$	&	$2.7\%$			&	$\ssbottom_2^* b $		&	$2.7\%$		\\
				&						&	$\ssdown_R \bar{d} (\ssstrange_R \bar{s})$	&	$2.6\%$	&	$\ssdown_R^* d (\ssstrange_R^* s ) $&	$2.6\%$	\\
				&						&	$\ssup_R \bar{u} (\sscharm_R \bar{c})$	&	$2.5\%$	&	$\ssup_R^* u (\sscharm_R^* c ) $&	$2.5\%$	\\
				&						&	$\ssup_L \bar{u} (\sscharm_L \bar{c})$	&	$1.3\%$	&	$\ssup_L^* u (\sscharm_L^* c ) $&	$1.3\%$	\\
				&						&	$\ssdown_L \bar{d} (\ssstrange_L \bar{s})$	&	$1.2\%$	&	$\ssdown_L^* d (\ssstrange_L^* s ) $&	$1.2\%$	\\
\hline
\end{tabular}
\caption{Same as Table~\ref{Tab:BE1}, but for the benchmark scenario BE3.}
\label{Tab:BE3}
\end{table}
}

We show in Tables~\ref{Tab:BE1}, \ref{Tab:BE2} and \ref{Tab:BE3} the
mass spectra and the dominant decays of the supersymmetric particles
of the benchmark points BE1, BE2 and BE3, respectively; see
Table~\ref{Tab:benchmarks} for a definition. Sparticle masses, that
are reduced by more than $5\gev$ (compared to the $R$-parity
conserving case) and $R$-parity violating decays are marked in
bold-face. Note that only the masses of those sparticles, which couple
directly to the $L_i L_j \bar E_k$ operator, are significantly
reduced, \cf Sec.~\ref{Sect:RGE}.  Therefore, in our benchmark models
(${\lambda_{231}}|_{\rm GUT}\not =0$) only the $\sse_R$, $\ssmu_L$,
$\ssnumu$, $\stau_2$ and $\ssnutau$ are affected.

These sparticles then also exhibit $R$-parity violating decays to SM
particles via $\lam_{231}$. For the $\ssnutau$ this can lead to a
striking peak in the electron-muon invariant mass distribution; \cf 
Fig.~\ref{Fig:mll_ssnu}. In addition, the $\stau_1$ can also decay via
the $\lam_{231}$ coupling, because of its (small) left-handed
component. This happens in particular in scenarios, where the $\stau_1$
is the NLSP and its mass is close to the LSP mass (as in BE1,
Table~\ref{Tab:BE1}), {\it i.e.} the $R$-parity conserving decay into
the LSP is phase-space suppressed. The $\sse_R$ LSP can only decay via
$R$-parity violating interactions: $\sse_R\to\mu\nutau$ and $\sse_R\to
\tau\numu$.

Common to all benchmark points is a rather light $\sstop_1$ (compared
to the other squarks).  For all benchmark points, the $\sstop_1$ mass
is around $450\gev$-$550\gev$ and the other squark masses are in the
range of $800\gev$-$1\tev$. Because of the large top Yukawa coupling,
the stop mass receives large negative contributions from RGE running,
especially for a negative $\azero$ with a large magnitude
\cite{Drees:1995hj,Ibanez:1984vq}; see Sec.~\ref{Sect:A0dependence}
for a similar case. Furthermore, the light stop mass is reduced by
large mixing between the left- and right-handed states. As one can see
in Tables~\ref{Tab:BE1}, \ref{Tab:BE2} and \ref{Tab:BE3}, the (mainly
right-handed) $\sstop_1$ dominantly decays into the (bino-like)
$\neut_1$ and a top quark, while the decay into the (wino-like)
lightest chargino, $\charge_1^+$, is subdominant.

The $\sse_R$, $\ssmu_R$, $\stau_1$ and $\neut_1$ always form the
lightest four sparticles in $\Bthree$ mSUGRA models with a $\sse_R$ or
$\ssmu_R$ LSP.  The scenario BE1, Table~\ref{Tab:BE1}, exhibits a
$\stau_1$ NLSP that is nearly degenerate in mass with the $\sse_R$
LSP. Thus, it undergoes the $R$-parity violating decay $\stau_1 \to e
\numu$, yielding high-$p_T$ electrons. The $\ssmu_R$ is the NNLSP and
decays into the $\sse_R$ or the $\stau_1$ via 3-body decays producing
in general two low-$p_T$ charged leptons due to the reduced
phase space. We calculate and discuss these decays in detail in
Appendix~\ref{App:sleptondecays}. The $\neut_1$ is the NNNLSP. Besides
the decay into the $\sse_R$ LSP and an electron ($47.6\%$), it also
decays to a sizable fraction ($42.0\%$) into the $\stau_1$ NLSP and a
$\tau$ lepton.

The benchmark scenario BE2, Table~\ref{Tab:BE2}, also has a $\stau_1$
NLSP. However, the $\stau_1$ NLSP is nearly mass degenerate with the
$\neut_1$ NNLSP. Therefore, it decays exclusively via 3-body decays
into the $\sse_R$ LSP, yielding a low-$p_T$ tau lepton and an
electron; \cf Appendix~\ref{App:sleptondecays}. The $\neut_1$ NNLSP
always decays into the $\sse_R$ LSP and an electron.

In contrast to BE1 and BE2, the NLSP in BE3, Table~\ref{Tab:BE3}, is
the $\neut_1$ which is roughly $3\gev$ heavier than the $\sse_R$
LSP. Therefore, the electrons from the $\neut_1$ decay into the LSP
are very soft. We have a $\stau_1$ NNLSP, which decays $R$-parity
conserving into the $\neut_1$ and a tau as well as via $R$-parity
violating decays into an electron and a neutrino. In both BE2 and BE3,
the $\ssmu_R$ is the NNNLSP and decays exclusively into the $\neut_1$
and a muon.

The remaining sparticle mass spectra and decays look very similar to
those of $R$-parity conserving mSUGRA \cite{Allanach:2002nj}.

\section{Cut-Flow for $\sqrt{s} = 14 \tev$}
\label{App:14TeV}

\begin{table*}[t]
\centering
\setlength{\tabcolsep}{1pc}
\begin{tabular*}{\textwidth}{@{\extracolsep{\fill}}lrrrrr}
\hline\hline
Sample	&	Before cuts	&	$N_\mathrm{lep}\ge 3$	&	$N_\mathrm{jet} \ge 2$		&	$M_\mathrm{OSSF}$	&	$M_\mathrm{eff}^\mathrm{vis} \ge 400\gev$	\\
\hline
top				&	$(5215 \pm 2) \cdot 10^3$& $553 \pm 21$	& $491 \pm 20$	& $397 \pm 19$	& $55.9 \pm 7.0$	\\
$Z +\mbox{jets}$	&	$(5601 \pm 2)\cdot 10^3$& $1980 \pm 41$	& $571 \pm 22$	& $48.7 \pm 6.4$	& $2.6 \pm 1.5$	\\
$W +\mbox{jets}$	&	$(9516\pm 9)\cdot 10^2$	& $4.8 \pm 2.0$	& $1.6 \pm 1.1$ 	& $\lesssim 1.0$	& $\lesssim 1.0$	\\
di-boson			&	$(7719\pm 8)\cdot 10^2$ & $2573 \pm 17$	& $605 \pm 11$	&$56.7 \pm 4.4$	& $6.1 \pm 1.1$ 	\\\hline
all SM			&	$(12540\pm 3) \cdot 10^3$	& $5110 \pm 49$	& $1669 \pm 32$	& $503 \pm 20$	& $64.7 \pm 7.2$	\\	
\hline
BE1				& $23040 \pm 47$		& $14412\pm 37$		& $13925 \pm 37$	& $12204\pm 34$	& $11854 \pm 34$	\\
BE2				& $30980 \pm 57$		& $13910 \pm 38$	& $13442 \pm 37$	& $12227 \pm 36$& $11569 \pm 35$	\\
BE3				& $31160 \pm 55$		& $9118\pm 30$		& $8700 \pm 29$		& $7807\pm 28$	& $7533 \pm 27$	\\
\hline\hline
\end{tabular*}
\caption{Number of SM background and signal events after each step in the event selection at 
$\sqrt{s}=14\tev$, scaled to an integrated luminosity of $10~\mathrm{fb}^{-1}$. We provide 
the results for the benchmark scenarios BE1, BE2 and BE3 (Table~\ref{Tab:benchmarks}). The uncertainties 
correspond to statistical fluctuations.}
\label{Tab:cutflow14TeV}
\end{table*}

We present in Table~\ref{Tab:cutflow14TeV} the cut flow of the signal
and SM background events for an integrated luminosity of
$10~\mathrm{fb}^{-1}$ at $\sqrt{s}=14\tev$. Although the benchmark
scenarios BE1, BE2 and BE3 (see Table~\ref{Tab:benchmarks}) are
already observable with very early LHC data, \cf
Sec.~\ref{Sect:cutflow}, we provide their expected event yields as a
reference in order to compare the signal efficiencies at
$\sqrt{s}=7\tev$ and $\sqrt{s}=14\tev$.

We apply the inclusive three-lepton analysis developed in
Sec.~\ref{Sect:cutflow}. After the three lepton requirement (third
column of Table~\ref{Tab:cutflow14TeV}), the expected SM background is
reduced to roughly $5110$ events. Already at this stage, the expected
signal event yield of the benchmark points BE1, BE2 and BE3 is
overwhelming, \textit{i.e.} a factor of 2--3 larger than the SM
backgrounds. The signal efficiency of this first cut is the same as
for the LHC at $\sqrt{s}=7\tev$.

The jet multiplicity requirement (fourth column of
Table~\ref{Tab:cutflow14TeV}) reduces the SM background to $1670$
events. It mainly originates from $Zj$ ($26\%$), $t\overline t$
($24\%$) and $WZ$ ($15\%$) production. Because sparton pair production
strongly dominates the signal for all benchmark scenarios at
$\sqrt{s}=14\tev$, \cf Table~\ref{Tab:signalMC}, almost every signal
event has at least two hard jets. Therefore, the signal efficiency of
this cut is large, \textit{i.e.} 95\% or higher for all benchmark
points.  This is higher than for the $\sqrt{s}=7\tev$ sample, \cf
Table~\ref{Tab:cutflow}.

The $Z$ veto (fifth column of Table~\ref{Tab:cutflow14TeV})
effectively reduces the $Z+\mbox{jets}$ and di-boson backgrounds,
leaving only a total SM background of roughly 500 events. The
background is now dominated by the $t\overline t$ production. The
number of signal events is only reduced by roughly 10\%.

Finally, after the requirement on the visible effective mass (last
column of Table~\ref{Tab:cutflow14TeV}), the SM background is reduced
to approximately $65$ events. At the same time, nearly all signal
events pass this cut. The signal to background ratio is now of
$\mathcal{O}(100)$. This justifies neglecting the SM
background events for the mass reconstruction, \cf
Sec.~\ref{Sect:mass_reco}.\\

\section{Kinematic Endpoints of Invariant Mass Distributions}
\label{App:endpoints}

Assuming the cascade decay of Fig.~\ref{Fig:generalcascade}, analytic
formulas for the (measureable) kinematic endpoints of the two-- and
three--particle invariant masses, Eq.~(\ref{Eqn:inv_masses}), can be
derived \cite{Miller:2005zp,Allanach:2000kt}
\begin{widetext}
\begin{align}
(m_{ba}^\mathrm{max})^2 =& (m_{C}^2-m_{B}^2)(m_{B}^2 -m_A^2)/m_{B}^2,\\
(m_{ca}^\mathrm{max})^2 =& (m_D^2 - m_C^2)(m_B^2-m_A^2)/m_B^2, 
\label{Eqn:mca}\\
(m_{cb}^\mathrm{max})^2 =& (m_D^2 - m_C^2)(m_C^2-m_B^2)/m_C^2, 
\label{Eqn:mcb}\\
(m_{cba}^\mathrm{max})^2 =& \left\{ \begin{array}{l}  
\max\Big[\frac{(m_D^2-m_C^2)(m_C^2-m_A^2)}{m_C^2},~\frac{(m_D^2-m_B^2)
(m_B^2-m_A^2)}{m_B^2},~\frac{(m_D^2m_B^2-m_C^2m_A^2)(m_C^2-m_B^2)}
{m_C^2m_B^2}\Big],\\
\mbox{or $(m_D - m_A)^2$ if $m_B^2 < m_A m_D < m_C^2$ and $m_A m_C^2 
< m_B^2 m_D$.}
\end{array} \right.\\
(m_{cba}^\mathrm{min})^2 =& \big[\,2m_B^2(m_D^2-m_C^2)(m_C^2 - m_A^2) 
+ (m_D^2+m_C^2)(m_C^2-m_B^2)(m_B^2-m_A^2) \nonumber\\
& - (m_D^2-m_C^2)\sqrt{(m_C^2+m_B^2)^2(m_B^2+m_A^2)^2 - 16 m_C^2 m_B^4
m_A^2}\, \big]/(4m_B^2m_C^2).\label{Eqn:mabc_min}
\end{align}
\end{widetext}
These equations can be solved for the unknown particle masses in the
decay chain.

\clearpage

\section{Three-Body Slepton Decays}
\label{App:sleptondecays}

As we have shown in Sec.~\ref{Sect:parameterspace}, some regions of the $\slepton_R$ 
LSP parameter space exhibit the SUSY mass hierarchies
\begin{align}
M_{\slepton_R} < M_{\stau_1} < M_{\slepton^{'}_R} < M_{\neut_1},
\end{align}
and
\begin{align}
M_{\slepton_R} < M_{\stau_1} <  M_{\neut_1},
\end{align}
where $\slepton'_R$ is a right-handed non-LSP slepton of the first or second generation.
In this case, the 3-body decays  
\begin{align}
\slepton'^{-}_R \to & \,\, \ell'^{-} \ell^\pm \slepton_R^\mp , \nonumber \\
\stau_1^- \to & \,\, \tau^- \ell^\pm \slepton_R^\mp ,
\label{Eq:new_decays}
\end{align}
can be the dominant decay modes of the $\slepton'_R$ and
$\stau_1$. This is for example the case in the benchmark scenario BE1
(BE2) for the $\ssmu_R$ ($\stau_1$), \cf Table~\ref{Tab:BE1}
(Table~\ref{Tab:BE2}).

In {\tt ISAJET7.64}, that we employ to calculate the 2- and 3-body
decays of the SUSY particles, the decays in Eq.~(\ref{Eq:new_decays})
are not implemented, because in most SUSY scenarios, the $\stau_1$ is
considered to be lighter than the other sleptons.

In this appendix, we fill this gap and calculate the missing 3-body
slepton decays of Eq.~(\ref{Eq:new_decays}). We show the resulting
squared matrix elements and give numbers for the respective branching
ratios. The phase-space integration is performed numerically within
{\tt ISAJET}.  We use the 2-component spinor techniques and notation
from Ref.~\cite{Dreiner:2008tw} for the calculation of the matrix
elements. To our knowledge, the calculation of the 3-body decays is
not yet given in the literature.

\subsection{Three-Body Slepton Decay $\slepton'^{-}_R \to \ell'^{-} \ell^\pm\slepton^\mp_R$}
\label{Sec:3bodydecay1}

We now calculate the 3-body slepton decays $\slepton_R^{'-} \rightarrow \ell^{'-} \ell^\pm\slepton_R^\mp$,
Eq.~(\ref{Eq:new_decays}), that are mediated by a virtual neutralino\footnote{In principle, there are also
3-body decays with virtual charginos. However, these decays are negligible due to the heavier propagators.
Furthermore, the right-handed sleptons can not couple to wino-like charginos.}. Because $\slepton_R$ and 
$\slepton'_R$ are sleptons of the first two generations, we can neglect contributions proportional 
to the ($R$-parity conserving) Yukawa couplings.

\begin{figure}[t]
\begin{center}
\includegraphics[scale=1.0]{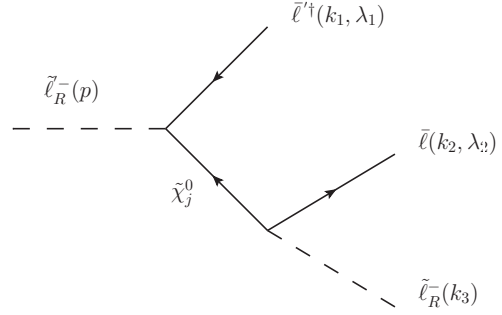}
\end{center}
\caption{Feynman diagram for the 3-body slepton decay $\slepton'^-_R \rightarrow \ell'^- \ell^+\slepton_R^-$.}
\label{feyn_LRtoLRPminus}
\end{figure}

The relevant Feynman diagram for the decay $\slepton'^-_R \rightarrow \ell'^- \ell^+\slepton_R^-$ 
is shown in Fig.~\ref{feyn_LRtoLRPminus}, where the momenta ($p,k_1,k_2,k_3$) 
and polarizations ($\lambda_1,\lambda_2$) of the particles 
are indicated. The neutralino mass eigenstates are denoted by $\neut_j$ with $j=1,\dots ,4$. Using 
the rules and notation of Ref.~\cite{Dreiner:2008tw}, we obtain for the amplitude 
\begin{align}
i\mathcal{M} = (-ia_j^*)(-ia_j)x_2^\dagger\frac{i(p-k_1)\cdot \bar\sigma}{(p-k_1)^2-\mj^2} y_1,
\label{Eq:amplitude1}
\end{align}
where $a_j \equiv \sqrt{2}g'N_{j1}$, and the spinor wave functions are
$y_1=y(\vec{k_1},\lambda_1)$ and $x_2^\dagger =
x^\dagger(\vec{k_2},\lambda_2)$. Squaring the amplitude then yields
\begin{align}
\left| \mathcal{M}\right|^2 = A\, x_2^\dagger(p-k_1)\cdot \bar\sigma y_1 y_1^\dagger (p-k_1)\cdot \bar \sigma x_2,
\label{Eq:amplitude1-sq}
\end{align}
with
\begin{align}
A \equiv  \sum_{j,k=1}^4 \frac{|a_j|^2}{(p-k_1)^2-\mj^2}\cdot \frac{ |a_k|^2 }{(p-k_1)^2-\mk^2}\,.
\end{align}
Summing Eq.~(\ref{Eq:amplitude1-sq}) over the spins leads to
\begin{align}
\sum_{\lambda_1, \lambda_2}\left| \mathcal{M}\right|^2 = A\,\left[m_{13}^2m_{23}^2 - p^2 k_3^2\right],\label{Eqn:slepR-_to_slepR-}
\end{align}
where
\begin{align}
m_{13}^2 \equiv (p - k_2)^2 = (k_1 + k_3)^2,\\
m_{23}^2 \equiv (p - k_1)^2 = (k_2 + k_3)^2.
\end{align}
Here, we have neglected the lepton masses, \textit{i.e.} $k_1^2, k_2^2 = 0$, in Eq.~\eqref{Eqn:slepR-_to_slepR-}.

\begin{figure}[t]
\begin{center}
\includegraphics[scale=1.0]{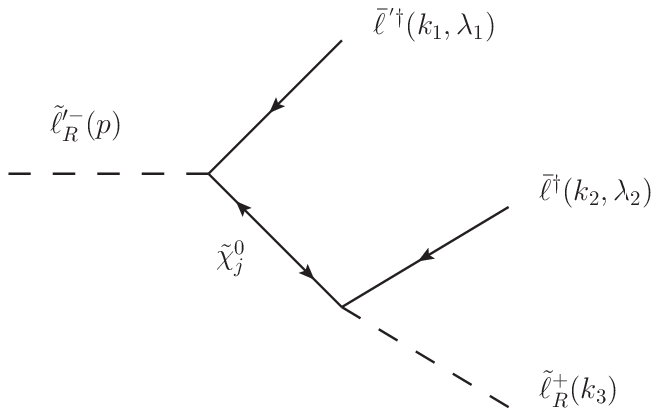}
\end{center}
\caption{Feynman diagram for the 3-body slepton decay $\slepton'^{-}_R \rightarrow \ell'^{-} \ell^-\slepton^+_R$.}
\label{feyn_LRtoLRPplus}
\end{figure}
\begin{figure*}[t]
\begin{center}
\includegraphics[scale=1.0]{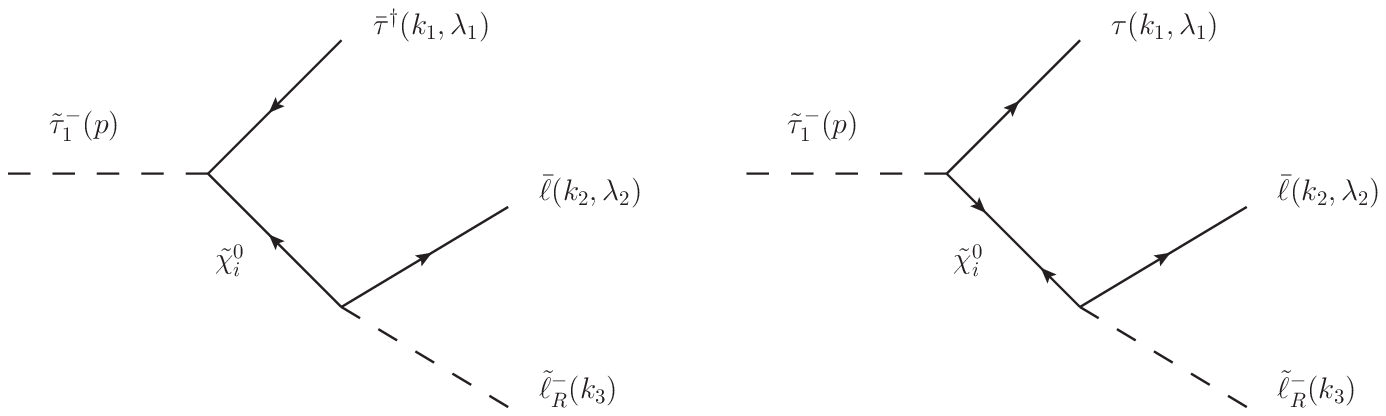}
\end{center}
\caption{Feynman diagrams for the three-body slepton decay $\sstau_1^- \rightarrow \tau^- \ell^+\slepton_R^-$.}
\label{feyn_STtoLR}
\end{figure*}
\begin{figure*}[t]
\begin{center}
\includegraphics[scale=1.0]{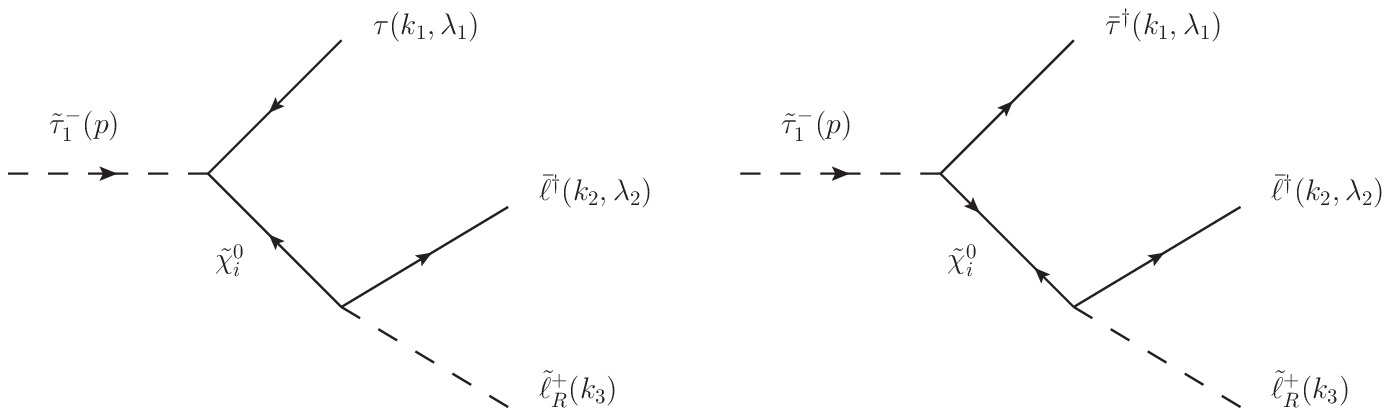}
\end{center}
\caption{Feynman diagrams for the three-body slepton decay $\sstau_1^- \rightarrow \tau^- \ell^-\slepton_R^+$.}
\label{feyn_STtoLRplus}
\end{figure*}

We now turn to the decay $\slepton'^{-}_R \rightarrow \ell'^{-}
\ell^-\slepton_R^+$. The respective Feynman diagram is given in
Fig.~\ref{feyn_LRtoLRPplus}. The amplitude is
\begin{align}
i\mathcal{M} = (-ia_j^*)(-ia_j) \frac{i \mj}{(p-k_1)^2-\mj^2} y_1 y_2\,,
\label{Eq:amplitude2}
\end{align}
which leads to the following expression for the total amplitude squared:
\begin{align}
\left| \mathcal{M}\right|^2 = B\, y_1y_2 y_2^\dagger y_1^\dagger,
\end{align}
with
\begin{align}
B \equiv  \sum_{j,k=1}^4 \frac{|a_j|^2 \mj}{(p-k_1)^2-\mj^2}\cdot\frac{ |a_k|^2\mk }{(p-k_1)^2-\mk^2}\,.
\end{align}
Summing Eq.~(\ref{Eq:amplitude2}) over the spins, we arrive at
\begin{align}
\sum_{\lambda_1, \lambda_2}\left| \mathcal{M}\right|^2= B(-m_{13}^2-m_{23}^2+p^2+k_3^2)\,. \label{Eqn:slepR-_to_slepR+}
\end{align}
Here, the proportionality to the neutralino mass, $m_{\neut_j}$,
in the amplitude, is due to the helicity flip of the
neutralino in Fig.~\ref{feyn_LRtoLRPplus}.

\newpage

\subsection{Three-Body Slepton Decay $\stau_1^-\rightarrow \tau^-\ell^\pm\slepton^\mp$}
\label{Sec:3bodydecay2}

In this section, we calculate the more complicated decays $\sstau_1^-\rightarrow \tau^-\ell^\pm\slepton_R^\mp$.
On the one hand, the $\sstau_1$ is a mixture of the left- and right handed eigenstates. On the other hand, 
we cannot neglect the Yukawa couplings for the third generation.

The Feynman diagrams for the decay $\sstau_1^- \rightarrow \tau^-
\ell^+\slepton_R^-$ are given in Fig. \ref{feyn_STtoLR} and the
respective matrix elements are \cite{Dreiner:2008tw}
\begin{align}
i\mathcal{M}_{\mathrm{I}} &=  (-i a_j^{\sstau})(-ia_j^{\slepton\,*}) \,x_2^\dagger \frac{i (p-k_1)\cdot \bar\sigma}{(p-k_1)^2-\mj^2} y_1,\\
i\mathcal{M}_{\mathrm{II}} &=  (ib_j^{\sstau})(-ia_j^{\slepton\,*}) \frac{i\mj}{(p-k_1)^2-\mj^2}\,x_2^\dagger x_1^\dagger,
\end{align}
with
\begin{align}
a_j^{\slepton} &\equiv \sqrt{2} g' N_{j1}, \\
a_j^{\sstau} &\equiv Y_\tau N_{j3} L_{\sstau_1}^* + \sqrt{2}g' N_{j1}R_{\sstau_1}^*,\\
b_j^{\sstau} &\equiv Y_\tau N_{j3}^* R_{\sstau_1}^* - \frac{1}{\sqrt{2}} (gN_{j2}^* + g'N_{j1}^*)L_{\sstau_1}^*.
\end{align}
The total amplitude squared is
\begin{align}
\left| \mathcal{M}\right|^2 =& \sum_{j,k=1}^4 C_{jk}\Big[
a_j^{\sstau}a_k^{\sstau \,*} x_2^\dagger(p-k_1) \cdot \bar\sigma y_1 y_1^\dagger (p-k_1)\cdot \bar\sigma x_2 \nonumber\\
& -\big[ a_j^{\sstau}b_k^{\sstau \, *}\mk + a_k^{\sstau}b_j^{\sstau \, *}\mj \big] x_2^\dagger(p-k_1)\cdot \bar\sigma y_1 x_1 x_2 \nonumber\\
&+ b_j^{\sstau} b_k^{\sstau \,*} \mj \mk x_2^\dagger x_1^\dagger x_1 x_2\Big],
\end{align}
where
\begin{align}
C_{jk} \equiv \frac{a_j^{\slepton\,*}}{(p-k_1)^2-\mj^2}\cdot \frac{a_k^{\slepton}}{(p-k_1)^2-\mk^2}.
\end{align}

Summing over the spins of the final state leptons, we obtain
\begin{widetext}
\begin{align}
\sum_{\lambda_1, \lambda_2}\left| \mathcal{M}\right|^2 =&\sum_{j,k=1}^4 C_{jk}\Big\{
a_j^{\sstau}a_k^{\sstau \,*} \left[(-m_{23}^2 +p^2 -k_1^2)(-m_{13}^2 +p^2) - (p^2+k_3^2-m_{13}^2 -m_{23}^2)(p^2-k_1^2)\right] \nonumber\\
& -( a_j^{\sstau}b_k^{\sstau \, *}\mk + a_k^{\sstau}b_j^{\sstau \, *}\mj ) m_\tau (m_{23}^2 -k_3^2)+ b_j^{\sstau} b_k^{\sstau \,*} \mj \mk (p^2 + k_3^2 - m_{13}^2 - m_{23}^2) \Big\},\label{2Msquared}
\end{align}
\end{widetext}
where we have neglected the mass of the first or second
generation lepton, \textit{i.e.} $k_2^2 = 0$.

We finally calculate the related decay $\sstau_1^- \rightarrow
\tau^- \ell^-\slepton_R^+$, where the the $\neut_j$ exhibits a
helicity flip, \cf Fig.~\ref{feyn_STtoLRplus}. The matrix elements for
these diagrams are
\begin{align}
i\mathcal{M}_{\mathrm{I}} &=  (-ib_j^{\sstau\,*}) (-i a_j^{\slepton\,*}) \,x_2^\dagger \frac{i (p-k_1)\cdot \bar\sigma}{(p-k_1)^2-\mj^2} y_1,\\
i\mathcal{M}_{\mathrm{II}} &=  (ia_j^{\sstau\,*})(-i a_j^{\slepton\,*}) \frac{i\mj}{(p-k_1)^2-\mj^2}\,x_2^\dagger x_1^\dagger.
\end{align}
The calculation of the squared amplitude is analogous to those for the decay 
$\sstau_1^- \rightarrow \tau^- \ell^+\slepton_R^-$ if one changes the coefficients 
$a_j^{\sstau} \leftrightarrow b_j^{\sstau \,*}$.

\subsection{Resulting Branching Ratios}

\begin{figure*}[t]
\centering
\subfigure[\,Branching ratio for the decay $\ssmu_R^- \to \mu^- e^- \sse_R^+$. The dotted 
and dashed contour lines correspond to the mass difference $M_{\ssmu_R}-M_{\neut_1}$ and 
$M_{\ssmu_R}-M_{\sse_R}$, respectively.\label{Fig:ssmuR_to_sseR}]{
	\begin{minipage}{0.46\linewidth}
	   	\includegraphics[scale=1.0]{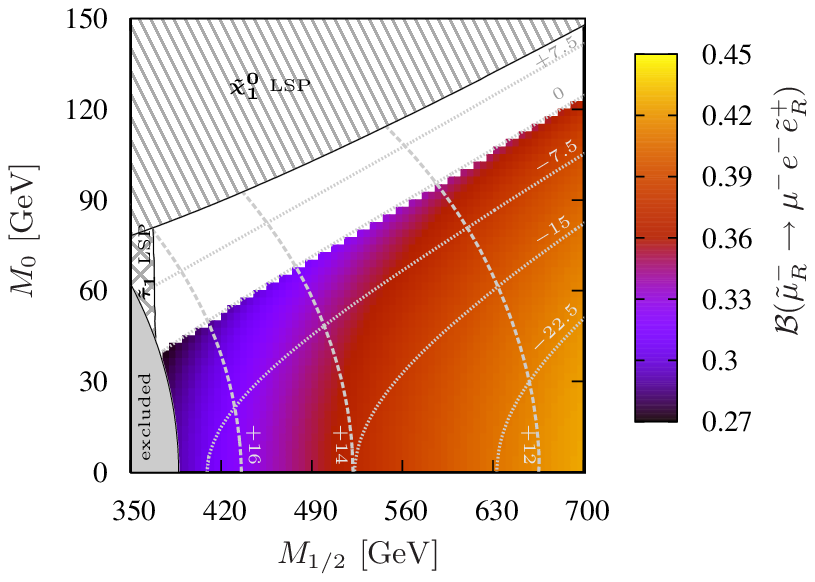}
	\vspace{0.0cm}
	\end{minipage}
}
\quad
\subfigure[\,Branching ratio for the decay  $\ssmu_R^- \to \mu^- \tau^- \sstau_1^+$. The dotted 
and dashed contour lines correspond to the mass difference $M_{\ssmu_R}-M_{\neut_1}$ and 
$M_{\ssmu_R}-M_{\sstau_1}$, respectively.\label{Fig:ssmuR_to_sstau1}]{
	\begin{minipage}{0.46\linewidth}
	   	\includegraphics[scale=1.0]{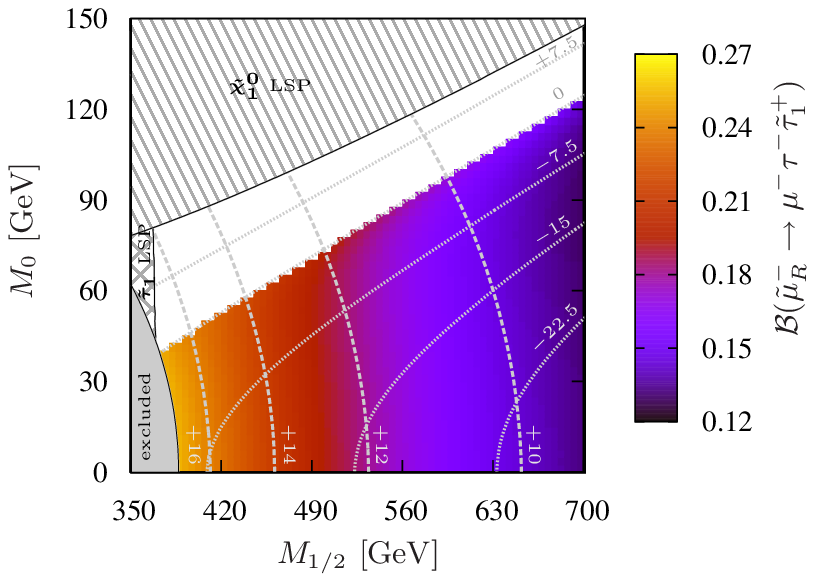}
	\vspace{0.0cm}
	\end{minipage}
}

\subfigure[\,Branching ratio for the decay $\sstau_1^- \to \tau^- e^- \sse_R^+$. 
The dotted and dashed contour lines correspond to the mass difference $M_{\sstau_1}-M_{\neut_1}$ 
and $M_{\sstau_1}-M_{\sse_R}$, respectively.\label{Fig:stau1_to_sseR}]{
	\begin{minipage}{0.46\linewidth}
	   	\includegraphics[scale=1.0]{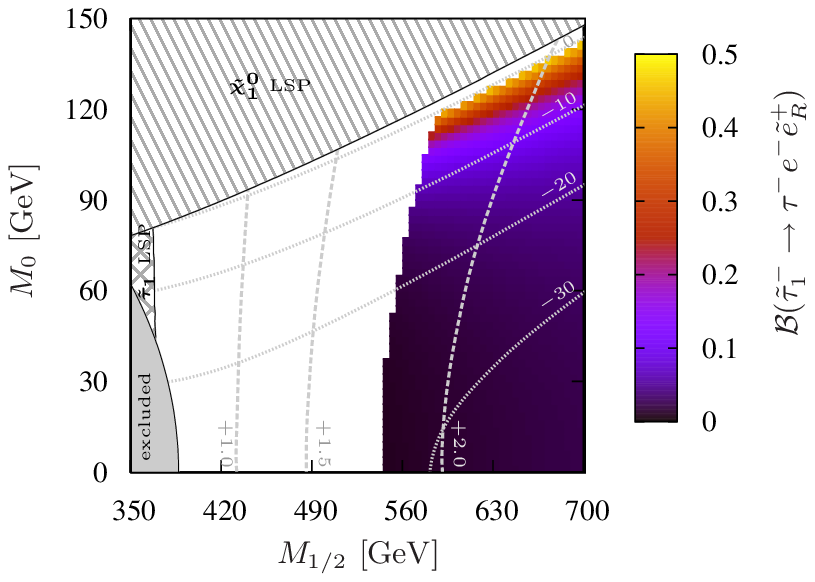}
	\vspace{0.0cm}
	\end{minipage}
}
\caption{Branching ratios for the 3-body slepton decays calculated in Sec.~\ref{Sec:3bodydecay1}
and Sec.~\ref{Sec:3bodydecay2} as a function of $\mhalf$ and $\mzero$. The other $\Bthree$ mSUGRA parameters 
are $\azero = -1250\gev$, $\tanb=5$, $\sgnmu=+$ and $\lam_{231}|_\mathrm{GUT} = 0.045$. In
the white region, the decays are kinematically not allowed or heavily suppressed.}
\label{Fig:threebodydecays}
\end{figure*}

We now briefly study the new 3-body slepton decays for the $\sse_R$ 
LSP parameter space in the $\mhalf-\mzero$ plane. In Fig.~\ref{Fig:threebodydecays} we show 
the same parameter region as for the LHC discovery in Fig.~\ref{Fig:discovery_7TeV}. 
Gray contour lines indicate sparticle mass differences (in GeV) that are relevant for 
the three-body slepton decays; see captions for more details. 

We show in Fig.~\ref{Fig:ssmuR_to_sseR} the branching ratio for the
decay $\ssmu_R^- \to \mu^- e^- \sse_R^+$. The dashed (dotted) gray
contour lines correspond to the mass difference between the $\ssmu_R$
and the $\sse_R$ LSP ($\neut_1$). In the white region, the $\ssmu_R$
is heavier than the $\neut_1$ and decays nearly exclusively via a
2-body decay into the $\neut_1$ and a muon. In the colored region in
Fig.~\ref{Fig:ssmuR_to_sseR}, the $\ssmu_R$ is more than 10 GeV
heavier than the $\sse_R$ LSP. Therefore, there is enough phase-space
for our decay $\ssmu_R^- \to \mu^- e^- \sse_R^+$ at a
significant rate. We observe that the branching ratio increases with
$\mhalf$ and is rather insensitive to $\mzero$. This increase is
due to the competing decay $\ssmu_R^- \to \mu^- \tau^- \sstau_1^+$,
Fig.~\ref{Fig:ssmuR_to_sstau1}, becoming relatively less
important with increasing $M_{1/2}$; see the discussion below.  The
decay $\ssmu_R^-\to \mu^- e^+ \sse_R^-$ behaves similarly to
the decay $\ssmu_R^- \to \mu^- e^- \sse_R^+$, although there are some
small differences due to the different results for the spin--summed
squared matrix element, \cf Eq.~\eqref{Eqn:slepR-_to_slepR-} and
Eq.~\eqref{Eqn:slepR-_to_slepR+}.

The branching ratio of the decay $\ssmu_R^- \to \mu^- \tau^-
\sstau_1^+$ is shown in Fig.~\ref{Fig:ssmuR_to_sstau1}. The decay
$\ssmu_R^- \to \mu^- \tau^+\sstau_1^-$ behaves similarly. The
dashed (dotted) gray contour lines correspond now to the mass
difference between the $\ssmu_R$ and the $\sstau_1$ ($\neut_1$). For
light $\sse_R$ LSP scenarios, \ie at $\mhalf \approx 380\gev$, the
$\ssmu_R$ decays with almost the same rate into the $\sstau_1$ and the
$\sse_R$ LSP, because both particles are nearly degenerate in
mass. However, the branching ratio $\mathcal{B}(\ssmu_R^- \to \mu^-
\tau^-
\sstau_1^+)$ decreases with increasing $\mhalf$, because the
$\sstau_1$ mass increases more rapidly with $M_{1/2}$ than the
$\sse_R$ mass due to the left-handed component of the
$\stau_1$. Therefore, at higher values of $\mhalf$ the $\ssmu_R$
prefers to decay into the $\sse_R$ LSP.

We finally present the branching ratio for the decay $\sstau_1^- \to
\tau^- e^- \sse_R^+$ in Fig.~\ref{Fig:stau1_to_sseR}. The dashed
(dotted) gray contour lines give the mass difference between the
$\sstau_1$ and the $\sse_R$ LSP ($\neut_1$). Since the $\sse_R$ and
$\sstau_1$ are nearly mass degenerate for small $M_{1/2}$, this decay
is only kinematically allowed for higher $\mhalf$ values, \cf the
colored region in Fig.~\ref{Fig:stau1_to_sseR}.  Here, the branching
ratio strongly depends on $\mzero$, \textit{i.e.} it
significantly increases with increasing $M_0$. This is because
there is also the competing $R$-parity violating decay $\sstau_1 \to e
\nu_\mu$ via $\lam_{231}$. Thus, only for scenarios with a low mass
difference between the $\neut_1$ and $\sstau_1$, \textit{i.e.} where
the $\neut_1$ in Fig.~\ref{feyn_STtoLR} and Fig.~\ref{feyn_STtoLRplus}
is nearly on-shell, the 3--body decays $\sstau_1^- \to \tau^- e^-
\sse_R^+$ (and $\sstau_1^- \to \tau^- e^+ \sse_R^-$) become
important.

	

\end{document}